\documentclass[%
 reprint,
 superscriptaddress,
 amsmath,amssymb,
 aps,
prb,
]{revtex4-2}

\usepackage{graphicx} 
\usepackage{dcolumn} 
\usepackage{bm}
\usepackage{hyperref}
\usepackage{tikz}
\usetikzlibrary{calc}

\usepackage[export]{adjustbox}

\usepackage{ellipsis} 

\usepackage{tikz}
\usetikzlibrary{calc}
\newcolumntype{M}[1]{>{\centering\arraybackslash}m{#1}}

\usepackage[final]{changes}

\hypersetup{
    colorlinks=true,
    linkcolor=red,
    filecolor=magenta,      
    urlcolor=cyan,
}
\usepackage{braket}
\usepackage{physics}
\usepackage{chemformula}
\newcommand{\rhocl}{\rho_{\text{cl}}}
\newcommand{\rhow}{\rho_{\text{w}}}
\newcommand{\responsevector}{\boldsymbol{r}}
\newcommand{\perturbationvector}{\boldsymbol{p}}
\newcommand{\Ow}{O_{\text{w}}}

\newcommand{\w}{\omega}

\usepackage{bm}

\newcommand{\bDfour}{\overset{(4)}{\bm D}}
\newcommand{\bDthree}{\overset{(3)}{\bm D}}

\newcommand{\secname}{{Sec.}}
\newcommand{\eqname}{{Eq.}}

\begin{document}

\preprint{APS/123-QED}

\title{Wigner Gaussian dynamics: simulating the anharmonic and quantum ionic motion}
\author{Antonio Siciliano}
\email[]{antonio.siciliano@uniroma1.it}
\affiliation{Dipartimento di Fisica, Università di Roma La Sapienza, Piazzale Aldo Moro 5, 00185 Roma, Italy}
\author{Lorenzo Monacelli}
\affiliation{Theory and Simulation of Materials (THEOS), and National Centre for Computational Design
and Discovery of Novel Materials (MARVEL), École Polytechnique Fédérale de Lausanne, 1015
Lausanne, Switzerland}
\author{Giovanni Caldarelli}
\affiliation{Dipartimento di Fisica, Università di Roma La Sapienza, Piazzale Aldo Moro 5, 00185 Roma, Italy}
\author{Francesco Mauri}
\affiliation{Dipartimento di Fisica, Università di Roma La Sapienza, Piazzale Aldo Moro 5, 00185 Roma, Italy}

\date{\today}

\begin{abstract}
The atomic motion controls important \replaced{properties}{features} of materials, such as thermal transport, phase transitions, and vibrational spectra. However, \replaced{simulating the ionic dynamics }{the simulation of ionic dynamics } is exceptionally challenging when quantum fluctuations are relevant (e.g., at low temperatures or with light atoms) and the energy landscape is anharmonic. In this work, we \replaced{present}{formulate} the Time-Dependent Self-Consistent Harmonic Approximation (TDSCHA) \deleted{\cite{TDSCHA_monacelli,LihmTDSCHA}} [\href{https://journals.aps.org/prb/abstract/10.1103/PhysRevB.103.104305}{L.\ Monacelli and F.\ Mauri, Phys. Rev. B 103, 104305 (2021)}] in the Wigner framework, paving the way for the efficient computation of the nuclear motion in systems with sizable quantum and thermal anharmonic fluctuations. 
Besides the improved numerical efficiency, the Wigner formalism unveils the classical limit of TDSCHA and provides a link with the many-body perturbation theory of Feynman diagrams. 

We further extend the method to account for the non-linear couplings between phonons and photons, responsible, e.g., for a nonvanishing Raman signal in high-symmetry Raman inactive crystals, firstly discussed by Rasetti and Fermi\deleted{\cite{rasetti_raman, fermi_rasetti_ramaneffect}}. We benchmark the method in phase III of high-pressure hydrogen \emph{ab initio}. The nonlinear photon-phonon coupling reshapes the IR spectra and explains the high-frequency shoulder of the \ch{H2} vibron observed in experiments\deleted{\cite{Goncharov_hydrogen}}. 

The Wigner TDSCHA is computationally cheap and derived from first principles\replaced{. It}{: it} is unbiased by assumptions on the phonon-phonon and phonon-photon scattering and does not depend on empirical parameters. Therefore, the method can be adopted in unsupervised high-throughput calculations. 
\end{abstract}

\maketitle


\section{Introduction}
\label{Introduction}
\replaced{The burst of computational resources accomplished during the last decades unlocked}{The progress in computational resources achieved over the last few decades has opened } the path to material design: we can synthesize \emph{in silico} new materials and \emph{measure} their properties before the experimental realization. High-throughput simulations are driving the discovery of new cathode materials for batteries \cite{Hautier_batteries}, superconductor hydrides \cite{Lilia2022}, and 2D materials \cite{exfol2D}, among others.
However, the \replaced{ available toolchain}{toolchain available} for high-throughput simulation fails when anharmonicity is strong. In these cases, the harmonic approximation and perturbative approaches are inadequate. This happens in hydrides where quantum fluctuations alter the free energy landscape \cite{H3S_SSCHA,LH10_SSCHA}, phase diagram of hydrogen-rich compounds like high-pressure ice \cite{Marco_ICE, monacelli2018pressure} \replaced{and}{,} solid hydrogen \cite{phase_diagram_hydrogen, Drummond_QMC_solid_hydrogen}, and materials undergoing displacive phase transition like charge density wave (CDW) transition metal dichalcogenides \cite{CDW_melt_TiSe2,CDW_NbSe2,CDW_NbS2,SSCHA_CDW_VSe2}.

Lattice anharmonicity influences the vibrational spectra, the optical properties, and the conductivity of a crystal \cite{caldarelli2022many}. \replaced{These} {Those} consist in the main experimental signatures of the atomic structure and phase transitions when diffraction is not possible due to small sample size or low cross-section\replaced{. Examples include }{, as happens in} solid hydrogen \cite{black_metal_hydrogen, Loubeyre_hydrogen_metalization_IR}, high-pressure water \cite{Bernasconi_iceVII_X}, and hydrides superconductors \cite{Capitani_H3S}.

Furthermore, accurate simulations of the ionic motion are fundamental in quantum paraelectric perovskites like \ch{KTaO3} \cite{ranalli2022temperature} and \ch{SrTiO3} \cite{verdi2022quantum}, in which the ferroelectric phase transition is hindered by quantum ionic fluctuations. These materials play a crucial role in applications of nonlinear phononics \cite{Spaldin_nonlinearphonics,Cavalleri_nonlinear_phononics}, where the anharmonic coupling between phonons induces transient structural changes and crystal-symmetry breaking upon light pumping \cite{Cavalleri_nonlinear_phononics,Cavalleri_manganite}. Their theoretical and computational investigation has been limited to models assuming specific patterns of phonon-phonon and phonon-photon interactions \cite{distinguish_nonlinear_path,Spaldin_nonlinearphonics}. Moreover, the lack of an unsupervised technique prevents the systematic and high-throughput search for better materials in nonlinear phononics.

The Time-Dependent Self-consistent Harmonic Approximation (TDSCHA) \cite{TDSCHA_monacelli,LihmTDSCHA} \replaced{provides an efficient numerical solution for finite-temperature nuclear dynamics with quantum and anharmonic fluctuations beyond perturbative approaches.}{tackles these complex problems by providing an efficient numerical solution for the finite-temperature nuclear dynamics with quantum and anharmonic fluctuations beyond perturbative approaches.} This method has been successfully applied to simulate Raman and IR spectra of high-pressure molecular hydrogen \cite{black_metal_hydrogen} and to characterize the quantum paraelectric transition in \ch{KTaO3} \cite{ranalli2022temperature}.

TDSCHA approximates the nuclear density matrix with the most general Gaussian. This leads to a more efficient technique than path-integral (PI) methods, where classical-like trajectories of different replicas are sampled \cite{CMD,RPMD,MatsubaraDYnamics,CeottoSemiclassicalSpectroscopic}. 
Instead, it shares some similarities with the Linearized Semi-Classical Initial Value Representation method (LSC-IVR) \cite{PleAnahrmonicTrjectory}, in which an approximate quantum initial condition for the ionic system \cite{BonellaPhaseIntegrationMethod,edgeworth_wigner_langevin,PoulsenWignerdistrSamplin,LocalHarmonic} is evolved subject to classical dynamics.
Among all these methods, TDSCHA is the computationally most efficient for medium-sized systems (containing hundreds of ions)\replaced{.}{with \emph{ab-initio}.}

However, the exceptional complexity of the original TDSCHA formulation hampers its physical interpretations, and many questions remain unanswered\replaced{. For example, }{:} how does it relate to other approaches employed in quantum chemistry \cite{PleAnahrmonicTrjectory}? Which phonon scattering mechanisms is the theory able to describe? What are the limitations of its applicability? And how does the theory behave in the classical limit?

In this work, we reformulate the TDSCHA in the Wigner formalism, where the density matrix is expressed as a function of position and momentum in a quantum-phase space \cite{wigner_original} (Sec.\ \ref{Nuclear dynamics}). 
In this way, the TDSCHA equations are simplified, and the nuclear evolution is governed only by the position-momentum averages and correlators at a fixed time (\secname~\ref{dynamical theory}).  The $\hbar$ vanishes from the propagator\replaced{,}{;} thus, the initial condition determines whether the evolution is quantum or classical, and they share the same computational cost.

The TDSCHA linear response in the Wigner formalism (\secname~\ref{Linear response}) becomes very intuitive and compact. \replaced{The propagator is reformulated }{We reformulate the propagator} in terms of Feynman diagrams, \replaced{establishing a simple connection }{providing a simple link} with other many-body approaches like the Self-Consistent Phonon (SCP) \replaced{technique}{approach} \cite{Tadano_nomodemixing, TadanoSrTiO3_polarizationmixing}. In \secname~\ref{Infrared and Raman spectra}, we extend the formalism \replaced{to allow}{allowing} for the simulation of nonlinear coupling between phonons and photons. Finally, we benchmark the theory on the IR spectrum of high-pressure hydrogen phase III (\secname~\ref{sec:hydrogen}), where \replaced{it is demonstrated }{we show} that the high-frequency overtone observed in \cite{Goncharov_hydrogen} is \replaced{a result of}{due to} the aforementioned nonlinear coupling between photons and phonons.

We start by reviewing the classical and quantum dynamics within the Wigner formalism in the next Section (\ref{Nuclear dynamics}).

\section{Nuclear dynamics}
\label{Nuclear dynamics}
Here, we review the Wigner formalism for the exact nuclear evolution and compare the quantum and classical dynamics of $N$ ions in a closed system. Nuclear spins and ionic exchange are neglected.

\subsection{Classical nuclear evolution}
\label{classical nuclear evolution}
In the classical limit, the nuclei behave like dimensionless particles \replaced{that move}{moving} according to the time-dependent Hamiltonian
\begin{equation}
\label{time dependent classical H}
    H(t) = \sum_{a=1}^{3N} \frac{P^2_a}{2m_a} + V^{(\text{tot})}(\bm R, t).
\end{equation}
For brevity, we indicate with $a=(i,\alpha)$ a composite index with the atomic index $i$ and Cartesian index $\alpha$, $m_a = m_i$ is the mass of atom $i$, $\small P_a=P_{i,\alpha}$ is the momentum of atom $i$ along the $\alpha$ direction  and $\small \bm R = \left(\bm R_1,..,\bm R_N\right)$ represents a configuration of atomic positions. The total potential can be divided into an internal static interaction and an external time-dependent perturbation\begin{equation}
\label{V tot classical definition}
    V^{(\text{tot})}(\bm R, t) = V^{(\text{BO})}(\bm R) + V^{(\text{ext})}(\bm R, t),
\end{equation}
where $\small V^{(\text{BO})}(\bm R)$ is the nuclear interaction mediated by the electrons within the Born-Oppenheimer approximation, i.e.\ the ground state electronic energy at fixed nuclear configurations. It can be evaluated as the total energy of an \emph{ab-initio} calculation like density-functional theory (DFT), or an appropriately parametrized force field. $ \small{V^{(\text{ext})}(\bm R, t)}$ is the external time-dependent potential. It encodes the interaction between the probe (usually an electromagnetic field) and the ions, mediated by the electrons.
 
\replaced{Physical properties}{The dynamics of physical properties} are computed as averages over the phase-space of the corresponding observable with the time-dependent probability distribution (normalized and positive-definite) $\rhocl(\bm R, \bm P, t)$:
\begin{equation}
\label{classical averages}
     \left\langle O \right\rangle_{\rhocl}  = \int d \bm R \int d\bm P O(\bm R,  \bm P) \rhocl(\bm R, \bm P, t).
\end{equation}
In a closed system, $\rhocl(\bm R, \bm P, t)$  evolves according to the Liouville equation
\begin{equation}
\label{Liouville equation classical}
\begin{aligned}
    & \frac{\partial}{\partial t}\rhocl(\bm R, \bm P, t) + i\mathcal{L}^{\text{cl}}\rhocl(\bm R, \bm P, t) = 0.
\end{aligned}
\end{equation}
The classical Liouville operator is defined as 
\begin{equation}
\label{L Lioville classical}
    i\mathcal{L}^{\text{cl}}\circ = -H(t)\overset{\leftrightarrow}{\Lambda} \circ
\end{equation}
and $\overset{\leftrightarrow}{\Lambda}$ is the Poisson brackets operator.
\begin{equation}
\label{Poission brackets def}
    \overset{\leftrightarrow}{\Lambda}= \sum_{a=1}^{3N}\left(\overset{\leftarrow}{\pdv{}{ R_a}} \overset{\rightarrow}{\pdv{}{ P_a}} -  \overset{\leftarrow}{\pdv{}{ P_a}} \overset{\rightarrow}{\pdv{}{ R_a}} \right),
\end{equation}
where the arrows indicate on which side the derivative is applied:
\begin{equation}
    A\overset{\leftrightarrow}{\Lambda} B = \{A,B\}= \sum_{a=1}^{3N}\left(\pdv{A}{ R_a} \pdv{B}{ P_a} - \pdv{A}{ P_a}\pdv{B}{ R_a} \right).
\end{equation}
Eq.\ \eqref{Liouville equation classical} preserves the phase-space volume, which behaves as an incompressible fluid, so probability can not be created nor destroyed leading to entropy conservation.

\subsection{Quantum nuclear evolution}
\label{Wigner formulation}
The Wigner formalism describes the quantum nuclear evolution in terms of position and momentum degrees of freedom, \replaced{in a way analogous}{similarly} to the classical Liouville propagation discussed in \secname~\ref{classical nuclear evolution}. \replaced{Thus}{In this way}, the differences between quantum and classical dynamics are clear.

\replaced{The Wigner quasi-distribution \cite{imre1967wigner}, defined as a Fourier transform of the Von Neumann density operator $\hat{\rho}(t)$, is the quantum analogue of the classical probability distribution in the phase space. It is given by}{The quantum equivalent to the classical probability distribution in the phase-space is the Wigner quasi-distribution \cite{imre1967wigner}, defined as a Fourier transform of the Von Neumann density operator $\hat{\rho}(t)$ as}
\begin{equation}
\small
\label{def rho Wigner}
\begin{aligned}
    & \rhow(\bm R, \bm P, t) \hspace{-0.05cm} = \hspace{-0.08cm}\int\hspace{-0.05cm} \frac{d\bm R'e^{-\frac{i}{\hbar}\bm P\cdot\bm R'}}{(2\pi\hbar)^{3N}} 
    \bra{\bm R + \frac{\bm R'}{2}}\hat{\rho}(t)\ket{\bm R - \frac{\bm R'}{2}} .
\end{aligned}
\end{equation}
We remark that Eq.\ \eqref{def rho Wigner} is normalized, as in the classical case, but, in general, it is not positive-definite \replaced{\cite{negative_wigner_doublewell}}{\cite{wigner_negative_def,negative_wigner_doublewell}}, hence it can not be interpreted as a probability distribution. Still, it encodes all the information on the system.

Similarly, the Wigner expression for an operator $\hat{O}$ is defined as
\begin{equation}
\small
\label{def operator Wigner}
    \Ow(\bm R, \bm P) =  \hspace{-0.05cm}\int\hspace{-0.05cm}d\bm R'e^{-\frac{i}{\hbar}\bm P\cdot\bm R'}
        \bra{\bm R + \frac{\bm R'}{2}}\hat{O}\ket{\bm R - \frac{\bm R'}{2}} ,
\end{equation}
so that quantum averages in the Wigner formalism have the same expression as the classical ones (Eq.\ \eqref{classical averages}):
\begin{equation}
\label{average Wigner}
    \left\langle \Ow \right\rangle_{\rhow}  = \int d\bm R \int d\bm P \Ow(\bm R, \bm P) \rhow(\bm R, \bm P, t).
\end{equation}
The Wigner-Liouville equation controls the time evolution of $\rhow(\bm R, \bm P, t)$
\begin{equation}
\label{Liouville equation for Wigner}
    \frac{\partial}{\partial t} \rhow(\bm R,\bm P, t)
     +  i\mathcal{L}\rhow(\bm R,\bm P,t) =0,
\end{equation}
where $i\mathcal{L}$ is unitary and contains a classical ($\text{cl}$) and a quantum ($\text{q}$) propagator:
\begin{equation}
\label{Wigner Liouville operator quantum}
    i\mathcal{L} = i\mathcal{L}^{\text{cl}} + 
    i\mathcal{L}^{\text{q}}.
\end{equation}
The classical propagator $i\mathcal{L}^{\text{cl}} $ coincides with Eq.\ \eqref{L Lioville classical}. 
On the other hand, the quantum part depends explicitly on $\hbar$:
\begin{equation}
\label{L Wigner quantum}
    \small
    i\mathcal{L}^{\text{q}} \circ = -\sum_{n=1}^{+\infty}\frac{(-\hbar^2)^n}{2^{2n}(2n + 1)!} H(t) \left(\overset{\leftrightarrow}{\Lambda}\right)^{2n + 1} \hspace{-0.1cm}\circ.
\end{equation}
Eq.\ \eqref{L Wigner quantum} gives quantum corrections to the dynamics as odd powers of the Poisson brackets operator.

Interestingly, any quadratic potential has $i\mathcal{L}^{\text{q}} = 0$, even when its coefficients are time-dependent, e.g.
\begin{equation}
\small
\label{general quadratic H time dep}
    H(t) =
    \sum_{a=1}^{3N}\frac{P^2_a}{2m_a} + \frac{1}{2}\sum_{ab=1}^{3N} (R -  R_0(t) )_a K_0(t)_{ab} (R -R_0(t) )_b.
\end{equation}
The density matrix propagates only according to $i\mathcal{L}^{\text{cl}}$, and the dynamic is independent on $\hbar$. The quantum/classical nature of the system is encoded only in the initial condition.

In the next Section, we will show that TDSCHA \cite{TDSCHA_monacelli} shares
the same feature since it is based on a Hamiltonian with the same form of Eq.\ \eqref{general quadratic H time dep}, where $\bm K_0(t)$ and $\bm R_0(t)$ are evaluated self-consistently from the distribution.

\section{Gaussian dynamics in the Wigner framework}
\label{dynamical theory}
The TDSCHA constrains the quantum density matrix as the most general time-dependent Gaussian \cite{TDSCHA_monacelli}. Since the Wigner transformation is equivalent to a Fourier transform, also the nuclear time-dependent Wigner quasi-distribution is a Gaussian
\begin{align} 
\small
    \widetilde{\rho}(\bm R, \bm P, t) \hspace{-0.1cm}=& \mathcal{N}(t)  \exp\biggl[ -\frac{1}{2}\sum_{ab=1}^{3N}(R - \mathcal{R}(t))_a\alpha(t)_{ab}  ( R - \mathcal{R}(t))_b\notag\\
    &\left. -\frac{1}{2}\sum_{ab=1}^{3N}(P- \mathcal{P}(t))_a \beta(t)_{ab} (P- \mathcal{P}(t))_b\notag \right.\\
    & +\sum_{ab=1}^{3N}(R - \mathcal{R}(t))_a  \gamma(t)_{ab} (P- \mathcal{P}(t))_b\biggl]\label{rho(R,P,t) gaussian TDSCHA},
\end{align}  
where $\mathcal{N}(t)$ is the normalization, defined such that
\begin{equation}
    \int d\bm R \int d\bm P \widetilde{\rho}(\bm R, \bm P, t) = 1.
\end{equation}
$\boldsymbol{\mathcal{R}}(t)$, $\boldsymbol{\mathcal{P}}(t)$, $\bm \alpha(t)$, $\bm \beta(t)$, $\bm \gamma(t)$ are the time-dependent free parameters of the distribution. \replaced{In contrast to the general Wigner distribution}{Differently from the general Wigner distribution}, \eqname~\eqref{rho(R,P,t) gaussian TDSCHA} is positive-definite and can be interpreted as the quantum probability distribution. Thanks to the Wigner transformation, it is evident that $\widetilde{\rho}(\bm R, \bm P, t)$ is the multidimensional generalization of the one reported in \cite{PoulsenGaussian} for the 1D case. The position and momentum centroids $\boldsymbol{\mathcal{R}}(t)$ and $\boldsymbol{\mathcal{P}}(t)$ are $3N$ real vectors and represent, respectively, the instantaneous average position and momentum of the ions:
\begin{equation}
\vspace{-0.25cm}
\label{self-consistent centroid dynamic}
    \boldsymbol{\mathcal{R}}(t)= \left\langle\bm R\right\rangle_{\widetilde{\rho}(t)}, \qquad \boldsymbol{\mathcal{P}}(t)= \left\langle\bm P\right\rangle_{\widetilde{\rho}(t)}.
\end{equation}
$\bm\alpha(t)$, $\bm\beta(t)$ and $\bm \gamma(t)$ are  $3N\times3N$ real tensors ($\bm\alpha(t)$ and $\bm\beta(t)$ are symmetric) and represent the instantaneous position-momentum correlators:
\begin{subequations}
\small
\label{correlators R-P}
\begin{align}
      \left\langle \delta \widetilde{\bm R}(t) \delta \widetilde{\bm R}(t) \right\rangle_{\widetilde{\rho}(t)} \hspace{-0.3cm} = & \left(\widetilde{\bm \alpha}(t) - \widetilde{\bm \gamma}(t) \cdot \widetilde{\bm \beta}^{-1}(t) \cdot \widetilde{\bm \gamma}^T(t) \right)^{-1}, \\
      \left\langle \delta \widetilde{\bm R}(t) \delta \widetilde{\bm P}(t) \right\rangle_{\widetilde{\rho}(t)}  \hspace{-0.3cm} = &
      -\left(\widetilde{\bm \gamma}^T(t) - \widetilde{\bm \beta}(t) \cdot \widetilde{\bm \gamma}^{-1}(t) \cdot\widetilde{\bm \alpha}(t) \right)^{-1}, \\
      \left\langle \delta \widetilde{\bm{P}}(t) \delta \widetilde{\bm{P}}(t) \right\rangle_{\widetilde{\rho}(t)}  \hspace{-0.3cm} = &
      \left(\widetilde{\bm \beta}(t) - \widetilde{\bm \gamma}^T(t)\cdot\widetilde{\bm \alpha}^{-1}(t)\cdot \widetilde{\bm \gamma}(t)\right)^{-1}\replaced{,}{.}
\end{align}
\end{subequations}
where \added{ $\delta \widetilde{\bm R}(t) = \widetilde{\bm R} - \widetilde{\boldsymbol{\mathcal{R}}}(t)$,} $\delta \widetilde{\bm P}(t) = \widetilde{\bm P} - \widetilde{\boldsymbol{\mathcal{P}}}(t)$ and the $\widetilde{\circ}$ indicates a mass-rescaled variable like
\begin{equation}
\label{mass rescaled R P}
    \widetilde{R}_a = \sqrt{m_a} R_a, \qquad \widetilde{P}_a = \frac{  P_a}{\sqrt{m_a}} .
\end{equation}
The detailed derivation of \eqname~\eqref{correlators R-P} is reported in \appendixname~\ref{Equations of motion}.

The dynamics of the Wigner distribution can be obtained \replaced{by transforming}{transforming} the time-dependent equation of the TDSCHA in the Wigner basis. We prove in \appendixname~\ref{Equivalence with Time-Dependent Self-consistent Harmonic Approximation} that this is equivalent to evolve \eqname~\eqref{rho(R,P,t) gaussian TDSCHA}
with a self-consistent Wigner-Liouville equation
\begin{equation}
\vspace{-0.25cm}
\label{Liouville equation for Wigner TDSCHA}
    \frac{\partial}{\partial t}\widetilde{\rho}(\bm R,\bm P, t) + 
     i \mathcal{L}^{\text{sc}}\widetilde{\rho}(\bm R,\bm P, t) = 0,
\end{equation}
where
\begin{equation}
\label{TDSCHA Liouvillian}
    i \mathcal{L}^{\text{sc}} \circ = -\mathcal{H}( \widetilde{\rho})\overset{\leftrightarrow}{\Lambda}\circ,
\end{equation}
and $\mathcal{H}(\widetilde{\rho})$ is a quadratic time-dependent Hamiltonian that depends self-consistently on $\widetilde{\rho}(t)$:
\begin{equation}
\label{self consistent H quadratic time dependent}
\begin{aligned}
    \mathcal{H}( \widetilde{\rho}) =&  \sum_{a=1}^{3N} \frac{P^{2}_a}{2m_a} +  \sum_{a=1}^{3N} \delta R(t)_a \left\langle \frac{\partial  V^{(\text{tot})}(\bm R, t)}{\partial R_a} \right\rangle_{\widetilde{\rho}(t)} 
    \\
    & + \frac{1}{2} \sum_{ab=1}^{3N}\delta R(t)_a  \left\langle \frac{\partial^{2} V^{(\text{tot})}(\bm R, t)}{\partial R_a \partial R_b} \right\rangle_{\widetilde{\rho}(t)} \hspace{-0.3cm}\delta R(t)_b \added{.}
\end{aligned}
\end{equation}
\deleted{with $\delta \bm R(t) = \bm R - \boldsymbol{\mathcal{R}}(t)$. }
Inserting the Wigner-TDSCHA distribution, \eqname~\eqref{rho(R,P,t) gaussian TDSCHA}, in \eqname~\eqref{Liouville equation for Wigner TDSCHA} and substituting the expression of the propagators of \eqname~\eqref{correlators R-P} (details in \appendixname~\ref{Equations of motion}), we get the equations of motion
\begin{subequations}
\label{dynamical eqs motion delta P delta R}
\begin{align}
    \frac{d}{d t} \left\langle \widetilde{\bm R} \right\rangle_{\widetilde{\rho}(t)} = &\left\langle  \widetilde{\bm P}\right\rangle_{\widetilde{\rho}(t)} ,\label{delta R  equation}\\
     \frac{d}{d t} \left\langle\widetilde{\bm P} \right\rangle_{\widetilde{\rho}(t)} = &- \left\langle \frac{\partial V^{(\text{tot})}}{\partial \widetilde{  \bm R}}\right\rangle_{\widetilde{\rho}(t)},\label{delta P equation}\\
   \frac{d}{d t}\left\langle \delta \widetilde{\bm R} \delta \widetilde{\bm R} \right\rangle_{\widetilde{\rho}(t)}   =&
    \left\langle \delta \widetilde{\bm R} \delta \widetilde{\bm P} \right\rangle_{\widetilde{\rho}(t)}  + \left\langle \delta \widetilde{\bm P} \delta \widetilde{\bm R} \right\rangle_{\widetilde{\rho}(t)}, \label{delta R delta R equation}\\ 
     \frac{d }{d t}\left\langle \delta \widetilde{\bm P} \delta \widetilde{ \bm P} \right\rangle_{\widetilde{\rho}(t)} = &- \left\langle\frac{\partial^2 V^{(\text{tot})}}{\partial \widetilde{\bm R} \partial \widetilde{\bm R}}\right\rangle_{\widetilde{\rho}(t)} \hspace{-0.3cm}\cdot \left\langle \delta \widetilde{\bm R} \delta \widetilde{\bm P}\right\rangle_{\widetilde{\rho}(t)}\label{delta P delta P equation}\\
     &- \left\langle\delta \widetilde{\bm P} \delta \widetilde{\bm R} \right\rangle_{\widetilde{\rho}(t)}\hspace{-0.3cm} \cdot \left\langle\frac{\partial^2 V^{(\text{tot})}}{\partial \widetilde{\bm R} \partial \widetilde{\bm R}}\right\rangle_{\widetilde{\rho}(t)} , \notag\\
    \frac{d}{d t}\left\langle \delta \widetilde{\bm R} \delta \widetilde{\bm P} \right\rangle_{\widetilde{\rho}(t)}   = &  \left\langle \delta \widetilde{\bm P} \delta \widetilde{\bm P} \right\rangle_{\widetilde{\rho}(t)}   \label{delta R delta P equation}\\
    &- \left\langle \delta \widetilde{\bm R} \delta \widetilde{\bm R} \right\rangle_{\widetilde{\rho}(t)} \hspace{-0.3cm} \cdot
    \left\langle \frac{\partial^2 V^{(\text{tot})}}{\partial \widetilde{\bm R} \partial \widetilde{\bm R}}\right\rangle_{\widetilde{\rho}(t)}, \notag
\end{align}
\end{subequations}
where to compact the notation we drop the explicit time and position dependence. 

The TDSCHA dynamics of \eqname~\eqref{dynamical eqs motion delta P delta R} are simpler than the original derivation using the standard formalism of operators in quantum mechanics.  As anticipated in \secname~\ref{Wigner formulation}, Eqs \eqref{dynamical eqs motion delta P delta R} do not contain $\hbar$, and the dynamic is the same for quantum and classical distribution. Nevertheless, quantum features are included in the initial condition and preserved during the dynamics. 
The classical limit of the TDSCHA dynamics was not evident in \cite{TDSCHA_monacelli}, as the off-diagonal parameters of the density matrix are ill-defined as $\hbar\rightarrow 0$.

This approach is exact in the case of a time-dependent harmonic oscillator since the Wigner distribution is a Gaussian. Thanks to self-consistency, \replaced{this theory goes }{ we go} beyond harmonic/perturbative methods. No approximations are made on the total potential itself so anharmonic effects are included in a non-perturbative way. The dynamics of \eqname~\eqref{dynamical eqs motion delta P delta R} satisfy the conservation of energy and entropy, as expected from a closed system (see \appendixname~\ref{Energy conservation}).

\subsection{Equilibrium SCHA in the Wigner formalism}
A particular solution of \eqname~\eqref{dynamical eqs motion delta P delta R} is the steady state equilibrium in absence of a time-dependent perturbation. From Eqs \eqref{dynamical eqs motion delta P delta R} we get
\begin{subequations}
\label{static solution}
\begin{align}
    \left\langle \pdv{V^{\text{BO}}}{\widetilde{\bm R}}\right\rangle_{(0)} = &\bm 0  \label{no force SCHA}\\
    \left\langle \widetilde{\bm P} \right\rangle_{(0)} =& \bm 0 
    \qquad \left\langle \delta \widetilde{\bm R} \delta\widetilde{\bm P} \right\rangle_{(0)} = \bm 0, \label{no force RP corr}\\
    \left\langle \delta \widetilde{\bm P}  \delta\widetilde{\bm P} \right\rangle_{(0)} =&  
    \left\langle \delta \widetilde{\bm R} \delta \widetilde{\bm R} \right\rangle_{(0)}  \cdot
    \left\langle \frac{\partial^2 V^{(\text{BO})}}{\partial \widetilde{\bm R} \partial \widetilde{\bm R}}\right\rangle_{(0)} , \label{equipartition theorem}
\end{align}
\end{subequations}
where $(0)$ indicates that the averages are performed on $\widetilde{\rho}^{(0)}$, the equilibrium Wigner distribution that solve self-consistently \eqname~\eqref{equipartition theorem}. Since the average momentum is zero (\eqname~\ref{no force RP corr}), we take $\delta{\bm P}= \bm P$. 
As the mixed position-momentum correlation vanishes at equilibrium, the Wigner distribution becomes
\begin{equation}
\small
\label{rho(R,P) gaussian SCHA}
\begin{aligned}
    \widetilde{\rho}^{(0)}(\bm R, \bm P) = & \mathcal{N}^{(0)}
    \exp\biggl[
    -\frac{1}{2} \widetilde{\bm P} \cdot\left\langle  \widetilde{\bm P}  \widetilde{\bm P} \right\rangle_{(0)}^{-1} \cdot \widetilde{ \bm P} \\
    & -\frac{1}{2} \delta \widetilde{\bm R} \cdot\left\langle \delta \widetilde{\bm R} \delta \widetilde{\bm R} \right\rangle_{(0)}^{-1}\cdot \delta \widetilde{\bm R} \biggl]
\end{aligned}
\end{equation}
where $\left\langle  \widetilde{\bm P}  \widetilde{\bm P} \right\rangle_{(0)}$ and $\left\langle \delta \widetilde{\bm R} \delta \widetilde{\bm R} \right\rangle_{(0)}$ solve Eq.\ \eqref{equipartition theorem}. Eq.\ \eqref{rho(R,P) gaussian SCHA} is a steady-state solution of the Wigner TDSCHA equations. Among all \replaced{}{steady-state} distributions, the equilibrium one satisfy
\begin{subequations}
\label{RR PP static}
\begin{align}
    \left\langle \delta \widetilde{R}_a \delta \widetilde{R}_b \right\rangle_{(0)}  =  & \sum_{\mu=1}^{3N} \frac{\hbar(1 + 2n_\mu)}{2\omega_\mu} e^a_\mu e^b_\mu, \\
    \left\langle \widetilde{P}_a \widetilde{P}_b \right\rangle_{(0)}  = & 
    \sum_{\mu=1}^{3N} \frac{\hbar\omega_\mu(1 + 2n_\mu)}{2} e^a_\mu e^b_\mu,
\end{align}
\end{subequations}
where $\{\omega_\mu\}$ and $\{\bm e_\mu\}$ define non-interacting phonons of a generalized dynamical matrix
\begin{equation}
\label{SCHA phonons}
     \overset{(2)}{D}_{ab}= \left\langle \frac{\partial^2 V^{(\text{BO})}}{\partial \widetilde{R}_a \partial \widetilde{R}_b} \right\rangle_{(0)}=\sum_{\mu=1}^{3N}\omega^2_\mu  e^a_\mu e^b_\mu,
\end{equation}
and $n_\mu$ is the Bose-Einstein distribution
\begin{equation}
    n_\mu = \frac{1}{e^{\beta\hbar\omega_\mu} - 1}.
\end{equation}
The equilibrium solution is the one with the minimum free energy at fixed temperature, and in Appendix \ref{Equivalence with Time-Dependent Self-consistent Harmonic Approximation} we show that it coincides with the Wigner transform of the Self-Consistent Harmonic Approximation (SCHA) density matrix \cite{SCHA_main,SCP1,SCPII}. Our initial condition is constrained to be a Gaussian so we do not have to employ a sampling of the Wigner quasi-distribution \cite{edgeworth_wigner_langevin,BonellaPhaseIntegrationMethod} as done in  Linearized Semi-Classical Initial Value Representation methods (LSC-IVR) \cite{linearized_methods_bonella}. These approaches compute time correlation functions approximating the quantum dynamics with a classical evolution (setting $i\mathcal{L}^{\text{q}} = 0$ in Eq.\ \eqref{Wigner Liouville operator quantum}) \cite{PleAnahrmonicTrjectory}, whereas the TDSCHA evolution is justified from the quantum action principle \cite{TDSCHA_monacelli}.

\subsubsection{Diagrammatic representation of the SCHA}
\replaced{We present a diagrammatic expansion of the SCHA in a quartic potential to clarify the anharmonic processes involved in the theory, and to highlight the differences between SCHA, TDSCHA, and other approaches such as Self-Consistent Phonon (SCP) \cite{Tadano_nomodemixing,TadanoSrTiO3_polarizationmixing} and perturbation theory.}{Here, we report the diagrammatic expansion of the SCHA in a quartic potential. This clarifies the anharmonic processes included in the theory, and the differences between SCHA, TDSCHA, and other approaches, such as Self-Consistent Phonon (SCP) \cite{Tadano_nomodemixing}\cite{TadanoSrTiO3_polarizationmixing}  and perturbation theory.}

We define the SCHA propagator through \eqname~\eqref{SCHA phonons}.
\begin{equation}
\label{scha static prop}
    \bm{\mathcal{G}}^{(0)}(\omega)^{-1}=\omega^2 - \overset{(2)}{\bm D}\overset{\omega = 0}{\longrightarrow} - \overset{(2)}{\bm D}.
\end{equation}
Since the SCHA is a static theory and it is defined through $\omega = 0$ quantities. Similarly, one can define the harmonic propagator as
\begin{equation}
\small
\label{static harm prop}
    \bm g^{(0)}(\omega)^{-1} =\omega^2 - \frac{\partial^2 V^{\text{(BO)}}}{\partial \widetilde{ \bm R}\partial \widetilde{ \bm R}} \biggl|_{\bm R = \boldsymbol{\mathcal{R}}_{\text{BO}}}  \hspace{-0.3cm}\overset{\omega = 0}{\longrightarrow} - \frac{\partial^2 V^{\text{(BO)}}}{\partial \widetilde{ \bm R}\partial \widetilde{ \bm R}} \biggl|_{\bm R = \boldsymbol{\mathcal{R}}_{\text{BO}}},
\end{equation}
where
$\boldsymbol{\mathcal{R}}_{\text{BO}}$ is the minimum of $ V^{\text{(BO)}}(\bm R)$
\begin{equation} 
\label{VBO = 0 harmonic}
    \frac{\partial V^{\text{(BO)}}}{\partial \widetilde{\bm R}} \biggl|_{\bm R = \boldsymbol{\mathcal{R}}_{\text{BO}}} = \bm 0,
\end{equation}
and the harmonic phonons are the eigenvalues of the second-order expansion of the BO potential around its minimum $\boldsymbol{\mathcal{R}}_{\text{BO}}$
\begin{equation}
\label{Harmonic phonons}
    \frac{\partial^2 V^{\text{(BO)}}}{\partial \widetilde{ R}_a\partial \widetilde{ R}_b} \biggl|_{\bm R = \boldsymbol{\mathcal{R}}_{\text{BO}}} \hspace{-0.08cm} = \sum_{\mu=1}^{3N} \Omega^2_\mu \epsilon^a_\mu \epsilon^b_\nu.
\end{equation}
In what follows we use $\bm g^{(0)}$ and $\boldsymbol{\mathcal{G}}^{(0)}$ to denote the static limit ($\omega = 0$) of the propagators introduced in Eqs.\ \eqref{scha static prop}, \eqref{static harm prop}.

To connect the SCHA with perturbation theory, we expand the BO energy landscape $V^{\text{(BO)}}(\bm R)$  in a Taylor series around the positions $\bm{\mathcal{R}}_\text{BO}$. All the SCHA equations contain averages of all BO potential derivatives 
\begin{equation}
\label{SCHA perturbative}
\begin{aligned}
    \left\langle \frac{\partial^k V^{(\text{BO})}}{\partial \widetilde{R}_{b_1} ..\partial \widetilde{R}_{b_k}} \right\rangle_{(0)}\hspace{-0.3cm}  & =
    \sum_{n = 0}^\infty \frac{1}{n!}\sum_{a_1..a_n=1}^{3N}\hspace{-0.3cm} \overset{(k + n)}{d}\hspace{-0.2cm}_{b_1 .. b_k a_1..a_{n}} \\
    &\left\langle  
    \left(\delta\widetilde R + \widetilde{\delta}\right)_{a_1}..
    \left(\delta\widetilde R+\widetilde{\delta}\right)_{a_n}\right\rangle_{(0)},
\end{aligned}
\end{equation}
where the anharmonic vertices in Eq.\ \eqref{SCHA perturbative} are evaluated at the minimum of the BO energy landscape
\begin{equation}
\label{anh vertex on harm position}
    \overset{(n)}{d}\hspace{-0.05cm}_{a_1 a_2..a_n} = \frac{\partial^n V^{(\text{BO})}}{\partial \widetilde{R}_{a_1} \partial \widetilde{R}_{a_2}...\partial \widetilde{R}_{a_n} } \biggl|_{\bm R = \boldsymbol{\mathcal{R}}_{\text{BO}}},
\end{equation}
and $\widetilde{\bm \delta}$ is the difference between the minimum of the BO potential and the equilibrium centroids of the SCHA
\begin{equation}
\label{delta definition shift of av position}
    \widetilde{\bm \delta}=  \boldsymbol{\widetilde{\mathcal{R}}}^{(0)}
    -\boldsymbol{\widetilde{\mathcal{R}}}_{\text{BO}} .
\end{equation}
The SCHA distribution is a Gaussian so the averages in \eqname~\eqref{SCHA perturbative} can be evaluated analytically up to any order by means of the Wick theorem.

In a perturbative expansion, we assume that each anharmonic vertex scales as
\begin{equation}
    \overset{(n)}{\bm d}  \sim \mathcal{O}( \lambda^{n-2})
\end{equation}
where $\lambda$ is the perturbative parameter and can be estimated as the ratio of the thermal length and the average bond distance \cite{Bianco, ScatteringNeutronsPhonons}. In what follows we truncate the BO potential to the \replaced{fourth order}{fourth-order}, setting
\begin{equation}
    \overset{(n)}{\bm d} = 0 \quad n\geq 5 .  
\end{equation}
By doing this we get the SCP equations as a limit case of SCHA ones.

Substituting the Taylor expansion into the first SCHA equation, Eq.\ \eqref{no force SCHA}, we obtain a self-consistent equation for $\widetilde{\bm \delta}$ that takes into account quantum and anharmonic effects on the atomic positions shift
\begin{equation}
\begin{aligned}
\label{SCHA eq delta}
    \delta_a  =  &\sum_{bcd=1}^{3N}\frac{g^{(0)}_{ab}}{2}\left[
    \left(\overset{(3)}{d}_{bcd} + \frac{1}{3}\sum_{e=1}^{3N}\overset{(4)}{d}_{bcde}\delta_e\right)\delta_{c}\delta_d \right.\\
    & \left.-\left(\overset{(3)}{d}_{bcd} + \sum_{e=1}^{3N}\overset{(4)}{d}_{bcde}\delta_e \right)
   \mathcal{G}^{(0)}(t=0^-)_{cd} \right],
\end{aligned}
\end{equation}
where, in analogy with many-body theory, $\boldsymbol{\mathcal{G}}^{(0)}(t=0^-)$ is proportional to the SCHA position-position correlator (Eq.\ \eqref{delta R delta R equation})
\begin{equation}
\label{flower definition SCHA}
    \mathcal{G}^{(0)}(t = 0^-)_{ab} = -
    \left\langle
    (\widetilde{R}  - \widetilde{\mathcal{R}}^{(0)})_a(\widetilde{R}  - \widetilde{\mathcal{R}}^{(0)})_b\right\rangle_{(0)},
\end{equation}
here $t=0^-$ is the many-body analytical continuation in time \cite{Mahan2000}.
Similarly, the second SCHA equation (Eq.\ \ref{SCHA phonons}) results in a self-consistent expression for the self-energy
\begin{equation}
\label{SCHA eq propagator}
    {\mathcal{G}^{(0)}}^{-1}_{ab} =  {g^{(0)}}^{-1}_{ab} - \Pi^{(0)}_{ab} ,
\end{equation}
\begin{equation}
\label{SCHA eq PI}
    \Pi^{(0)}_{ab} =  
    \sum_{c=1}^{3N}\overset{(3)}{d}_{abc}\delta_c
    -\sum_{cd=1}^{3N}\frac{\overset{(4)}{d}_{abcd}}{2}\left(\mathcal{G}^{(0)}(t=0^-)_{cd} - \delta_c \delta_d\right).
\end{equation}

Self-consistent phonon (SCP) methods \cite{Tadano_nomodemixing, TadanoSrTiO3_polarizationmixing} solves Eqs \eqref{SCHA eq delta} \eqref{SCHA eq PI} \eqref{SCHA eq propagator} by fitting the anharmonic force constants \cite{Tadano_anhIFC}. Often, \eqname~\eqref{SCHA eq delta} is ignored, and $\widetilde{\bm \delta}$ is assumed to be zero, which is true only if all the atomic coordinates are constrained by symmetry \cite{lazzeri2003anharmonic}. On the contrary, the SCHA can go beyond the SCP method including all the anharmonic vertices with $n\geq 5$ and polarization mixing effects. These effects are automatically incorporated thanks to a stochastic sampling of the potential \cite{SCHA_main}. We remark that the SCHA and SCP method are static theories: the self-energy is real and the phonons defined in Eq.\ \eqref{SCHA eq propagator} are non-interacting excitations with an infinite lifetime.
\begin{figure}[!htb]
    \centering
    \begin{minipage}[c]{1.0\linewidth}
    \includegraphics[width=1.0\textwidth]{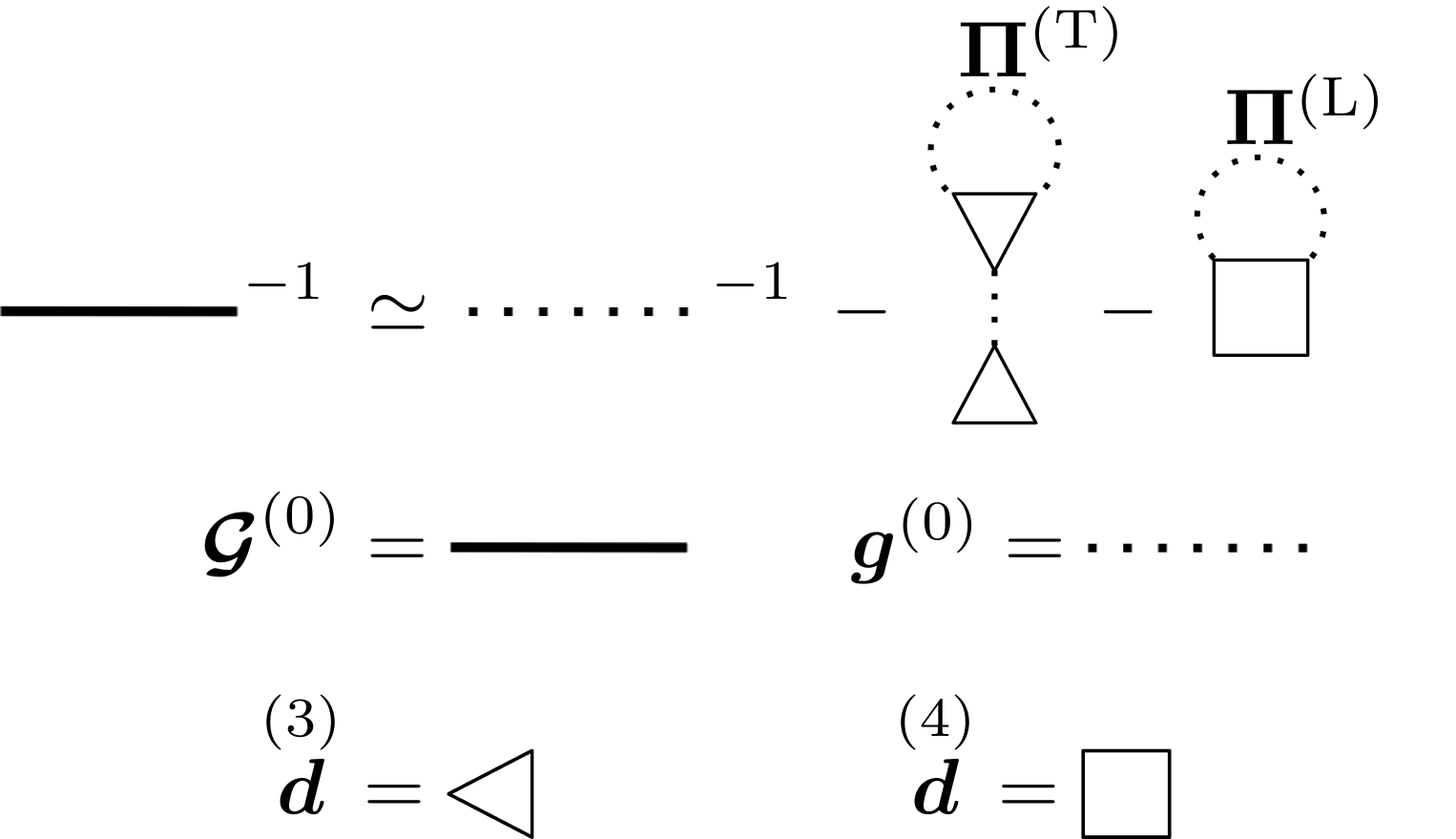}
    \end{minipage}
    \caption{Diagrammatic expression for the SCHA one-phonon propagator at lowest perturbative order $\lambda^2$ (see Ref.\ \cite{Bianco}). The single solid line is the static SCHA propagator $ \boldsymbol{\mathcal{G}}^{(0)}$ Eq.\ \eqref{scha static prop}. The thin dotted line is the static harmonic propagator $\bm g^{(0)}$ Eq.\ \eqref{static harm prop}. The scattering vertices are defined in Eq.\ \eqref{anh vertex on harm position} with $n = 3$ (triangle) and $n = 4 $(square). The tadpole and loop diagram are defined in Eq.\ \eqref{SCHA self-energy}}
    \label{fig:SCHA diagrams}
\end{figure}

Expressing both the SCHA propagator and the position shift as a series of $\mathcal{O}(\lambda^n)$ corrections
\begin{subequations}
\begin{align}
    \boldsymbol{\mathcal{G}}^{(0)} &= {\boldsymbol{\mathcal{G}}^{(0)}}^{\lambda=0} +
    {\boldsymbol{\mathcal{G}}^{(0)}}^{\lambda=1} + .. \hspace{-0.2cm}\quad      {\boldsymbol{\mathcal{G}}^{(0)}}^{\lambda=n} \sim \mathcal{O}(\lambda^n)\\
     \widetilde{\bm \delta}& =  \widetilde{\bm \delta}^{\lambda=0} + \widetilde{\bm \delta}^{\lambda=1} + ..  \quad \widetilde{\bm \delta}^{\lambda=n} \sim \mathcal{O}(\lambda^n)
\end{align}
\end{subequations}
one can solve order by order in $\lambda$ Eqs \eqref{SCHA eq delta} \eqref{SCHA eq PI} \eqref{SCHA eq propagator} to systematically \replaced{obtain}{get} all the corrections to the SCHA propagator with a cubic-quartic potential. Ref.\ \cite{Bianco} solved Eqs.\ \eqref{SCHA eq delta}, \eqref{SCHA eq PI}, \eqref{SCHA eq propagator} up to $\mathcal{O}(\lambda^2)$ and showed that
\begin{equation}
\label{Raffaello expansion for G SCHA}
   \left(\mathcal{G}^{(0)}\right)^{-1}_{ab} \simeq \left( g^{(0)}\right)^{-1}_{ab} -  \left(\Pi^{(\text{L})}_{ab} +  \Pi^{(\text{T})}_{ab}\right)
\end{equation}
where $\bm \Pi^{(\text{L})}$ and $\bm \Pi^{(\text{T})}$ are respectively the loop (L) and tadpole (T) diagram (cf.\ Fig.\ \ref{fig:SCHA diagrams}),
\begin{subequations}
\label{SCHA self-energy}
\begin{align}
     \Pi^{(\text{L})}_{ab} = & -\frac{1}{2}\sum_{cd=1}^{3N}\overset{(4)}{ d}_{abcd} g^{(0)}(t=0^-)_{cd}, \label{loop}\\
     \Pi^{(\text{T})}_{ab}= & -\frac{1}{2}\sum_{cdef=1}^{3N}\overset{(3)}{ d}_{abc}  g^{(0)}_{cd} \overset{(3)}{ d}_{def}    g^{(0)}(t=0^-)_{ef} \label{tadpole} \:,
\end{align}
\end{subequations}
here $\bm g^{(0)}(t=0^-)$ is the harmonic counterpart of $\boldsymbol{\mathcal{G}}^{(0)}(t = 0^-)$, Eq.\ \eqref{flower definition SCHA}. The loop diagram, Eq.\ \eqref{loop}, comes from quantum/anharmonic fluctuations at fixed positions by setting  $\widetilde{\bm \delta}=\bm 0$ in Eq.\ \eqref{SCHA eq PI}. On the other hand, the tadpole diagram, Eq.\ \eqref{tadpole}, comes from the renormalization of atomic positions, Eq.\ \eqref{SCHA eq delta}. 

\section{Linear response}
\label{Linear response}
The static SCHA corrects the bare phonon propagator with a real self-energy, thus only renormalizing the phonon frequency without introducing a finite lifetime of phonons.
In this Section, we revise the fully-dynamical TDSCHA linear response within the Wigner formalism and show how new diagrams with a nonvanishing imaginary part emerge.

\subsection{Linearized equations of motion and general response function}
\label{Linearized equations of motion and general response function}
\replaced{When an external time-dependent potential is coupled to phonons without causing irreversible changes in the material, the system is in the linear response regime.}{When the external time-dependent potential coupled to the phonons does not cause irreversible changes in the material, we are in the linear response regime.} This scenario is relevant, for example, when the ionic degrees of freedom are probed with electromagnetic fields (X-ray or Raman scattering and IR absorption) or with neutrons. In these cases the external perturbation $V^{(\text{ext})}(\bm R, t)$ is in the form
\begin{equation}
\label{V ext definition}
    V^{(\text{ext})}(\bm R, t) = \mathcal{B}(\bm R) \mathcal{V}(t).
\end{equation}
In this regime, the SCHA distribution $\widetilde{\rho}^{(0)}(\bm R)$ (see Appendix \ref{Expansion of the probability distribution}) is perturbed by $\widetilde{\rho}^{(1)}(\bm R, t)$
\begin{equation}
    \label{rho = rho 0 + rho 1}
    \widetilde{\rho}(\bm R, t) = \widetilde{\rho}^{(0)}(\bm R) + \widetilde{\rho}^{(1)}(\bm R, t).
\end{equation}
\replaced{The probability distribution change leads to the emergence of time-dependent correction (denoted by the $(1)$) to the static correlators, Eqs \eqref{correlators R-P},}{As the probability distribution changes, the correlators, Eqs \eqref{correlators R-P}, are the sum of the equilibrium ones, Eqs \eqref{RR PP static}, and a time-dependent correction (denoted by the $(1)$)}
\begin{subequations}
\label{perturbed free parameters}
\begin{align}
    \widetilde{\mathcal{R}}(t)_\mu &= \widetilde{\mathcal{R}}^{(0)}_\mu +\widetilde{\mathcal{R}}^{(1)}_\mu,\label{R 1}\\
     \left\langle \delta \widetilde{R}_\mu \delta \widetilde{R}_\nu\right\rangle_{\widetilde{\rho}(t)} & =  \left\langle \delta \widetilde{R}_\mu \delta \widetilde{R}_\nu \right\rangle_{(0)}+
      \left\langle \delta \widetilde{R}_\mu \delta \widetilde{R}_\nu \right\rangle_{(1)}, \label{delta R delta R 1}\\
     \left\langle \delta \widetilde{P}_\mu \delta \widetilde{P}_\nu\right\rangle_{\widetilde{\rho}(t)} & = \left\langle  \widetilde{P}_\mu  \widetilde{P}_\nu\right\rangle_{(0)} + \left\langle \delta \widetilde{P}_\mu \delta \widetilde{P}_\nu \right\rangle_{(1)} \label{delta P delta P 1},
\end{align}
\end{subequations}
where we express the tensors in the polarization basis $\{\bm e_\mu\}$ of Eq.\ \eqref{SCHA phonons}. In Appendix \ref{Derivation of the linear response system}, we demonstrate that the dynamics of the average momentum $\small\widetilde{\boldsymbol{\mathcal{P}}}^{(1)}$ and of the mixed correlator $\small{\left\langle \delta \widetilde{ \bm R} \delta \widetilde{ \bm P}\right\rangle_{(1)}}$ can be reabsorbed in those of the variables of Eqs \eqref{perturbed free parameters}, which define unambiguously the state of the system.

The linearized equations of motion are obtained by plugging the perturbed correlators, Eqs \eqref{perturbed free parameters}, into the equations of motion, Eq. \eqref{dynamical eqs motion delta P delta R}, and using Eq. \eqref{rho = rho 0 + rho 1} when computing averages of the total potential. This (see Appendix \ref{Derivation of the linear response system} for details) leads to
\begin{equation}
\label{linearized equations of motion}
\begin{aligned}
    &\left(\boldsymbol{\mathcal{L}}' + \omega^2\right) \cdot
    \begin{bmatrix}
    \widetilde{\boldsymbol{\mathcal{R}}}^{(1)} \\
    \left\langle \delta \widetilde{\bm R} \delta \widetilde{\bm R} \right\rangle_{(1)}\\
    \left\langle \delta \widetilde{\bm P} \delta \widetilde{\bm P} \right\rangle_{(1)}
    \end{bmatrix} = 
    \perturbationvector' \mathcal{V}(\omega),
\end{aligned}
\end{equation}
which defines the linearized TDSCHA equations in frequency domain.
In Eq.\ \eqref{linearized equations of motion}, $\omega^2$ comes from the Fourier transform of second-order time derivatives, and $\boldsymbol{\mathcal{L}}'$ is the linearized Wigner-Liouville operator Eq.\ \eqref{TDSCHA Liouvillian}. In general, we can separate $\boldsymbol{\mathcal{L}}'$ in two terms
\begin{equation}
\label{L = L_harm + L_anh}
    \boldsymbol{\mathcal{L}}' = \added{\boldsymbol{\mathcal{L}}'^\dagger =}\boldsymbol{\mathcal{L}}'_{\text{harm}}  + \boldsymbol{\mathcal{L}}'_{\text{anh}}.
\end{equation}
The harmonic part $\boldsymbol{\mathcal{L}}'_{\text{harm}}$ describes the free evolution of the SCHA phonons defined in Eq.\ \eqref{SCHA phonons}. 
The scattering, hence the interaction between these phonons \replaced{arises}{come} from the anharmonic part $\boldsymbol{\mathcal{L}}'_{\text{anh}}$. The perturbation vector $\perturbationvector'$ depends, in general, on equilibrium averages of first and second derivatives of $\mathcal{B}(\bm R)$.

The external potential modifies the non-interacting equilibrium state, defined by Eq.\ \eqref{rho(R,P) gaussian SCHA}. The response function encodes how the observable $\mathcal{A}(\bm R)$ changes when the system is out of equilibrium. In the linear response regime, this modification depends only on equilibrium quantities.

The TDSCHA response function is obtained expanding $ \small{\left\langle \mathcal{A} \right\rangle_{\widetilde{\rho}^{(0)} + \widetilde{\rho}^{(1)}}}$ (see Eq.\ \eqref{rho = rho 0 + rho 1}) in the perturbed parameters of Eqs \eqref{perturbed free parameters} around the equilibrium value $ \small{\left\langle \mathcal{A} \right\rangle_{(0)}}$. In Appendix \ref{Derivation of the linear response system}, we show that the first-order correction in the frequency domain is a simple scalar product in the space of the correlators (Eq.\ \eqref{perturbed free parameters})
\begin{equation}
\label{response code 2 T finite main}
\begin{aligned}
    & \left\langle \mathcal{A} \right\rangle_{(1)}(\omega) = \responsevector'^\dagger \cdot
   \begin{bmatrix}
    \widetilde{\boldsymbol{\mathcal{R}}}^{(1)} \\
    \left\langle \delta \widetilde{\bm R} \delta \widetilde{\bm R} \right\rangle_{(1)}\\
    \left\langle \delta \widetilde{\bm P} \delta \widetilde{\bm P} \right\rangle_{(1)}
    \end{bmatrix},
\end{aligned}
\end{equation}
where the response vector $\responsevector'$, similarly to $\perturbationvector'$, contains equilibrium averages of position-derivatives of $\mathcal{A}(\bm R)$.

Finally, inverting the linearized equations of motion, Eq.\ \eqref{linearized equations of motion}, we get
\begin{equation}
   \left\langle \mathcal{A} \right\rangle_{(1)}(\omega) = \responsevector'^\dagger \cdot \left(\boldsymbol{\mathcal{L}}' + \omega^2\right)^{-1}  \cdot \perturbationvector' \mathcal{V}(\omega),
\end{equation}
where $\circ^{-1}$ denotes the inverse. 
The general response function of an observable $\mathcal{A}(\bm R)$ to an external perturbation $\mathcal{B}(\bm R)$ is
\begin{equation}
\label{general response formula}
    \chi(\omega)_{\mathcal{A},\mathcal{B}} =  \responsevector'^\dagger \cdot \left(\boldsymbol{\mathcal{L}}' + \omega^2\right)^{-1}  \cdot \perturbationvector'.
\end{equation}
This expression has the same form as the one presented in Ref.\ \cite{TDSCHA_monacelli} where the standard description of quantum mechanics is adopted. 

In Appendix \ref{appendix: Lanczos algorithm} we show how to compute the general response function (Eq.\ \eqref{general response formula}) with the Lanczos algorithm following the original work on TDSCHA \cite{TDSCHA_monacelli}. The algorithm generates a basis in which  $\boldsymbol{\mathcal{L}}'+\omega^2$ is tridiagonal. As shown in Appendix \ref{appendix: Lanczos algorithm}, from this form it is easy to get in one shot the response function for all values of $\omega$. The Lanczos basis is generated from $N_{\text{steps}}$ sequential applications of $\boldsymbol{\mathcal{L}}'$ to a starting vector. Each iteration corresponds to free propagations ($\boldsymbol{\mathcal{L}}'_{\text{harm}}$) and scattering ($\boldsymbol{\mathcal{L}}'_{\text{anh}}$). In this way, we build the full anharmonic propagators (see next Section \ref{Interpretation of linear response}). In addition, we \replaced{leverage}{exploit} the properties of the Wigner formulation that $\boldsymbol{\mathcal{L}}'=\boldsymbol{\mathcal{L}}'^\dagger$ and that, if $\mathcal{A} = \mathcal{B}$, $\perturbationvector'=\responsevector'$. \replaced{These features imply that a symmetric Lanczos algorithm is sufficient to compute Eq.\ \eqref{general response formula}, effectively speeding up the original code by a factor of two \cite{TDSCHA_monacelli}.}{These features imply that the calculation of Eq.\ \eqref{general response formula} requires a symmetric Lanczos algorithm speeding up the original code \cite{TDSCHA_monacelli} by a factor of two.}

\begin{table*}[]		
\begin{tabular}{ M{2cm}  c  M{5cm}  M{7cm}  M{1cm}}
\hline\hline
Notation & Diagram & Formula & Description & Eq.\  \\ \hline 
    $\mathcal{G}^{(0)}(\w)_{\mu\nu}$&
			\begin{minipage}{3cm}
				 \centering{
                    \begin{tikzpicture}
                        \node (image) at (0,0) {\includegraphics[width=0.35\textwidth]{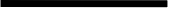}};
                        \node (mu) at ($(image.west)$) {$\scriptstyle\mu$};\node (nu) at ($(image.east)$) {$\scriptstyle\nu$};
                        \end{tikzpicture} }
    		    \end{minipage}
			& $ \frac{\delta_{\mu\nu}}{\omega^2 - \omega^2_\mu}$
			& Bare one-phonon SCHA propagator & \eqref{scha static prop}
			\\
   ${g}^{(0)}(\w)_{\mu\nu}$&
			\begin{minipage}{3cm}
				 \centering{
                    \begin{tikzpicture}
                        \node (image) at (0,0) {\includegraphics[width=0.35\textwidth]{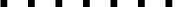}};
                        \node (mu) at ($(image.west)$) {$\scriptstyle\mu$};\node (nu) at ($(image.east)$) {$\scriptstyle\nu$};
                        \end{tikzpicture} }
    		    \end{minipage}
			& $ \frac{\delta_{\mu\nu}}{\omega^2 - \Omega^2_\mu}$
			& Bare perturbative one-phonon  propagator&\eqref{static harm prop}
			\\
			$\chi^{(0)}(\w)_{\mu\nu\eta\lambda}$&
			\begin{minipage}{3cm}
				 \centering{
                    \begin{tikzpicture}
                    \node (image) at (0,0) {\includegraphics[width=0.45\textwidth]{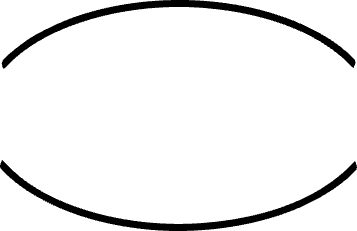}};
                    \node (mu) at ($0.7*(image.north west)+ 0.3*(image.south west) $)    {$\scriptstyle\mu$}; \node (nu) at ($0.3*(image.north west)+ 0.7*(image.south west) $) {$\scriptstyle\nu$};
                    \node (eta) at ($0.7*(image.north east)+ 0.3*(image.south east) $) {$\scriptstyle\eta$}; \node (lambda) at ($0.3*(image.north east)+ 0.7*(image.south east) $) {$\scriptstyle\lambda$};
                    \end{tikzpicture}}
		    \end{minipage}
			& $\chi^{(0)}_-(\w)_{\mu\nu\eta\lambda} - \chi^{(0)}_+(\w)_{\mu\nu\eta\lambda}$
			& Bare two-phonon SCHA propagator&\eqref{2 ph free propagator}
			\\
   $\chi^{(0)}_-(\w)_{\mu\nu\eta\lambda}$&
			\begin{minipage}{3cm}
				 \centering{
                    \begin{tikzpicture}
                    \node (image) at (0,0) {\includegraphics[width=0.45\textwidth]{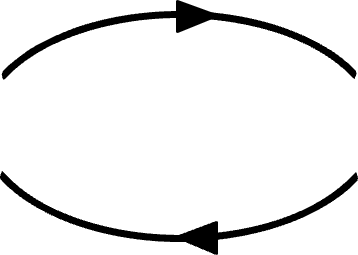}};
                    \node (mu) at ($0.7*(image.north west)+ 0.3*(image.south west) $)    {$\scriptstyle\mu$}; \node (nu) at ($0.3*(image.north west)+ 0.7*(image.south west) $) {$\scriptstyle\nu$};
                    \node (eta) at ($0.7*(image.north east)+ 0.3*(image.south east) $) {$\scriptstyle\eta$}; \node (lambda) at ($0.3*(image.north east)+ 0.7*(image.south east) $) {$\scriptstyle\lambda$};
                    \end{tikzpicture}}
		    \end{minipage}
			& $\delta_{\mu\eta}\delta_{\nu\lambda} \frac{\hbar\left[\omega_\mu - \omega_\nu\right]\left[n_\mu - n_\nu\right]}{4\omega_\mu\omega_\nu[(\omega_\mu -\omega_\nu)^2 - \omega^2]}$
			& Resonant two-phonon SCHA propagator&\eqref{chi - main}
			\\
   $\chi^{(0)}_+(\w)_{\mu\nu\eta\lambda}$&
			\begin{minipage}{3cm}
				 \centering{
                    \begin{tikzpicture}
                    \node (image) at (0,0) {\includegraphics[width=0.45\textwidth]{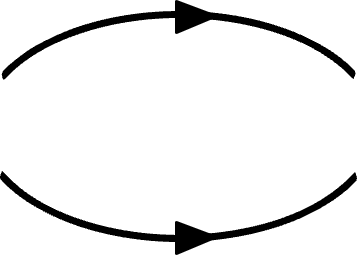}};
                    \node (mu) at ($0.7*(image.north west)+ 0.3*(image.south west) $)    {$\scriptstyle\mu$}; \node (nu) at ($0.3*(image.north west)+ 0.7*(image.south west) $) {$\scriptstyle\nu$};
                    \node (eta) at ($0.7*(image.north east)+ 0.3*(image.south east) $) {$\scriptstyle\eta$}; \node (lambda) at ($0.3*(image.north east)+ 0.7*(image.south east) $) {$\scriptstyle\lambda$};
                    \end{tikzpicture}}
		    \end{minipage}
			& $\delta_{\mu\eta}\delta_{\nu\lambda}
    \frac{\hbar\left[\omega_\mu + \omega_\nu\right]\left[1+n_\mu + n_\nu\right]}{4\omega_\mu\omega_\nu [(\omega_\mu + \omega_\nu)^2 - \omega^2]}$
			& Antiresonant two-phonon SCHA propagator&\eqref{chi + main}
			\\
			 $\overset{(4)}{D}_{\mu\nu\eta\lambda}$&
			\begin{minipage}{3cm}\centering{
			    \begin{tikzpicture}
                        \node (image) at (0,0) {\includegraphics[width=0.3\textwidth]{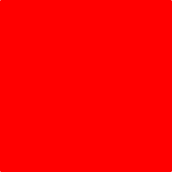}};
                        \node (mu) at ($0.93*(image.north west)$) {$\scriptstyle\mu$};\node (nu) at ($0.93*(image.south west)$) {$\scriptstyle\nu$};
                        \node (eta) at ($0.93*(image.north east)$) {$\scriptstyle\eta$};\node (lambda) at ($0.93*(image.south east)$){$\scriptstyle\lambda$};
                    \end{tikzpicture}  }
                \end{minipage}
			& $\left\langle \frac{\partial^4 V^{(\text{BO})}}{\partial \widetilde{R}_\mu \partial \widetilde{R}_\nu \partial \widetilde{R}_\eta \partial \widetilde{R}_\lambda} \right\rangle_{(0)}$
			& Four-phonon SCHA scattering vertex&\eqref{def D4 main}
			\\ 
   $\overset{(4)}{D}_{\mu\nu\eta\lambda} \hspace{-.7
   cm}\textcolor{white}{,}^{(0)}$&
			\begin{minipage}{3cm}\centering{
			    \begin{tikzpicture}
                        \node (image) at (0,0) {\includegraphics[width=0.3\textwidth]{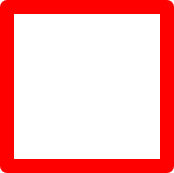}};
                        \node (mu) at ($0.93*(image.north west)$) {$\scriptstyle\mu$};\node (nu) at ($0.93*(image.south west)$) {$\scriptstyle\nu$};
                        \node (eta) at ($0.93*(image.north east)$) {$\scriptstyle\eta$};\node (lambda) at ($0.93*(image.south east)$){$\scriptstyle\lambda$};
                    \end{tikzpicture}  }
                \end{minipage}
			& $\frac{\partial^{4} V^{\text{(BO)}}}{\partial \widetilde{R}_{\mu}\partial \widetilde{R}_{\nu}\partial \widetilde{R}_{\eta} \partial \widetilde{R}_{\lambda}}\biggl|_{\bm R =\boldsymbol{\mathcal{R}}^{(0)}} $
			& Perturbative four-phonon scattering vertex evaluated at the SCHA equilibrium positions &\eqref{bare anharmonic tensor}
			\\ 
   $\overset{(4)}{d}_{\mu\nu\eta\lambda}$&
			\begin{minipage}{3cm}\centering{
			    \begin{tikzpicture}
                        \node (image) at (0,0) {\includegraphics[width=0.3\textwidth]{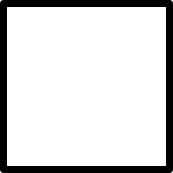}};
                        \node (mu) at ($0.93*(image.north west)$) {$\scriptstyle\mu$};\node (nu) at ($0.93*(image.south west)$) {$\scriptstyle\nu$};
                        \node (eta) at ($0.93*(image.north east)$) {$\scriptstyle\eta$};\node (lambda) at ($0.93*(image.south east)$){$\scriptstyle\lambda$};
                    \end{tikzpicture}  }
                \end{minipage}
			& $\frac{\partial^{4} V^{\text{(BO)}}}{\partial \widetilde{R}_{\mu}\partial \widetilde{R}_{\nu}\partial \widetilde{R}_{\eta} \partial \widetilde{R}_{\lambda}}\biggl|_{\bm R =\boldsymbol{\mathcal{R}}_{\text{BO}}} $
			& Perturbative four-phonon scattering vertex &\eqref{anh vertex on harm position}
			\\ 
			 $\overset{(3)}{D}_{\mu\nu\eta}$&
			\begin{minipage}{3cm}\centering{
			  \begin{tikzpicture}
                    \node (image) at (0,0) {\includegraphics[width=0.3\textwidth]{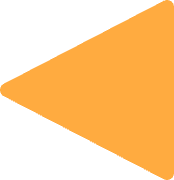}};
                    \node (mu) at($(image.west)$) {$\scriptstyle\mu$};\node (nu) at ($0.93*(image.north east)$) {$\scriptstyle\nu$};
                    \node (eta) at ($0.93*(image.south east)$) {$\scriptstyle\eta$};
                    \end{tikzpicture}   }
                \end{minipage}
			& $\left\langle \frac{\partial^3 V^{(\text{BO})}}{\partial \widetilde{ R}_\mu \partial \widetilde{ R}_\nu \partial \widetilde{ R}_\eta} \right\rangle_{(0)}$
			& Three-phonon SCHA scattering vertex   &\eqref{def D3 main}
   \\
			 $\overset{(3)}{D}_{\mu\nu\eta}\hspace{-.55
   cm}\textcolor{white}{,}^{(0)}$&
			\begin{minipage}{3cm}\centering{
			  \begin{tikzpicture}
                    \node (image) at (0,0) {\includegraphics[width=0.3\textwidth]{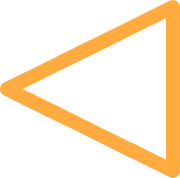}};
                    \node (mu) at($(image.west)$) {$\scriptstyle\mu$};\node (nu) at ($0.93*(image.north east)$) {$\scriptstyle\nu$};
                    \node (eta) at ($0.93*(image.south east)$) {$\scriptstyle\eta$};
                    \end{tikzpicture}   }
                \end{minipage}
			& $ \frac{\partial^3 V^{(\text{BO})}}{\partial \widetilde{ R}_\mu \partial \widetilde{ R}_\nu \partial \widetilde{ R}_\eta}\biggl|_{\bm R =\boldsymbol{\mathcal{R}}^{(0)}} $
			& Perturbative three-phonon scattering vertex evaluated at SCHA equilibrium positions &\eqref{bare anharmonic tensor}
   \\
			 $\overset{(3)}{d}_{\mu\nu\eta}$&
			\begin{minipage}{3cm}\centering{
			  \begin{tikzpicture}
                    \node (image) at (0,0) {\includegraphics[width=0.3\textwidth]{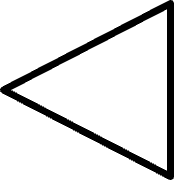}};
                    \node (mu) at($(image.west)$) {$\scriptstyle\mu$};\node (nu) at ($0.93*(image.north east)$) {$\scriptstyle\nu$};
                    \node (eta) at ($0.93*(image.south east)$) {$\scriptstyle\eta$};
                    \end{tikzpicture}   }
                \end{minipage}
			& $ \frac{\partial^3 V^{(\text{BO})}}{\partial \widetilde{ R}_\mu \partial \widetilde{ R}_\nu \partial \widetilde{ R}_\eta}\biggl|_{\bm R =\boldsymbol{\mathcal{R}}_{\text{BO}}} $
			& Perturbative three-phonon scattering vertex &\eqref{anh vertex on harm position}
   \\
          \hline
	   \end{tabular}
    \caption{Collection of symbols frequently used in the main text. First column: the notation used. Second column: graphical expression with indexes. Third column: mathematical definition. Fourth column: description of the symbol. Fifth column: first labeled equation where the symbol appears. }
    \label{Tab:summary}
\end{table*}

\subsection{Diagrammatic interpretation of linear response}
\label{Interpretation of linear response}
In this Section, we provide \replaced{an}{a physical} interpretation in terms of Feynman diagrams of Eq.\ \eqref{general response formula}.
To do this, we introduce a new basis (see Appendix \ref{Symbolic inversion} and Appendix \ref{Derivation of the interacting Green function} for details), a linear combination of the position and momentum correlators, Eqs \eqref{delta R delta R 1} \eqref{delta P delta P 1}. In this basis, the general response function (Eq. \eqref{general response formula}) takes the form
\begin{equation}
\label{response formula for calculations}
    \chi(\omega)_{\mathcal{A},\mathcal{B}} = \responsevector \cdot \boldsymbol{\mathcal{L}}(\omega)^{-1} \cdot \perturbationvector,
\end{equation}
where the response vector $\responsevector$ and the perturbation vector $\perturbationvector$ have a simple expression in terms of equilibrium averages of $\mathcal{A}(\bm R)$ and $\mathcal{B}(\bm R)$
\begin{equation}
\begin{aligned}
\label{perturbation response vector}
    &  \perturbationvector= 
    \begin{bmatrix}
    \left\langle \frac{\partial \mathcal{B}}{\partial \widetilde{R}_\mu}\right\rangle_{(0)}\\
    \left\langle \frac{\partial^2 \mathcal{B}}{\partial \widetilde{R}_\mu \partial \widetilde{R}_\nu}\right\rangle_{(0)} \\
     \left\langle \frac{\partial^2 \mathcal{B}}{\partial \widetilde{R}_\mu \partial \widetilde{R}_\nu}\right\rangle_{(0)}
    \end{bmatrix}, \qquad
    \responsevector=
    \begin{bmatrix}
    \left\langle \frac{\partial \mathcal{A}}{\partial \widetilde{R}_\mu}\right\rangle_{(0)}\\
    \left\langle \frac{\partial^2 \mathcal{A}}{\partial \widetilde{R}_\mu \partial \widetilde{R}_\nu}\right\rangle_{(0)}  \\
     \left\langle \frac{\partial^2 \mathcal{A}}{\partial \widetilde{R}_\mu \partial \widetilde{R}_\nu}\right\rangle_{(0)} 
    \end{bmatrix}.
\end{aligned}
\end{equation}

$\boldsymbol{\mathcal{L}}(\omega)$ propagates the perturbation caused by $\bm p$ in the system and encodes the information on how the latter affects $\bm r$ which defines how the observable changes. Indeed, its multiplication by a vector representing the status of the system, as $\bm r$ or $\bm p$ (\eqname~\ref{perturbation response vector}), gives the anharmonic scattering processes that further dress the SCHA Green's function \replaced{introducing}{and introduce} a finite lifetime (see Section \ref{Green functions}).

\begin{figure}[!htb]
    \centering
    \begin{minipage}[c]{1.0\linewidth}
    \includegraphics[width=1.0\textwidth]{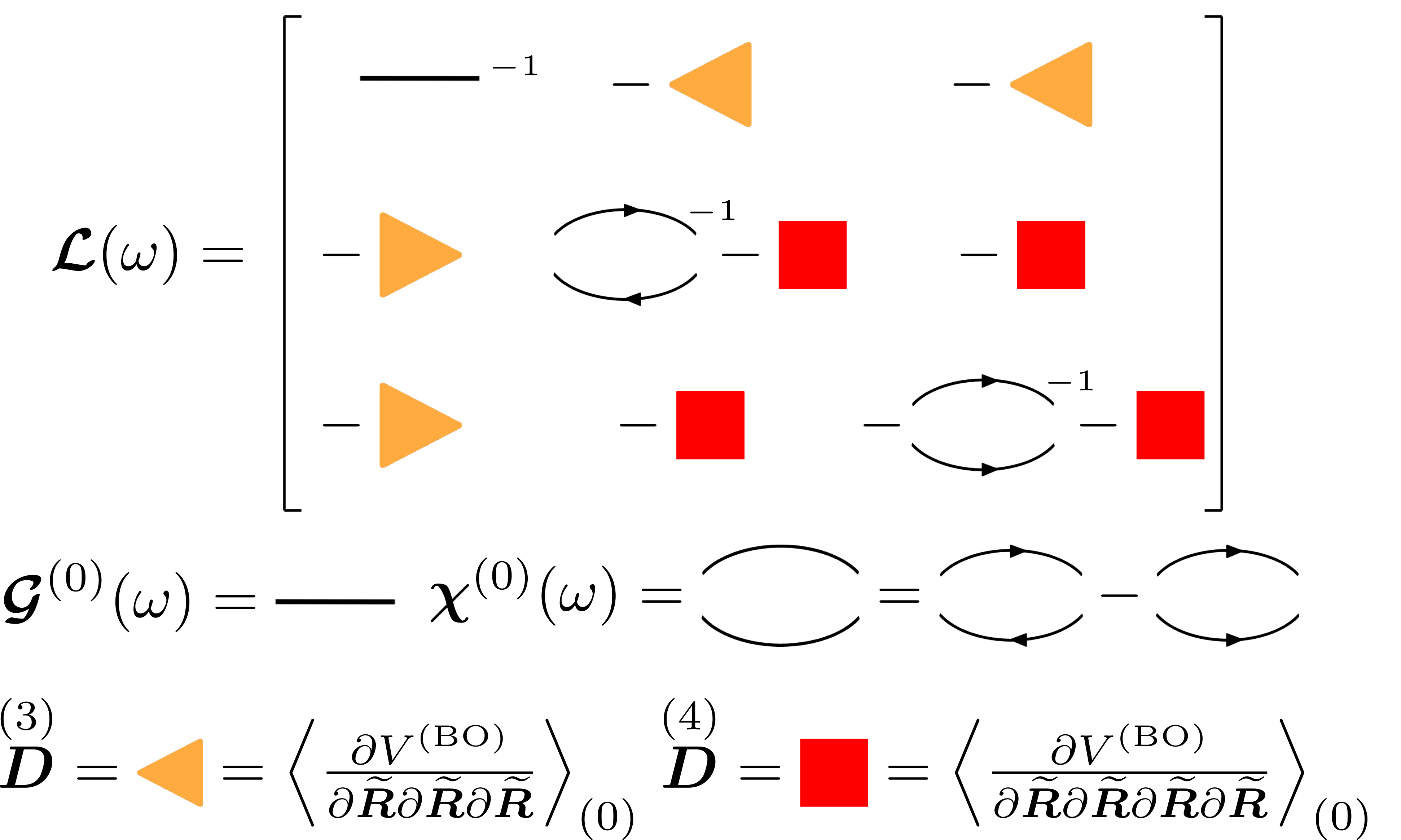}
    \end{minipage}
    \caption{Graphical expression for Eq. \eqref{L + omega2 easy to understand} in terms of the free SCHA propagators (single and double solid line Eqs \eqref{1 ph free propagator} \eqref{2 ph free propagator}) and the third and fourth order scattering vertex (orange triangle and red square defined in Eqs \eqref{def D3 main} \eqref{def D4 main}).}
    \label{fig:L operator}
\end{figure}

In this basis, $\boldsymbol{\mathcal{L}}(\omega)$ of Eq.\ \eqref{response formula for calculations} has a simple symmetric form
\begin{equation}
\label{L + omega2 easy to understand}
    \boldsymbol{\mathcal{L}}(\omega) \hspace{-0.1cm}= \hspace{-0.1cm}\begin{bmatrix}
    \boldsymbol{\mathcal{G}}^{(0)}(\omega)^{-1} & -\overset{(3)}{\bm D} &
    -\overset{(3)}{\bm D} \\ 
    -\overset{(3)}{\bm D}  &
     \bm \chi^{(0)}_{-}(\omega)^{-1} -\overset{(4)}{\bm D}   & 
    -\overset{(4)}{\bm D} \\ 
    -\overset{(3)}{\bm D} 
    & -\overset{(4)}{\bm D} 
    &   -\bm \chi^{(0)}_{+}(\omega)^{-1}-\overset{(4)}{\bm D}
    \end{bmatrix},
\end{equation} 
\begin{equation}
    \boldsymbol{\mathcal{L}}(\omega) = \boldsymbol{\mathcal{L}}(\omega)^\dagger.
\end{equation}
The harmonic and anharmonic contributions to $\boldsymbol{\mathcal{L}}(\omega)$ are
\begin{equation}
\label{L + omega2 easy to understand harm}
    \boldsymbol{\mathcal{L}}_{\text{harm}}(\omega) \hspace{-0.1cm}= \hspace{-0.1cm}\begin{bmatrix}
    \boldsymbol{\mathcal{G}}^{(0)}(\omega)^{-1} & \bm 0 & \bm 0 \\ 
    \bm 0  & \bm \chi^{(0)}_{-}(\omega)^{-1}    & \bm 0 \\ 
    \bm 0  & \bm 0  &   -\bm \chi^{(0)}_{+}(\omega)^{-1}
    \end{bmatrix},
\end{equation} 
\begin{equation}
\label{L + omega2 easy to understand anh}
    \boldsymbol{\mathcal{L}}_{\text{anh}}(\omega) \hspace{-0.1cm}= \hspace{-0.1cm}\begin{bmatrix}
    \bm 0 & -\overset{(3)}{\bm D} &
    -\overset{(3)}{\bm D} \\ 
    -\overset{(3)}{\bm D}  &
      -\overset{(4)}{\bm D}   & 
    -\overset{(4)}{\bm D} \\ 
    -\overset{(3)}{\bm D} 
    & -\overset{(4)}{\bm D} 
    &   -\overset{(4)}{\bm D}
    \end{bmatrix}.
\end{equation} 
We report in Fig.\ \ref{fig:L operator} a graphical expression for Eq.\ \eqref{L + omega2 easy to understand}. To construct a diagrammatic representation, we associate each tensor in $\boldsymbol{\mathcal{L}}(\omega)$ with a symbol that possesses a number of extremities equal to the rank of the tensor.

The single solid line in Fig.\ \ref{fig:L operator} represents the equilibrium SCHA Green's function $\boldsymbol{\mathcal{G}}^{(0)}(\omega)$
\begin{equation}
\label{1 ph free propagator}
    \mathcal{G}^{(0)}(\omega)_{\mu\nu} = 
    \frac{\delta_{\mu\nu}}{\omega^2 - \omega^2_\mu}, 
\end{equation}
where $\{\omega_\mu\}$ are the self-consistent auxiliary frequencies defined in Eq.\ \eqref{SCHA phonons}.
The double single solid line in Fig.\ \ref{fig:L operator} is the two-phonon SCHA propagator $\bm \chi^{(0)}(\omega)$
\begin{equation}
    \label{2 ph free propagator}
    \chi^{(0)}(\omega)_{\mu\nu\eta\lambda} = \chi^{(0)}_{-}(\omega)_{\mu\nu\eta\lambda} -\chi^{(0)}_{+}(\omega)_{\mu\nu\eta\lambda} 
\end{equation}
which contains a resonant  $\bm \chi^{(0)}_-(\omega)$ and an anti resonant term $\bm \chi^{(0)}_+(\omega)$
\begin{subequations}
\label{chi + - main}
\begin{align}
   \chi^{(0)}_{-}(\omega)_{\mu\nu\eta\lambda} =& \delta_{\mu\eta}\delta_{\nu\lambda} \frac{\hbar\left[\omega_\mu - \omega_\nu\right]\left[n_\mu - n_\nu\right]}{4\omega_\mu\omega_\nu[(\omega_\mu -\omega_\nu)^2 - \omega^2]} ,\label{chi - main}\\
    \chi^{(0)}_{+}(\omega)_{\mu\nu\eta\lambda}  =&  \delta_{\mu\eta}\delta_{\nu\lambda}
    \frac{\hbar\left[\omega_\mu + \omega_\nu\right]\left[1+n_\mu + n_\nu\right]}{4\omega_\mu\omega_\nu [(\omega_\mu + \omega_\nu)^2 - \omega^2]} .\label{chi + main}
\end{align}
\end{subequations}
The anti resonant part $\bm \chi^{(0)}_+(\omega)$  describes the absorption/emission processes of a phonon pair (double solid line with both arrows in the same directions in Fig.\ \ref{fig:L operator}) while the 
resonant $\bm \chi^{(0)}_-(\omega)$  the case in which one phonon is absorbed and the other one is emitted (double solid line with arrows pointing in  opposite directions in Fig.\ \ref{fig:L operator}).

The first row of Eq.\ \eqref{L + omega2 easy to understand} relates the propagation of single phonon excitation  $\boldsymbol{\mathcal{G}}^{(0)}(\omega)$ to two-phonon processes $\bm \chi^{(0)}_{\pm}(\omega)$. This is mediated by the three-phonon scattering vertex (orange triangle of Fig.\ \ref{fig:L operator})
\begin{equation}
\label{def D3 main}
    \overset{(3)}{ D}_{\mu\nu\eta} = \left\langle \frac{\partial^3 V^{(\text{BO})}}{\partial \widetilde{ R}_\mu \partial \widetilde{ R}_\nu \partial \widetilde{ R}_\eta} \right\rangle_{(0)}\hspace{-0.2cm}.
\end{equation}
The second and third rows of Eq.\ \eqref{L + omega2 easy to understand} show that double excitations interact with each other via the fourth-order scattering vertex (red square of Fig.\ \ref{fig:L operator})
\begin{equation}
\label{def D4 main}
    \overset{(4)}{D}_{\mu\nu\eta\lambda} = \left\langle \frac{\partial^4 V^{(\text{BO})}}{\partial \widetilde{R}_\mu \partial \widetilde{R}_\nu \partial \widetilde{R}_\eta \partial \widetilde{R}_\lambda} \right\rangle_{(0)}\hspace{-0.2cm},
\end{equation}
or can decay in a single phonon via the three-phonon scattering vertex, Eq.\ \eqref{def D3 main}. As a guide for the reader, in Table \ref{Tab:summary} we report a summary of all the symbols used.

So, by just looking at the expression of $\boldsymbol{\mathcal{L}}(\omega)$, we understand that in TDSCHA only single and double excitation are dressed by anharmonicity and that only single phonons can decay in a higher-order phonon propagation. 

\begin{figure}[!htb]
    \centering
    \begin{minipage}[c]{1.0\linewidth}
    \includegraphics[width=1.0\textwidth]{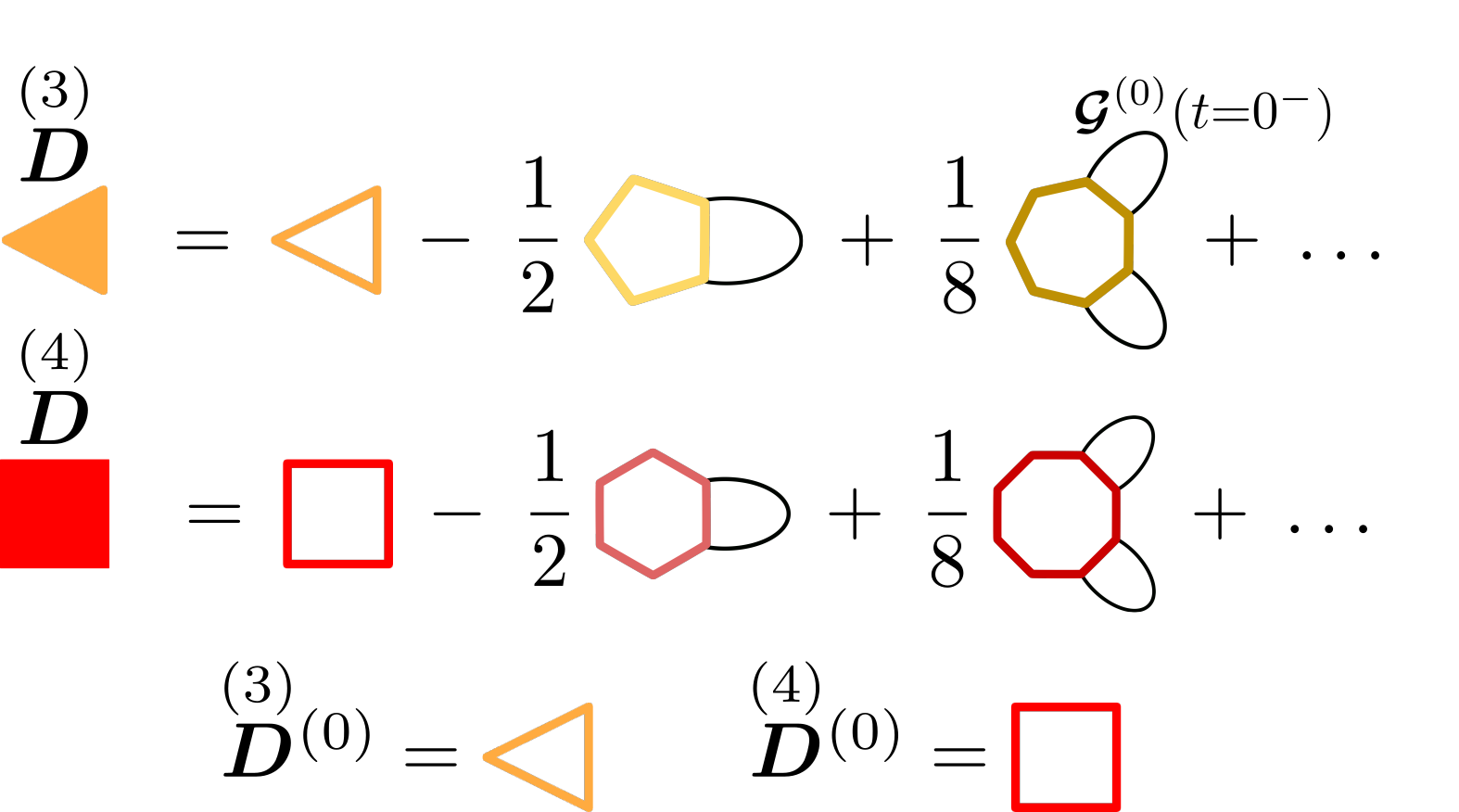}
    \end{minipage}
    \caption{Diagrammatic expression for the SCHA scattering tensors Eqs \eqref{def D3 main} \eqref{def D4 main} as presented in Eqs \eqref{3-4 scatt vertices diagrams}. Each SCHA propagator is contracted with a higher order derivative of the anharmonic tensor Eq.\ \eqref{bare anharmonic tensor} which are evaluated at the SCHA positions, i.e. they do not coincide with Eq.\ \eqref{anh vertex on harm position}. These are represented as $n=3,5,7..$ and $n=4,6,8..$ regular polygons.}
    \label{fig:d3 d4 flower diagrams}
\end{figure} 

The scattering vertices, Eqs \eqref{def D3 main} \eqref{def D4 main}, included in the dynamical response have an interesting diagrammatic expression (see also Ref.\ \cite{flowers1968}). \replaced{In general, they do not coincide}{They do not coincide in general} with the derivatives of the BO potential evaluated at the equilibrium SCHA positions $\small\boldsymbol{\mathcal{R}}^{(0)}$
but they contain extra terms due to quantum-thermal fluctuations. All of these terms are included in the TDSCHA. Expanding Eqs \eqref{def D3 main} \eqref{def D4 main} in  $\small\bm R - \small\boldsymbol{\mathcal{R}}^{(0)}$, we get the following series, as reported in Appendix \ref{Scattering vertices},
\begin{subequations}
\label{3-4 scatt vertices diagrams}
\begin{align}
    \overset{(3)}{D}_{\mu\nu\eta} \hspace{-0.1cm}=  & \sum_{n=0}^{+\infty} \frac{(-1)^n}{2^{n}n!}\sum_{\alpha_1..\alpha_{2n}=1}^{3N}
    \overset{(3 +2n)}{D}_{\mu\nu\eta\alpha_1..\alpha_{2n-1}\alpha_{2n}}\hspace{-2.4cm}\textcolor{white}{,}^{(0)} \notag\\
    &\mathcal{G}^{(0)}(t=0^-)_{\alpha_1\alpha_2}..\mathcal{G}^{(0)}(t=0^-)_{\alpha_{2n-1}\alpha_{2n}} ,\\
    \overset{(4)}{D}_{\mu\nu\eta\lambda} \hspace{-0.1cm}=  & \sum_{n=0}^{+\infty}
    \frac{(-1)^n}{2^{n}n!}\sum_{\alpha_1..\alpha_{2n}=1}^{3N}
    \overset{(4 +2n)}{D}_{\mu\nu\eta\lambda\alpha_1\alpha_2..\alpha_{2n-1}\alpha_{2n}} \hspace{-2.9cm}\textcolor{white}{,}^{(0)}\notag\\ 
    &\mathcal{G}^{(0)}(t=0^-)_{\alpha_1\alpha_2}..\mathcal{G}^{(0)}(t=0^-)_{\alpha_{2n-1}\alpha_{2n}} ,
\end{align}
\end{subequations}
where the anharmonic vertices in the series are
\begin{equation}
\label{bare anharmonic tensor}
    \overset{(n)}{D}_{\alpha_1..\alpha_{n}} 
    \hspace{-.95cm}\textcolor{white}{,}^{(0)}\quad = \frac{\partial^{n} V^{\text{(BO)}}}{\partial \widetilde{R}_{\alpha_1}..\partial \widetilde{R}_{\alpha_{n}}}\biggl|_{\bm R =\boldsymbol{\mathcal{R}}^{(0)}} .
\end{equation}
These differ in general from \replaced{}{the vertices } $ \overset{(n)}{\bm d}$, see Eq.\ \eqref{anh vertex on harm position}, since the minimum of the Born-Oppenheimer potential $\bm{\mathcal{R}}_{\text{BO}}$ does not coincide with the SCHA centroid $\bm{\mathcal{R}}^{(0)}$.
In Fig.\ \ref{fig:d3 d4 flower diagrams} we report the diagrammatic expansion for Eqs \eqref{3-4 scatt vertices diagrams}. 
Each anharmonic tensor in Eq.\ \eqref{bare anharmonic tensor} with $n>3,4$ has a pair of indexes contracted with a SCHA propagator $\boldsymbol{\mathcal{G}}^{(0)}(t=0^-)$ (Eq.\ \eqref{bare anharmonic tensor}). \replaced{This indicates that quantum-thermal fluctuations result in the renormalization of the anharmonic vertices.}{This means that the anharmonic vertices are renormalized by quantum-thermal fluctuations.}

\subsection{Anharmonic propagators}
\label{Green functions}
In this Section, we discuss the TDSCHA interacting propagators. Specifically, we present two-phonon processes that have been neglected in previous works \cite{TDSCHA_monacelli,LihmTDSCHA}.

In TDSCHA, the one-phonon $\mathcal{G}(\omega)_{\mu\nu}$, two-phonon $ \chi(\omega)_{\mu\nu\eta\lambda}$, and the one-two phonon $\Gamma(\omega)_{\mu\eta\lambda}$ interacting propagators are obtained as response functions by setting in $\chi(\omega)_{\mathcal{A},\mathcal{B}}$ (Eq.\ \eqref{response formula for calculations})
\begin{subequations}
\label{A B Green functions}
\begin{align}
    & \mathcal{A}_{ \mathcal{G}} = \delta\widetilde{R}^{(0)}_\mu  
    & \mathcal{B}_{\mathcal{G}} = & \delta\widetilde{R}^{(0)}_\nu \label{1 ph pertubation},\\
    & \mathcal{A}_{\chi} =  \frac{1}{2}\delta\widetilde{R}^{(0)}_\mu\delta\widetilde{R}^{(0)}_\nu 
    & \mathcal{B}_{\chi} = &  \frac{1}{2}\delta\widetilde{R}^{(0)}_\eta \delta\widetilde{R}^{(0)}_\lambda\label{2 ph pertubation} ,\\
    &  \mathcal{A}_{\Gamma} = \delta\widetilde{R}^{(0)}_\mu 
    &  \mathcal{B}_{\Gamma} = &\frac{1}{2}\delta\widetilde{R}^{(0)}_\eta \delta\widetilde{R}^{(0)}_\lambda\label{1 2 ph pertubation}.
\end{align}
\end{subequations}
The response and perturbation vectors (see Eqs \eqref{perturbation response vector}) corresponding to Eqs \eqref{A B Green functions} are as follows
\begin{subequations}
\label{r p Green functions}
\begin{align}
    \bm r_{\mathcal{G}}  =& 
    \begin{bmatrix}
    \bm \delta_{\mu} \\
    \bm 0 \\
    \bm 0
    \end{bmatrix} 
    &\bm p_{\mathcal{G}} = 
    \begin{bmatrix}
    \bm \delta_{\nu} \\
    \bm 0 \\
    \bm 0
    \end{bmatrix},\label{r p 1 ph}\\
    \bm r_{\chi} =& \begin{bmatrix}
    \bm 0 \\
    \bm S_{\mu\nu} \\
    \bm S_{\mu\nu}
    \end{bmatrix}
    &\bm p_{\chi} = \begin{bmatrix}
    \bm 0 \\
    \bm S_{\eta\lambda} \\
    \bm S_{\eta\lambda}
    \end{bmatrix}
    ,\label{r p 2 ph}\\
    \bm r_{\Gamma} = &  \begin{bmatrix}
    \bm \delta_{\mu} \\
    \bm 0 \\
    \bm 0
    \end{bmatrix} 
    &\bm p_{\Gamma} =  \begin{bmatrix}
    \bm 0 \\
    \bm S_{\eta\lambda} \\
    \bm S_{\eta\lambda}
    \end{bmatrix},\label{r p 1 2 ph}
\end{align}
\end{subequations}
where $\bm \delta_{\mu}$ is a $3N$ vector with 1 in the mode index $\mu$ and zero elsewhere and $\bm S_{\eta\lambda}$ is a $3N\times 3N$ matrix with $1/2$ on the mode indexes $\eta$ and $\lambda$  and zeros elsewhere.

The choice of $\mathcal{A}/\mathcal{B}$, as in Eqs \eqref{A B Green functions}, is not arbitrary. In the non-interacting case, $\boldsymbol{\mathcal{L}}(\omega)$ is diagonal, so the response function \replaced{simplifies to}{is simply}
\begin{equation}
\label{non interacting response}
\begin{aligned}
     \chi(\omega)_{\mathcal{A},\mathcal{B}} =& \responsevector_i \begin{bmatrix}
    \boldsymbol{\mathcal{G}}^{(0)}(\omega) & \bm 0 &
    \bm 0 \\ 
    \bm 0  &
     \bm \chi^{(0)}_{-}(\omega)    & 
    \bm 0 \\ 
    \bm 0 
    & \bm 0 
    &   -\bm \chi^{(0)}_{+}(\omega)
    \end{bmatrix} \perturbationvector_i \\
     i = &\mathcal{G}, \mathcal{\chi}, \Gamma \replaced{.}{,}
\end{aligned}
\end{equation} 
From Eq.\ \eqref{non interacting response} we recover the one and two phonon free propagators (Eqs \eqref{1 ph free propagator} \eqref{2 ph free propagator}) and no cross terms connecting single to double excitations.  These results are consistent with the standard linear response theory, in the non-interacting case, treated with the many-body formalism. \replaced{This shows that Eqs \eqref{r p Green functions} recover physically relevant quantities.}{This proves that with Eqs \eqref{r p Green functions} we recover the physically relevant quantities.}

To get the interacting Green's function, we plug $\responsevector$ and $\perturbationvector$ of Eqs \eqref{r p Green functions} in the expression of $\chi(\omega)_{\mathcal{A},\mathcal{B}}$, Eq.\ \eqref{response formula for calculations}, and we invert $\boldsymbol{\mathcal{L}}(\omega)$ following Refs \cite{TDSCHA_monacelli,MachedaPRL,MachedaNew} (see also Appendix \ref{Symbolic inversion}). To do this, we consider $\boldsymbol{\mathcal{L}}(\omega)$, Eq.\ \eqref{L + omega2 easy to understand}, as a $3\times 3$ block-matrix (as represented in Fig. \ref{fig:L operator}) where each block itself is a tensor. The same applies to $\bm r$ and $\bm p$, Eqs \eqref{perturbation response vector},  which are understood as $3$ components vectors. From the non-interacting case (Eq.\ \eqref{non interacting response}), we learn that the first component of $\perturbationvector/\responsevector$ controls the single-mode propagation while the second and third the two-phonon channel.

The inversion process mixes the matrix elements of $\bm{\mathcal{L}}(\omega)$ adding interactions to the free propagators which are expressed as diagrammatic series. The representation of Fig.\ \ref{fig:L operator} is a graphical aid to visualize the building blocks of these diagrams.
In Appendix \ref{Derivation of the interacting Green function} we report the details of all the results presented. 

In our calculations we always reduce the inversion of $\boldsymbol{\mathcal{L}}(\omega)$ to a $2\times2$ block-matrix with the following form
\begin{equation}
     \begin{bmatrix}
     \bm A & \bm C\\ 
     \bm C^\dagger & \bm B
    \end{bmatrix}
\end{equation}
which can be easily inverted (see Appendix \ref{Symbolic inversion})
\begin{equation}
\small
\label{main: inverse of L symbolic 3x3}
    \begin{bmatrix}
    \bm D^{-1} & \quad  -  \bm D^{-1}\cdot \bm C\cdot  \bm B^{-1} \\
    -\bm B^{-1} \cdot \bm C^\dagger \cdot \bm D^{-1} &
    \quad \bm B^{-1}\cdot\left(\bm1  +    \bm C^\dagger \cdot \bm D^{-1}\added{\cdot}\bm C \cdot \bm B^{-1}\right)
    \end{bmatrix}
\end{equation}
where $\bm D = \bm A -  \bm C \cdot\bm B^{-1}\cdot \bm C^\dagger$. 

First, we discuss the one-phonon propagator. Because only the first component of both $\perturbationvector_{\mathcal{G}}$ and $\responsevector_{\mathcal{G}}$, Eq.\ \eqref{r p 1 ph}, are non-zero, the response calculation is simplified. In particular, we compact the $2\times 2$ two-phonon sector of $\boldsymbol{\mathcal{L}}(\omega)$ (i.e.\ $\boldsymbol{\mathcal{L}}(\omega)_{ij}$ with $i,j>1$) in a $1\times 1$ block matrix. As shown in Appendix \ref{Derivation of the interacting Green function}, $\boldsymbol{\mathcal{L}}(\omega)$ is reduced to a $2\times2$ block-matrix $\boldsymbol{\mathcal{L}}_{\text{1ph}}(\omega)$, so that the one-phonon propagator is given by
\begin{equation}
\label{one ph Green function from 2x2 matrix}
\begin{aligned}
    \boldsymbol{\mathcal{G}}(\omega) = & \responsevector_{\mathcal{G}}\cdot  \boldsymbol{\mathcal{L}}(\omega)^{-1} \cdot \perturbationvector_{\mathcal{G}}= \left(\boldsymbol{\mathcal{L}}_{\text{1ph}}(\omega)^{-1}\right)_{11}  
\end{aligned} 
\end{equation}
where
\begin{equation}
    \boldsymbol{\mathcal{L}}_{\text{1ph}}(\omega) = 
    \begin{bmatrix}
    \boldsymbol{\mathcal{G}}^{(0)}(\omega)^{-1} &- \overset{(3)}{\bm D} \\
    -\overset{(3)}{\bm D} & \bm \chi^{(0)}(\omega)^{-1} -  \overset{(4)}{\bm D}  
    \end{bmatrix}.
\end{equation}
Note that $\left(\boldsymbol{\mathcal{L}}_{\text{1ph}}(\omega)\right)_{22}$ represents the anharmonic two-phonon channel in which single modes can decay through the three-phonon vertex, $\left(\boldsymbol{\mathcal{L}}_{\text{1ph}}(\omega)\right)_{12}$.
Graphically Eq.\ \eqref{one ph Green function from 2x2 matrix} corresponds to 
\begin{equation}
\begin{split}
   \includegraphics[width= 0.08\textwidth,valign=c]{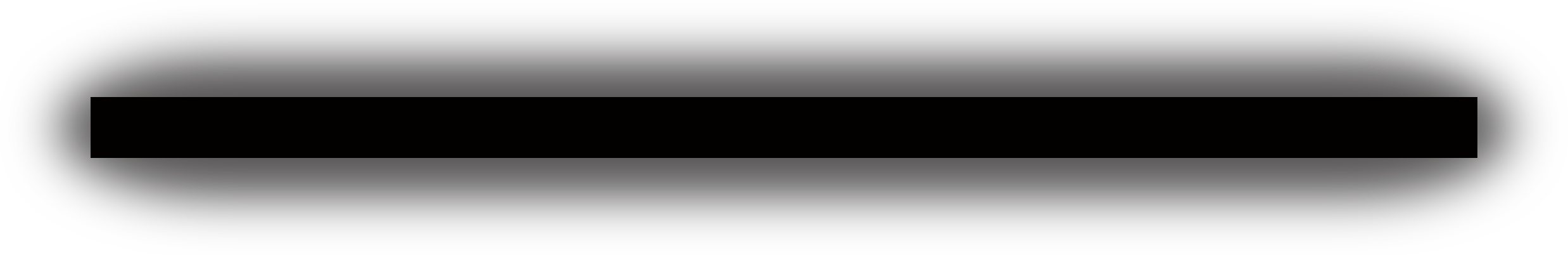} &=  \mqty[ \includegraphics[width=.05\textwidth,valign=c]{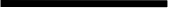}^{-1}& -\includegraphics[height = 0.5 cm,valign=c]{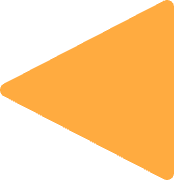}    \\-\scalebox{-1}[1]{\includegraphics[height = 0.5 cm,valign=c]{SIMBOLI/d3.png} }  & \includegraphics[width=1 cm,valign=c]{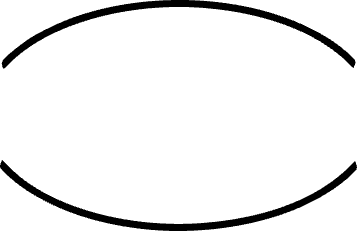}^{-1} -\includegraphics[height=0.5cm,valign=c]{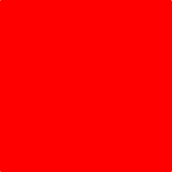}]^{-1}_{\:\: 11} .
\end{split}
\end{equation}
In Eq.\ \eqref{one ph Green function from 2x2 matrix} we apply the general result of Eq.\ \eqref{main: inverse of L symbolic 3x3} to obtain the interacting Green's function
\begin{equation}
\label{1 ph Green function}
  \boldsymbol{\mathcal{G}}(\omega) =\boldsymbol{\mathcal{G}}^{(0)}(\omega) +  \boldsymbol{\mathcal{G}}^{(0)}(\omega)\cdot\bm \Pi(\omega)\cdot\boldsymbol{\mathcal{G}}(\omega),
\end{equation}
\deleted{here $\small{\bm A : \bm B =\sum_{\mu\nu=1}^{3N} A_{..\mu\nu} B_{\mu\nu..}}$,}where the self-energy $\bm \Pi(\omega)$ coincides with the one reported in Refs \cite{TDSCHA_monacelli,LihmTDSCHA,Bianco}
\begin{equation} 
\label{Bianco self energy}
    \bm \Pi(\omega) = \overset{(3)}{\bm D} :\left(\bm 1 -\bm \chi^{(0)}(\omega) :\overset{(4)}{\bm D}\right)^{-1} \hspace{-0.2cm}: \bm \chi^{(0)}(\omega):\overset{(3)}{\bm D}\added{,}
\end{equation}
\added{here $\small{\bm A : \bm B =\sum_{\mu\nu=1}^{3N} A_{..\mu\nu} B_{\mu\nu..}}$.}
Our definition of the non-interacting two-phonon propagator $\bm \chi^{(0)}(\omega)$ (Eq.\ \eqref{2 ph free propagator}) is one reported by Ref.\ \cite{Bianco} in Eq.\ (72) multiplied by $-1/2$ so that all the definitions are consistent.

Physical phonon frequencies and lifetimes are \replaced{determined}{given} by real and imaginary parts of $\boldsymbol{\mathcal{G}}(\omega +i0^+)$, Eq.\ \eqref{1 ph Green function}, as discussed in Ref.\ \cite{SCHA_main}. In addition, we remark that the polarization vectors can also change when adding dynamical effects so polarization-mixing is automatically included in Eq.\ \eqref{1 ph Green function}.

\replaced{The bubble diagram is the lowest-order approximation for the self-energy, as described in Eq.\ \eqref{Bianco self energy}, and is incorporated in many self-consistent phonon (SCP) calculations within the improved SCP (ISCP) framework \cite{TadanoSrTiO3_polarizationmixing,Tadano_SCF3_ISCP,ISCP,Werthamer_ISCP}. TDSCHA represents a theoretical approach that justifies this from the least action principle and provides a path to move beyond the bubble approximation.}{The lowest-order approximation for the self-energy Eq.\ \eqref{Bianco self energy} is the bubble diagram which is included by many SCP calculations in the improved SCP (ISCP) framework \cite{TadanoSrTiO3_polarizationmixing}\cite{Tadano_SCF3_ISCP}\cite{ISCP}\cite{Werthamer_ISCP}. TDSCHA represents a theoretical approach that justifies this from the least action principle and paves the way to go beyond the bubble approximation.}
\begin{figure*}
    \includegraphics[width=1.0\textwidth]{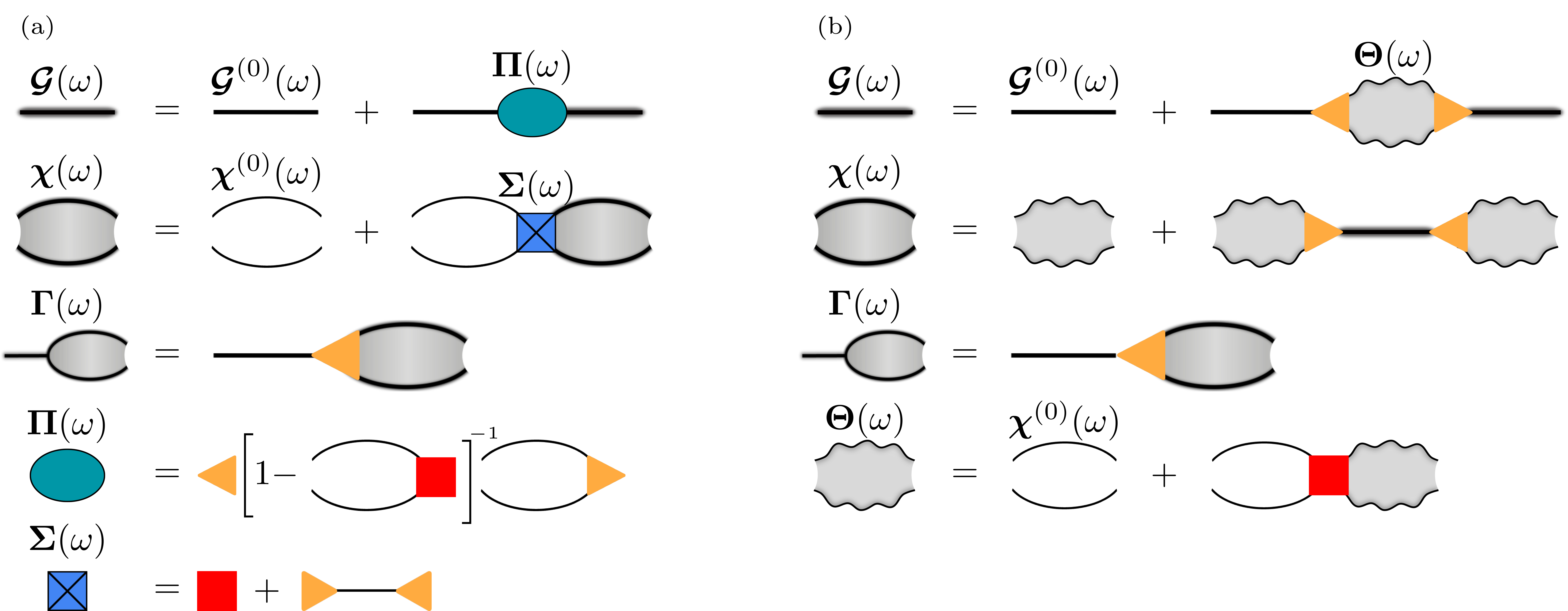}
    \caption{\replaced{In a) we report the diagrammatic expression of the interacting one, two and one-two phonon Green's functions, Eqs \eqref{1 ph Green function} \eqref{2 ph Green function} \eqref{1 2 ph Green function}. In b)  we show Eqs \eqref{G chi Gamma Theta compact}.}{Diagrammatic expression of the interacting one, two and one-two phonon Green's functions, Eqs \eqref{1 ph Green function} \eqref{2 ph Green function} \eqref{1 2 ph Green function}.} Thinner solid lines represent the non-interacting SCHA propagators $\boldsymbol{\mathcal{G}}^{(0)}(\omega)$ and $\boldsymbol{\chi}^{(0)}(\omega)$, defined in Eqs \eqref{1 ph Green function} \eqref{2 ph Green function}. The three/four-phonon scattering vertices are defined in Eqs \eqref{def D3 main} \eqref{def D4 main} and represented as orange triangles and red squares.}
    \label{fig:dyson equation for propagators ab panel}
\end{figure*}

\begin{figure}[!htb]
    \centering
    \begin{minipage}[c]{1.0\linewidth}
    \includegraphics[width=1.0\textwidth]{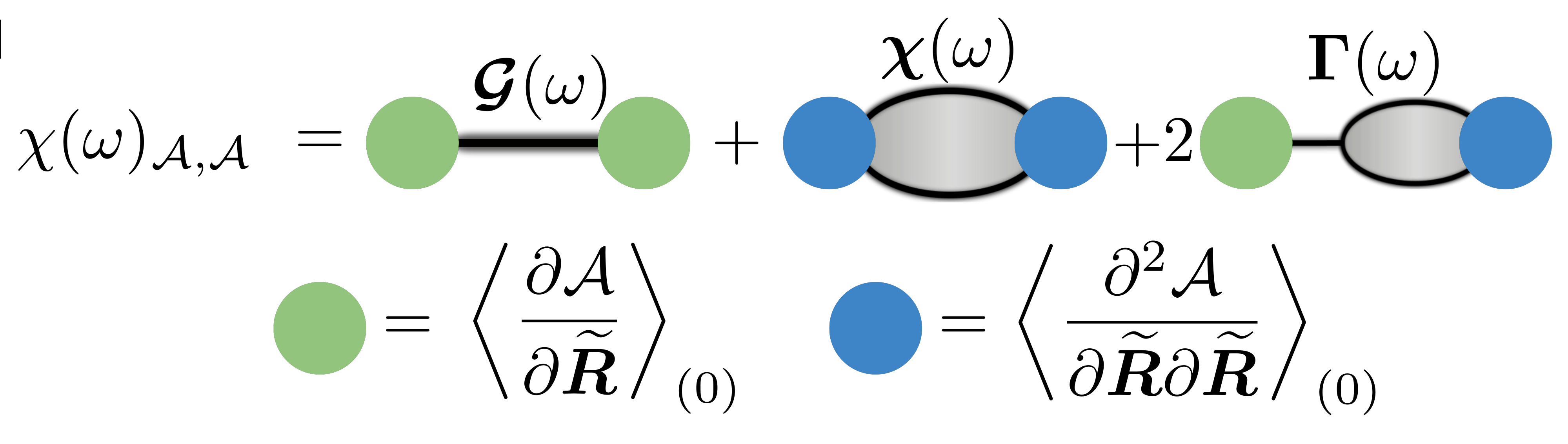}
    \end{minipage}
    \caption{Diagrammatic expression of the processes included in the fully interacting response, Eq.\ \eqref{response formula for calculations}, if $\mathcal{A} = \mathcal{B}$. The interacting TDSCHA Green's functions are reported in Eqs \eqref{1 ph Green function} \eqref{2 ph Green function} \eqref{1 2 ph Green function}. The green vertex is related to the first entry of $\perturbationvector$/$\responsevector$ while the blue vertex to the second and third entries of $\perturbationvector$/$\responsevector$, see Eqs. \eqref{perturbation response vector}. }
    \label{fig:response diagrams}
\end{figure} 

For the two-phonon case, we proceed as before. We use the definition of $\responsevector_\chi/\perturbationvector_\chi$ (Eq.\ \eqref{r p 2 ph}) to reabsorb the one-phonon sector of $\boldsymbol{\mathcal{L}}(\omega)$ (i.e.\ the row $\boldsymbol{\mathcal{L}}(\omega)_{1j}$ and column $\boldsymbol{\mathcal{L}}(\omega)_{j1}$ with $j=1,2,3$) into the two-phonon sector. So we solve
\begin{equation}
\label{two ph green fucntion 2x2 matrix}
    \bm \chi(\omega) = \responsevector_\chi \cdot \boldsymbol{\mathcal{L}}(\omega)^{-1} \cdot \perturbationvector_\chi = \sum_{ij}^{1,2}\left(\boldsymbol{\mathcal{L}}_{\text{2ph}}(\omega) ^{-1}\right)_{ij}
\end{equation}
where $\boldsymbol{\mathcal{L}}_{\text{2ph}}(\omega)$ is a $2\times 2 $ block-matrix
\begin{equation}
\label{L + omega2 2 ph}
    \hspace{-0.25cm}\boldsymbol{\mathcal{L}}_{\text{2ph}}(\omega) \hspace{-0.05cm} = \hspace{-0.15cm}
    \begin{bmatrix}
     \bm \chi^{(0)}_{-}(\omega)^{-1}
    -\bm \Sigma(\omega)
    &-\bm \Sigma(\omega)\\ 
    -\bm \Sigma(\omega) & -\bm \chi^{(0)}_{+}(\omega)^{-1} 
    -\bm \Sigma(\omega)
    \end{bmatrix}.
\end{equation}
$\bm \Sigma(\omega)$ is the two-phonon self-energy
\begin{equation}
\label{2 ph self energy}
    \bm \Sigma(\omega) = \bm \Sigma(\omega)^\dagger = \overset{(4)}{\bm D} + \overset{(3)}{\bm D}\cdot \boldsymbol{\mathcal{G}}^{(0)}(\omega)\cdot \overset{(3)}{\bm D}
\end{equation}
where single phonon excitations enter in $\bm \Sigma(\omega)$ via the three-phonon vertex.

Graphically, Eq.\ \eqref{two ph green fucntion 2x2 matrix} corresponds to
\begin{equation}
\begin{split}
    \includegraphics[width= 1.5cm,keepaspectratio,valign=c]{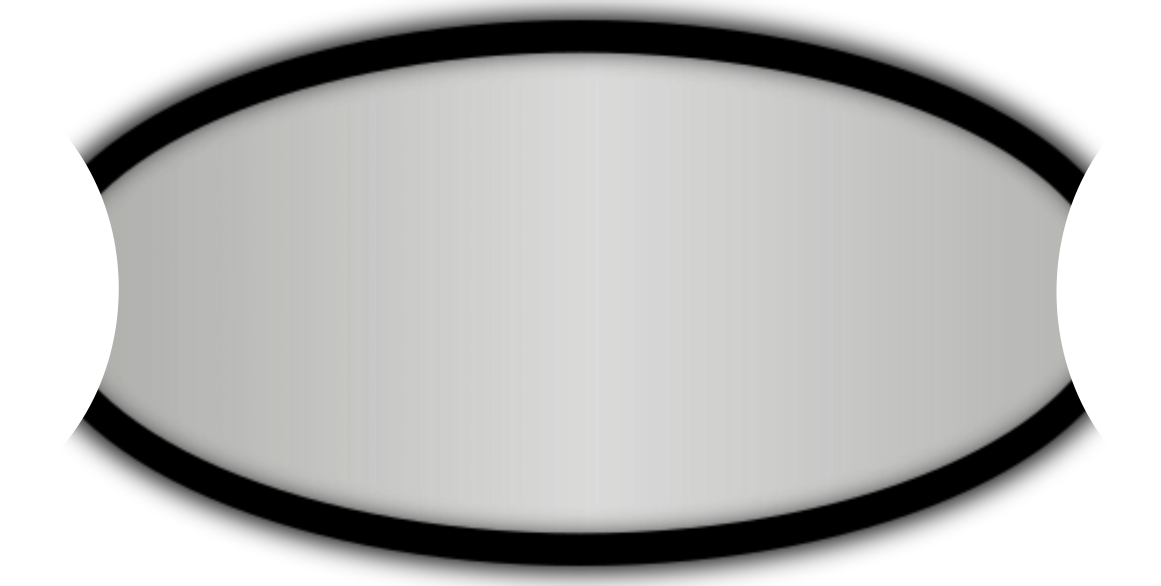} \hspace{-0.15cm} & = \hspace{-0.2cm} 
    \sum_{ij}^{1,2}\hspace{-0.1cm}
    \mqty[\includegraphics[width= 1cm,keepaspectratio,valign=c]{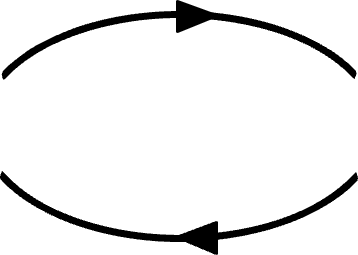}^{-1}\hspace{-0.2cm}-\includegraphics[width= 0.5cm,keepaspectratio,valign=c]{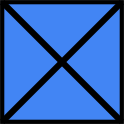}  &-\includegraphics[width= 0.5cm,keepaspectratio,valign=c]{SIMBOLI/sigmathin.png} \\ -\includegraphics[width= 0.5cm,keepaspectratio,valign=c]{SIMBOLI/sigmathin.png} &-\includegraphics[width= 1cm,keepaspectratio,valign=c]{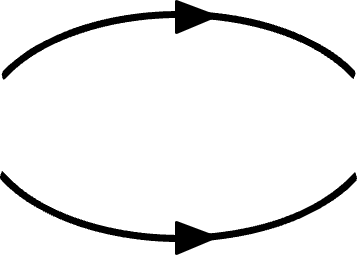}^{-1\hspace{-0.2cm}}-\includegraphics[width= 0.5cm,keepaspectratio,valign=c]{SIMBOLI/sigmathin.png}]^{-1}_{ij}    \\
    \bm \Sigma(\omega) & = \includegraphics[width= 0.5cm,keepaspectratio,valign=c]{SIMBOLI/sigmathin.png} = \includegraphics[width= 0.5cm,keepaspectratio,valign=c]{SIMBOLI/d4.png} + \includegraphics[height=0.5cm,keepaspectratio,valign=c]{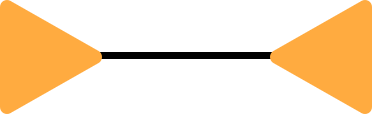}     .
\end{split}
\end{equation}

Again we use Eq.\ \eqref{main: inverse of L symbolic 3x3} to invert Eq.\ \eqref{L + omega2 2 ph} and we end up with the interacting two-phonon propagator
\begin{equation}
\label{2 ph Green function}
    \bm \chi(\omega) = \bm \chi^{(0)}(\omega) + \bm \chi^{(0)}(\omega) : \bm \Sigma(\omega) : \bm \chi(\omega)
\end{equation}
with $\bm \Sigma(\omega)$, Eq.\ \eqref{2 ph self energy}, being the TDSCHA two-phonon  self-energy.
We find that\replaced{, in the two-phonon propagation,}{in the two-phonon propagation} there is the possibility of either decay in a single phonon through the third-order scattering vertex  or in another pair through the fourth-order scattering vertex. There are no high-order decay processes. 

For the mixed propagator, we proceed as before. \replaced{The form of $\perturbationvector_\Gamma/\responsevector_\Gamma$ (Eq.\ \eqref{r p 1 2 ph}) allows single phonon excitations to be triggered by two-phonon propagation via the three-phonon vertex, leading to a non-zero one-two phonon propagation}{Thanks to the form of $\perturbationvector_\Gamma/\responsevector_\Gamma$ (Eq.\ \eqref{r p 1 2 ph}), single phonon excitations can be triggered by two-phonon propagation via the three-phonon vertex. This leads to a non-zero one-two phonon propagation}
\begin{equation}
\label{1 2 ph Green function}
    \bm \Gamma (\omega) =  \boldsymbol{\mathcal{G}}^{(0)}(\omega)\cdot  \overset{(3)}{\bm D} : \bm \chi(\omega).
\end{equation}

\added{We demonstrate that Eqs. \eqref{1 ph Green function} \eqref{2 ph Green function} are the fundamental components of the TDSCHA response. The relationship between these two propagators becomes clear when expressed in terms of the partially screened two-phonon propagator $\boldsymbol{\Theta}(\omega) = \bm{\chi} (\w)\eval_{\overset{(3)}{\boldsymbol{D}}=0}$ in which a phonon pair propagates only through the four phonon scattering vertex. Note that $\boldsymbol{\Theta}(\omega)$ generalizes in the dynamical regime Eq.\ (26) of Ref.\ \cite{Bianco}. As proved in Appendix \ref{Derivation of the interacting Green function}, the new expressions for the propagators are
\begin{subequations}
\small
\label{G chi Gamma Theta compact}
\begin{align}
    \boldsymbol{\mathcal{G}}(\omega) & =\boldsymbol{\mathcal{G}}^{(0)}(\omega) + \boldsymbol{\mathcal{G}}^{(0)}(\omega) \cdot \overset{(3)}{\boldsymbol{D}} :\boldsymbol{\Theta}(\omega) : \overset{(3)}{\boldsymbol{D}} \cdot  \boldsymbol{\mathcal{G}}(\omega)  , \\
    \boldsymbol{\chi} (\w) & = \boldsymbol{\Theta} (\w) + \boldsymbol{\Theta}(\w) : \overset{(3)}{\boldsymbol{D}} \cdot  \boldsymbol{\mathcal{G}}(\w) \cdot \overset{(3)}{\boldsymbol{D}} : \boldsymbol{\Theta}(\w), \label{chi = w + w g w} \\
    \boldsymbol{\Theta}(\omega) & = \boldsymbol{\chi}^{(0)}(\omega) + \boldsymbol{\chi}^{(0)}(\omega) : \overset{(4)}{\boldsymbol{D}} : \boldsymbol{\Theta}(\omega) \label{W def}.
\end{align}
\end{subequations}
Now, only $\boldsymbol{\mathcal{G}}(\omega)$ and $\boldsymbol{\Theta}(\omega)$ have a Dyson form, whereas $\boldsymbol{\chi}(\omega)$ has a different structure where single and double propagations are disentangled.}

\replaced{Fig.\ \ref{fig:dyson equation for propagators ab panel} summarizes the diagrammatic expressions for the interacting propagators. In panel a) we report Eqs \eqref{1 ph Green function} \eqref{2 ph Green function} \eqref{1 2 ph Green function}, while panel b) shows Eqs \eqref{G chi Gamma Theta compact}.}{Fig.\ \ref{fig:dyson equation for propagators ab panel} summarizes the diagrammatic expressions for the interacting propagators, Eqs \eqref{1 ph Green function} \eqref{2 ph Green function} \eqref{1 2 ph Green function}.}

By computing all the TDSCHA interacting propagators, we gain a full comprehension of the diagrammatic expression, Fig.\ \ref{fig:response diagrams}, introduced in Ref.\ \cite{TDSCHA_monacelli} for the TDSCHA response function $\chi(\omega)_{\mathcal{A},\mathcal{A}}$ (Eq.\ \eqref{response formula for calculations}). In fact, Eq.\ \eqref{response formula for calculations} can be decomposed into the interacting propagators.

One-phonon processes $\boldsymbol{\mathcal{G}}(\omega)$ are coupled to first-order position derivatives of the perturbation, first entry of $\perturbationvector$/$\responsevector$ Eqs \eqref{perturbation response vector}. On the other hand, two-phonon excitations $\boldsymbol{\chi}(\omega)$ and $\boldsymbol{\Gamma}(\omega)$ are triggered by non-zero second-order position derivatives of the perturbation, second and third entries of $\perturbationvector$/$\responsevector$ Eqs. \eqref{perturbation response vector}. 

We emphasize that the Lanczos algorithm includes the effect of the third and fourth-order scattering vertex, Eqs \eqref{def D3 main} \eqref{def D4 main}, in a non-perturbative way \cite{phase_diagram_hydrogen} (see Appendix \ref{appendix: Lanczos algorithm}). TDSCHA evolves ab-initio all the phonon modes in a given supercell without free parameters. This feature is interesting for applications in non-linear phononics where is crucial to comprehend relaxation pathways of coherent phonon oscillations \cite{distinguish_nonlinear_path,SumFrequencyIonicRamanScattering,UdinaKerr,Cavalleri_anharmonic_potential}. 

\subsubsection{Momentum Green function} 
In Wigner-TDSCHA, the ionic momentum is controlled directly, which was not possible in the original formulation.  Here we discuss the TDSCHA momentum-momentum Green's function $\small\boldsymbol{\mathcal{G}}_\text{p}(\omega)$. In our theory, this is computed setting
\begin{equation}
    \mathcal{A} = \widetilde{P}_\mu \quad \mathcal{B} = \widetilde{P}_\nu.
\end{equation}
The SCHA momentum Green's function $\boldsymbol{\mathcal{G}}^{(0)}_\text{p}(\omega)$ is proportional to $\boldsymbol{\mathcal{G}}^{(0)}(\omega)$ (Eq.\ \eqref{1 ph Green function}) since the equation for position and momentum are coupled
\begin{equation}
    i\omega\widetilde{\boldsymbol{\mathcal{P}}}(\omega) = \widetilde{\boldsymbol{\mathcal{R}}}(\omega) \longrightarrow
    \mathcal{G}^{(0)}_\text{p}(\omega)_{\mu\nu} = \omega_\mu^2
    \mathcal{G}^{(0)}(\omega)_{\mu\nu}.
\end{equation}
The free propagators are the building blocks for the interacting theory. Hence the interacting momentum Green's function $\boldsymbol{\mathcal{G}}_\text{p}(\omega)$ satisfies a perturbative expansion that is proportional to the one of $\boldsymbol{\mathcal{G}}(\omega)$ once the TDSCHA diagrams are selected, i.e. those from Eq.\ \eqref{1 ph Green function}.
So we have that, see Appendix \ref{Momentum Green function} for details,
\begin{equation}
\label{momentum Green function TDSCHA}
    \boldsymbol{\mathcal{G}}_\text{p}(\omega) = -\bm 1 + \omega^2 \boldsymbol{\mathcal{G}}(\omega).
\end{equation}
Thus, a Lanczos calculation also provides access to the TDSCHA momentum Green's function.

\subsection{Multiple excitations in TDSCHA}
\label{Multiple excitations in TDSCHA}
The Gaussian approximation defines a hierarchy of diagrams that is truncated at the two-phonon level. In this Section, we demonstrate that, in TDSCHA, all higher-order phonon propagators are
related to the Green's functions of Eqs \eqref{1 ph Green function} \eqref{2 ph Green function} \eqref{1 2 ph Green function}.

\begin{figure}[!htb]
    \centering
    \begin{minipage}[c]{0.8\linewidth}
    \includegraphics[width=1.0\textwidth]{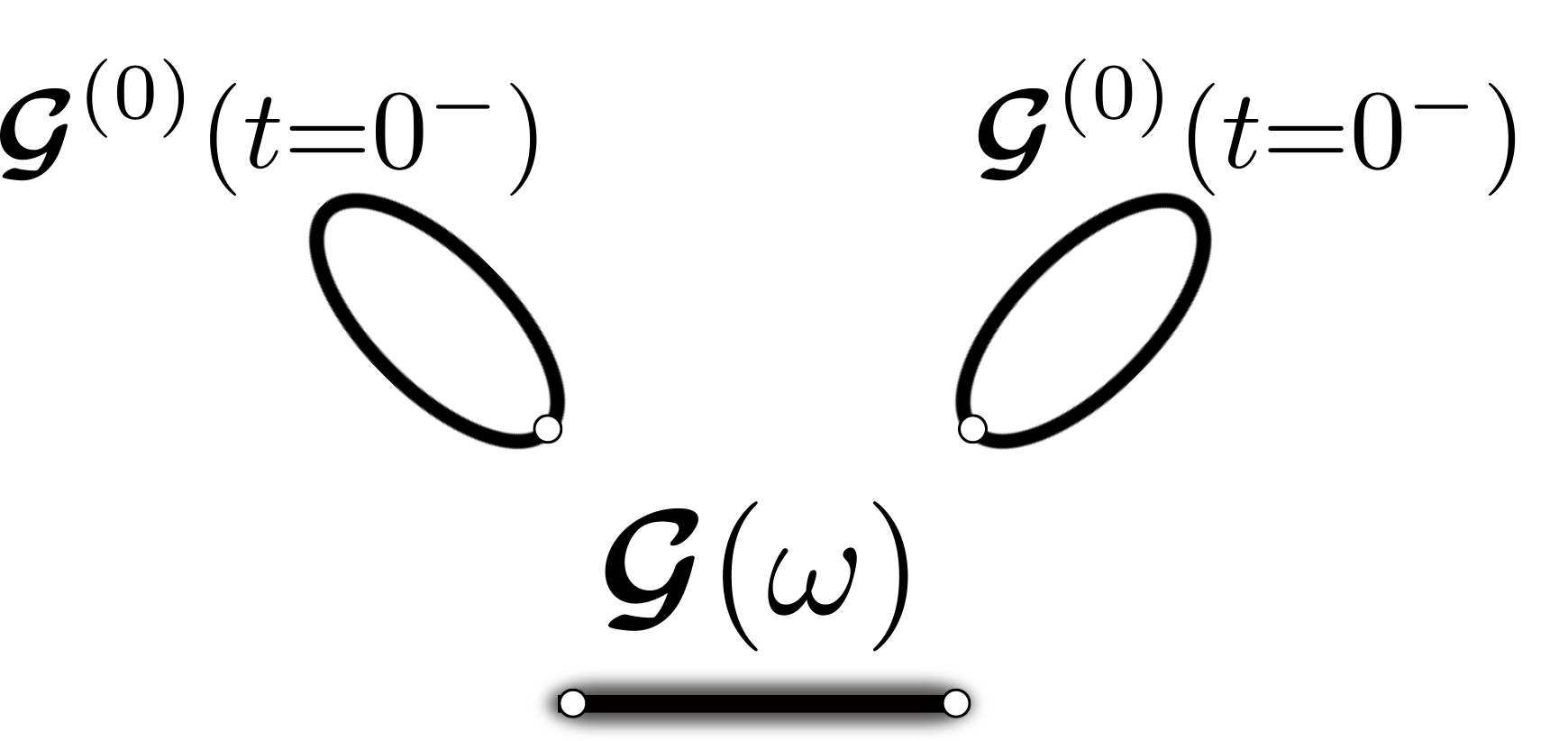}
    \end{minipage}
        \caption{Diagrammatic expression for the TDSCHA interacting three-phonon Green's function obtained as a response to a cubic perturbation, Eq.\ \eqref{three phonon A B}. In TDSCHA, the tree-phonons propagator is a disconnected diagram.}
    \label{fig:bunny}
\end{figure}

For example, the three-phonon propagator is obtained setting in Eq.\ \eqref{response formula for calculations} a tensor-like perturbation/response functions
\begin{equation}
\label{three phonon A B}
    \mathcal{A} = \delta\widetilde{R}^{(0)}_{\alpha} \delta\widetilde{R}^{(0)}_{\beta} \delta\widetilde{R}^{(0)}_{\gamma} \qquad
    \mathcal{B} = \delta\widetilde{R}^{(0)}_{\mu}\delta\widetilde{R}^{(0)}_{\nu}\delta\widetilde{R}^{(0)}_{\eta}.
\end{equation}
In this case, only the first entries of $\perturbationvector/\responsevector$, i.e. $\small\left\langle\partial\mathcal{A}/\partial\widetilde{\bm R}\right\rangle_{(0)}$, are non-zero.
This means that in the case of Eq.\ \eqref{three phonon A B} we have a one-phonon response, as the one obtained for $\boldsymbol{\mathcal{G}}(\omega)$ (Eq.\ \eqref{r p 1 ph}).
As computed in Appendix \ref{Derivation of the interacting Green function}, the three-phonon response is
\begin{equation}
\begin{aligned}
   &\chi_{\mu\nu\eta}^{\alpha\beta\gamma}(\omega) =    \mathcal{G}^{(0)}(t = 0^-)_{\beta\gamma}
    \mathcal{G}^{(0)}(t = 0^-)_{\nu\eta}\mathcal{G}(\omega)_{\alpha\mu} \\
    &
    +\text{permutations of $(\alpha\beta\gamma)$ and $(\mu\nu\eta)$}
\end{aligned}
\end{equation}
and in Fig.\ \ref{fig:bunny} we report its diagrammatic structure. This contains the one-phonon Green's function $\boldsymbol{\mathcal{G}}(\omega)$, Eq.\ \eqref{1 ph Green function}, and a disconnected part, $\boldsymbol{\mathcal{G}}^{(0)}(t=0^-)$ which comes from the averages $\left\langle\delta \widetilde{\bm R} \delta \widetilde{\bm R}\right\rangle_{(0)}$ (Eq.\ \eqref{flower definition SCHA}) in $\small\left\langle\partial\mathcal{A}/\partial\widetilde{\bm R}\right\rangle_{(0)}$.

In this case, the SCHA correction, $\boldsymbol{\mathcal{G}}^{(0)}(t=0^-)$, does not enter the phonon propagation but it dresses the interaction with the external probe. This means that if we take a scalar perturbation
\begin{equation}
    \mathcal{A} = \mathcal{B} = \frac{1}{3}\sum_{\alpha\beta\gamma=1}^{3N} K_{\alpha\beta\gamma}
    \delta\widetilde{R}^{(0)}_{\alpha} \delta\widetilde{R}^{(0)}_{\beta} \delta\widetilde{R}^{(0)}_{\gamma}, 
\end{equation}
with $K_{\alpha\beta\gamma}$ a tensor that does not depend on atomic positions, $\boldsymbol{\mathcal{G}}^{(0)}(t=0^-)$ is contracted with $\bm K$.

Similarly, all the higher-order propagators, i.e. those obtained with
\begin{equation}
    \mathcal{A} \sim \left(\delta\widetilde{R}^{(0)}\right)^{n -2} \quad
    \mathcal{B} \sim \left(\delta\widetilde{R}^{(0)}\right)^{m -2} \quad m,n > 2,
\end{equation}
give disconnected diagrams. In all these cases we will get a $\chi_{\mathcal{A},\mathcal{B}}(\omega)$ that contains only one of the TDSCHA propagators (Eqs \eqref{1 ph Green function} \eqref{2 ph Green function} \eqref{1 2 ph Green function}) \replaced{along with}{plus} a disconnected part that depends only on $\boldsymbol{\mathcal{G}}^{(0)}(t=0^-)$, Eq.\ \eqref{flower definition SCHA}. 

\begin{figure}[!htb]
    \centering
    \begin{minipage}[c]{0.9\linewidth}
    \includegraphics[width=0.5\textwidth]{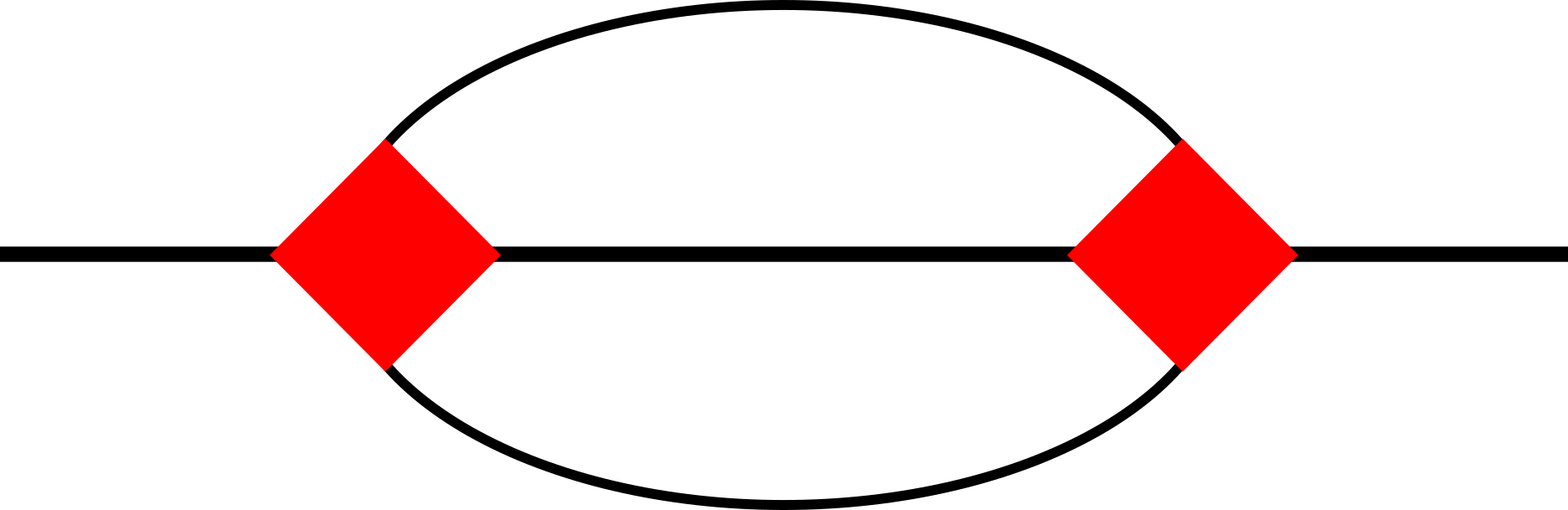}
    \end{minipage}
        \caption{Diagrammatic expression for the Saturn diagram with a three SCHA phonon propagation (solid lines are the propagators of Eq.\ \eqref{1 ph free propagator}). The red vertex is the four-phonon scattering vertex Eq.\ \eqref{def D4 main} which leads to three phonon excitations. This class of diagrams is missed by TDSCHA where we can not connect a single SCHA line to the four-phonon scattering vertex.}
    \label{fig:saturn}
\end{figure}

So TDSCHA can not capture processes beyond a two-phonon mechanism: the propagators of Eqs \eqref{1 ph Green function} \eqref{2 ph Green function} \eqref{1 2 ph Green function} \replaced{serve as}{are} the building blocks of the response. This means that there are general rules in the symbolic inversion of $\boldsymbol{\mathcal{L}}(\omega)$ (see Eq.\ \eqref{L + omega2 easy to understand}). In Fig.\ \ref{fig:L operator} the solid line (one-phonon SCHA propagator of Eq.\ \eqref{2 ph free propagator}) is always attached to one extremity of the orange triangle (Eq.\ \eqref{def D3 main}). The double solid line (two-phonon SCHA propagator Eq.\ \eqref{2 ph free propagator}) is connected to two extremities either of the red square (four phonon vertex Eq.\ \eqref{def D4 main}) or of the orange triangle (three phonon vertex Eq.\ \eqref{def D3 main}). 

We do not get three or more SCHA phonon resonances. One example is the 'Saturn' diagram (Fig.\ \ref{fig:saturn}) which is missed by our method. This diagram would correspond to a single SCHA propagator attached to the four-phonon vertex and this is not contained in TDSCHA.

\replaced{Notably, the TDSCHA diagrams arise from the stationary action principle of quantum mechanics \cite{TDSCHA_monacelli}, ensuring that there is no double counting and that the theory is consistent. The inclusion of new scattering mechanisms, such as Fig.\ \ref{fig:saturn}, must be approached with extreme care to avoid compromising the internal coherence and overcounting some anharmonic processes.}{Notably, the TDSCHA diagrams emerge from the stationary action principle of quantum mechanics \cite{TDSCHA_monacelli}. This guarantees that there is no double counting and that the theory is consistent. The inclusion of new scattering mechanisms, as Fig.\ \ref{fig:saturn}, must be done with extreme care to avoid spoiling the internal coherence and overcounting some anharmonic processes.}

\section{Nonlinear phonon-photon coupling: Infrared and Raman}
\label{Infrared and Raman spectra}
\replaced{In this section, we provide an overview of the infrared (IR) and Raman response in TDSCHA, with a particular emphasis on the two-phonon effect.}{In this Section, we review the infrared (IR) and Raman response in TDSCHA. In particular, we focus on the two-phonon effect.}

\replaced{IR experiments involve the absorption of infrared light by normal modes, which are associated with a variation in dipole moment and, in crystals, these are optical phonons.}{IR experiments are based on the absorption of infrared light by normal modes associated with a dipole moment variation which, in a crystal, are optical phonons.} The IR signal is proportional to the imaginary part of the dipole-dipole response function, $\Im[\chi(\omega+i0^+)_{\mathrm{p}_\alpha,\mathrm{p}_\beta}]$, along two Cartesian directions $\alpha$ and $\beta$ hence
\begin{equation}
    \mathcal{A}(\bm R) = \mathrm{p}_\alpha(\bm R), \qquad \mathcal{B}(\bm R) = \mathrm{p}_\beta(\bm R),
\end{equation}
where the dipole $\mathrm{p}_\alpha$ is per unit volume.

To get the response function we need the response and perturbation vector $\responsevector'$ and $\perturbationvector'$ (see Section \ref{Linearized equations of motion and general response function}). The first component of these vectors contains equilibrium averages of the effective charges:
\begin{equation}
\label{Ir 1 component}
\small
    \left\langle \frac{\partial \mathrm{p}_{\alpha}(\bm R)}{\partial R_a } \right \rangle_{(0)} = \left\langle Z^*(\bm R)_{a,\alpha} \right\rangle_{(0)},
\end{equation}
where $a$ is a supercell index and $\bm Z^*(\bm R)$ is the effective charges tensor. This vertex is the coupling for one phonon process.

The second and third components of $\responsevector'/\perturbationvector'$ contain the first derivatives of the effective charges, the second-order dipole moment:
\begin{equation}
\label{Ir 2 component}
    \left\langle \frac{\partial^2 \mathrm{p}_{\alpha}(\bm R)}{\partial R_a  \partial R_b} \right \rangle_{(0)} 
    =   \left\langle \frac{\partial Z^*(\bm R)_{a,\alpha}}{\partial R_b} \right\rangle_{(0)}.
\end{equation}

A Raman process consists in the scattering of light (usually visible) by zone-center phonons that induce a change in polarizability. The Raman cross-section contains the imaginary part of polarizability-polarizability response, $\Im[\chi_{\alpha_{\mu\nu},\alpha_{\eta\lambda}}(\omega + i0^+)]$, obtained with
\begin{equation}
    \mathcal{A}(\bm R) = \alpha(\bm R)_{\mu\nu} \qquad \mathcal{B}(\bm R) = \alpha(\bm R)_{\eta\lambda},
\end{equation}
where $\mu,\nu,\eta,\lambda$ are Cartesian directions. In a \added{non-resonant Stokes} Raman process phonons and photons scatter so we take into account the quantization of the electromagnetic field by 
 multiplying $\Im[\chi_{\alpha_{\mu\nu},\alpha_{\eta\lambda}}(\omega + i0^+)]$ by $1 + n(\omega)$ where $n(\omega)$ is the Bose-Einstein distribution for the photons. The first component of $\responsevector$ and $\perturbationvector$ gives one-phonon processes and contains:
\begin{equation}
\label{RAMAN 1 component}
    \left\langle \frac{\partial \alpha(\bm R)_{\mu\nu}}{\partial R_a} \right \rangle_{(0)} = \left\langle \Xi(\bm R)_{a,\mu\nu} \right\rangle_{(0)},
\end{equation}
where $\bm \Xi(\bm R)$ is the Raman tensor. As before, the other components of $\responsevector'$ and $\perturbationvector'$ depends on the second-order Raman polarizability:
\begin{equation}
\label{RAMAN 2 component}
    \left\langle \frac{\partial^2 \alpha(\bm R)_{\mu\nu}}{\partial R_a \partial R_b} \right \rangle_{(0)} =  \left\langle
    \frac{\partial \Xi(\bm R)_{a,\mu\nu}}{\partial R_b}
     \right\rangle_{(0)}.
\end{equation}

Eq.\ \eqref{Ir 2 component} and Eq.\ \eqref{RAMAN 2 component} trigger second-order IR/Raman processes \cite{rasetti_raman, fermi_rasetti_ramaneffect} exiting two phonons in the system, see Fig. \ref{fig:response diagrams}.
In principle, higher-order processes are possible, such as three-phonon etc. However, TDSCHA can not account for them as we showed that the three-phonon propagator is a disconnected diagram (see \secname~\ref{Multiple excitations in TDSCHA}).

\replaced{A two-phonon process, observable in both IR and Raman spectra, involves the scattering of photons and phonons while conserving both energy and momentum.}{A two-phonon process, both in IR and Raman spectra, is a scattering mechanism between photons and phonons that conserves both energy and momentum.} The long-wavelength electromagnetic field can either absorb or generate two phonons. Another possibility is that one phonon is absorbed and the other one is emitted interacting with photons. This involves pairs of phonons with opposite momentum in the Brillouin zone forming a continuum signal overlapped to the sharper peaks of one phonon process. 

This \replaced{phenomenon}{mechanism} is found both in harmonic and anharmonic systems. \replaced{ In systems like Si and Ge, which lack IR-active phonons due to inversion symmetry, two-phonon processes are essential for explaining the IR spectra \cite{IR_2_ph_Si_Ge}. Additionally, anharmonic systems such as liquid water \cite{water_ir_two} exhibit features resulting from effective charge position modulations. Two-phonon effects also play a significant role in many Raman spectra, including those of diamond and SiC \cite{SiC_raman_second, Diamond_raman_spectra}, as well as BaTiO$_3$ \cite{perovskite_raman}.}{The IR spectra of both Si and Ge can only be explained with these processes since there are no IR-active phonons due to inversion symmetry \cite{IR_2_ph_Si_Ge}. Also, anharmonic systems, such as liquid water \cite{water_ir_two}, present features due to effective charge position modulations. Two phonons effects play a central role in many Raman spectra: from diamond and SiC \cite{SiC_raman_second}\cite{Diamond_raman_spectra} to BaTiO$_3$ \cite{perovskite_raman}. } 

The most common approximation is
\begin{subequations}
\label{eff charges raman tensor harmonic}
\begin{align}
    & Z^*(\bm R)_{a,\alpha} \simeq Z^*(\boldsymbol{\mathcal{R}}^{(0)})_{a,\alpha},\\
    & \Xi(\bm R)_{a,\mu\nu} \simeq
    \Xi(\boldsymbol{\mathcal{R}}^{(0)})_{a,\mu\nu},
\end{align}
\end{subequations}
which suppresses all two phonon processes.

We use integration by parts and a Monte Carlo sampling, as proposed in \cite{TDSCHA_monacelli}, to compute all the components of $\responsevector'$ and $\perturbationvector'$ in an efficient and non-perturbative way using only effective charges and Raman tensor:
\begin{subequations}
\label{TDSCHA second order eff charges raman tensor}
\begin{align}
    & \left\langle \frac{\partial^2 \mathrm{p}(\bm R)_{\alpha}}{\partial \widetilde{R}_a  \partial \widetilde{R}_b} \right \rangle_{(0)} \hspace{-0.4cm}
    =  -\sum_{c=1}^{3N} \mathcal{G}^{(0)}(t=0^-)_{ac}\left\langle \delta \widetilde{R}^{(0)}_c \frac{Z_{b,\alpha}(\bm R)}{\sqrt{m_b}} \right\rangle_{(0)},\\
    & \left\langle \frac{\partial^2 \alpha(\bm R)_{\mu\nu}}{\partial \widetilde{R}_a  \partial \widetilde{R}_b} \right \rangle_{(0)} \hspace{-0.4cm}
    = -\sum_{c=1}^{3N} \mathcal{G}^{(0)}(t=0^-)_{ac}\left\langle \delta \widetilde{R}^{(0)}_c \frac{\Xi_{b,\mu\nu}(\bm R)}{\sqrt{m_b}} \right\rangle_{(0)}.
\end{align}
\end{subequations}
We remark that TDSCHA is the only method that computes second-order Raman tensors or effective charges with full position dependence without the need for higher-order DFT response. In Appendix \ref{Prepare IR/Raman spectra calculation} we report in detail how to prepare a IR/Raman calculation.

\section{Infrared spectra of high-pressure hydrogen}
\label{sec:hydrogen}
\begin{figure}[!htb]
    \centering
    \begin{minipage}[c]{1.0\linewidth}
    \includegraphics[width=1.0\textwidth]{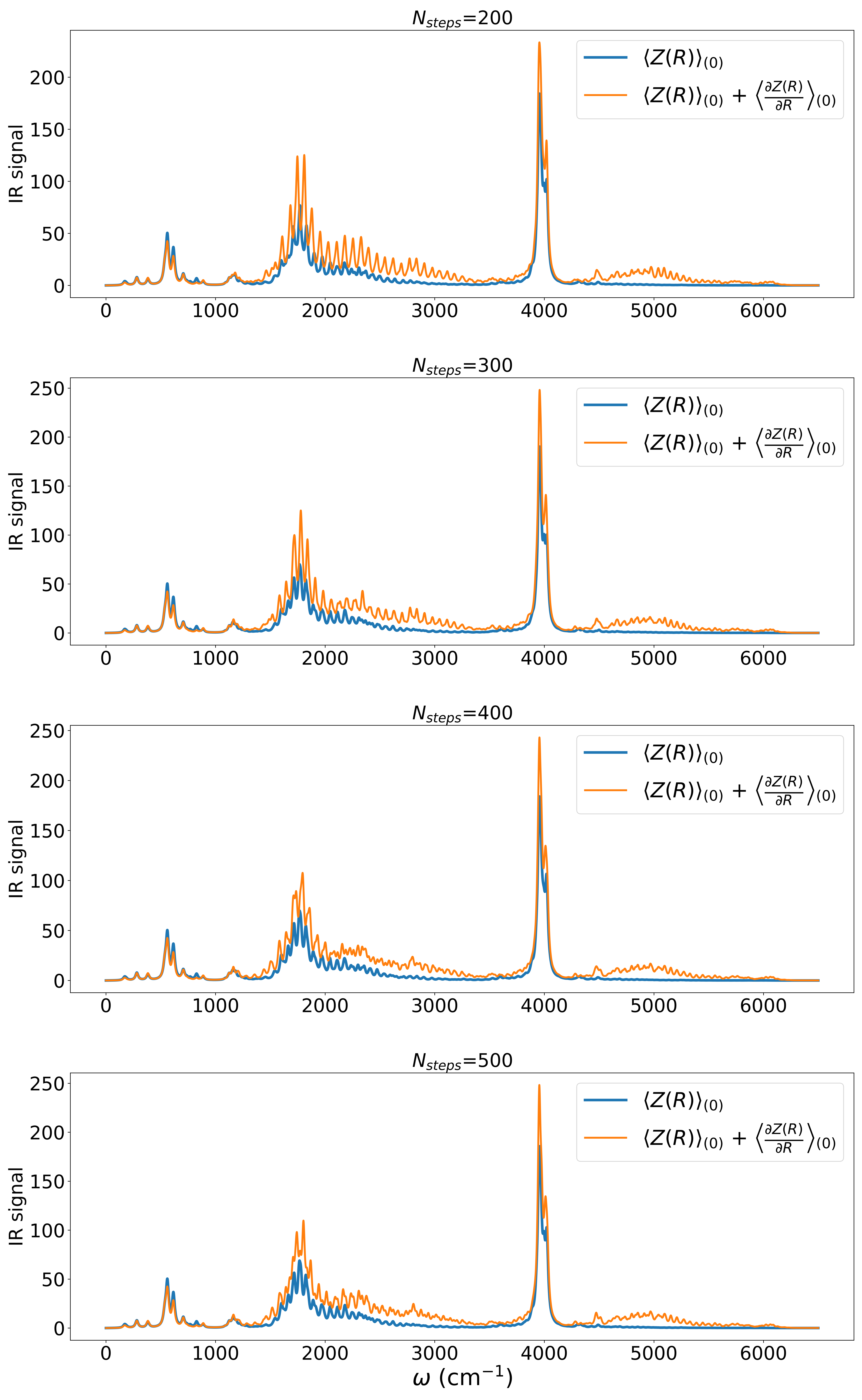}
    \end{minipage}
        \caption{Convergence of the IR signal with $\scriptstyle{\overset{(4)}{\bm D},\overset{(3)}{\bm D}\neq\bm 0}$  as a function of the Lanczos steps. The smearing $\delta$ is $15$ cm$^{-1}$.}
    \label{fig:hydrogen IR convergence}
\end{figure}

In this Section, we show the relevance of two phonon effects in a strongly anharmonic system such as high-pressure hydrogen phase III (C2c/24). We apply our new TDSCHA implementation on the infrared spectra of high-pressure hydrogen at $P=250$ GPa and $T= 0$ K including the effect of second-order effective charges, Eq.\ \eqref{Ir 2 component}. We employ $40000$ energy/forces and $2000$ effective charges calculations, on a $2\times2\times1$ supercell, to converge the anharmonic vertices, Eqs \eqref{def D3 main} \eqref{def D4 main}, and the IR overtone. Energies, forces and effective charges were computed using the BLYP functional \cite{BLYP} on a $4\times 4 \times 4$ k-grid (energy cutoff of $60$ Ry and $240$ Ry on the charge density)  as implemented in QUANTUM ESPRESSO \cite{Giannozzi2009,Giannozzi2017}, with a
plane wave basis set and a norm-conserving pseudopotential
from the PSEUDO DOJO library \cite{Pseudodojo}.
\begin{figure}[!htb]
    \centering
    \begin{minipage}[c]{1.0\linewidth}
    \includegraphics[width=1.0\textwidth]{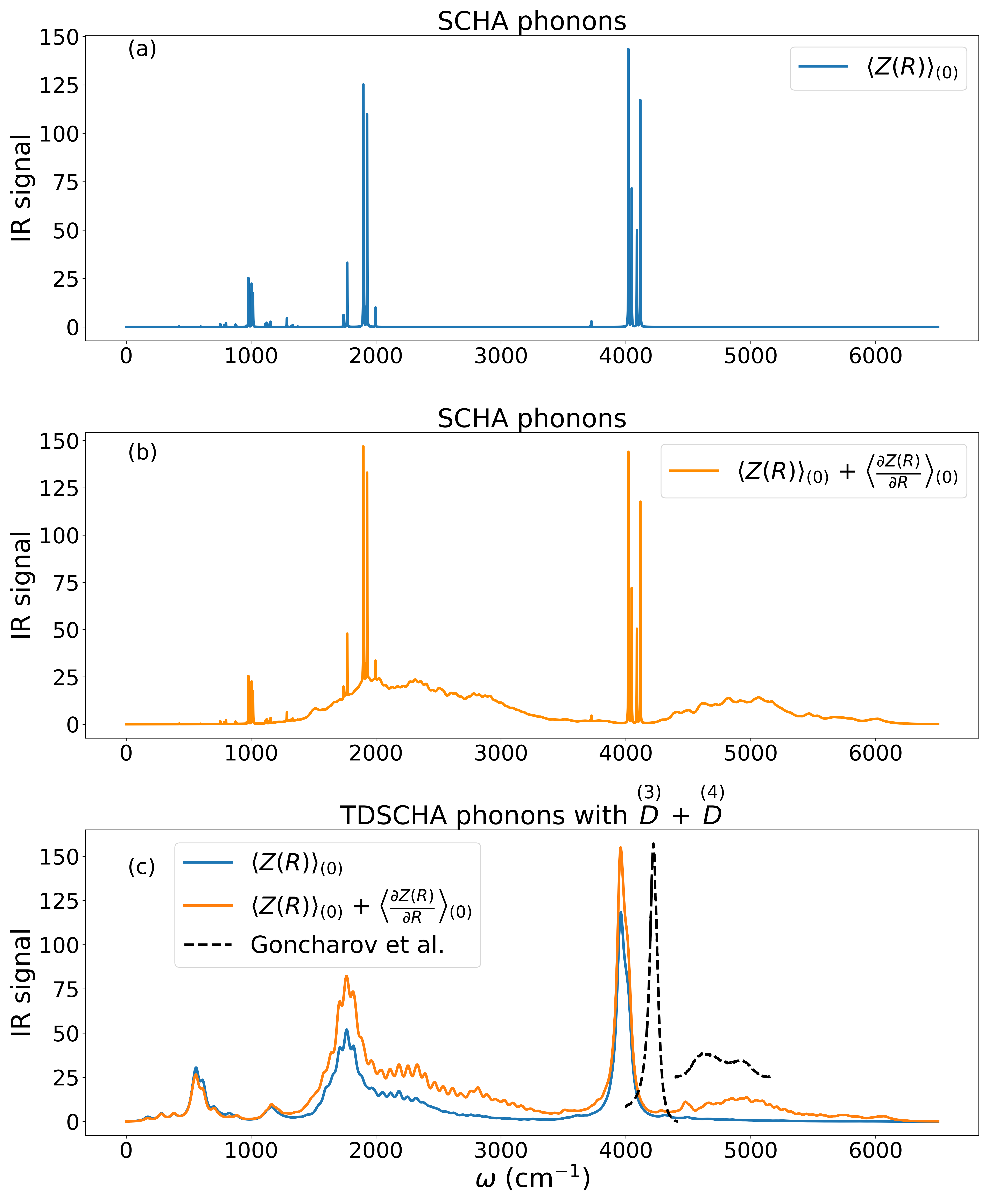}
    \end{minipage}
        \caption{Infrared signal for high pressure hydrogen at $P=250$ GPa and $T=0$ K.
        Panel (a) reports the IR spectra obtained with the SCHA phonons of Eq.\ \eqref{SCHA phonons} without two-phonon effects. These are included at the SCHA level in panel (b).
        Panel (c) reports the spectra setting both $\scriptstyle{\overset{(4)}{\bm D},\overset{(3)}{\bm D}\neq\bm 0}$ compared with the data extracted from Ref.\ \cite{Goncharov_hydrogen} at $248$ GPa and $20$ K. The smearing $\delta$ is $30$ cm$^{-1}$.}
    \label{fig:hydrogen IR}
\end{figure} \\
In Figure \ref{fig:hydrogen IR} we plot the IR signal using different approximations defined as
\begin{equation}
    \frac{1}{3}\sum_{\alpha}^{x,y,z}\Im\left[\chi(\omega+i\delta)_{\mathrm{p}_\alpha,\mathrm{p}_\alpha}\right]
\end{equation}
where $\delta$ is the smearing. In Fig.\ \ref{fig:hydrogen IR convergence} we plot the IR signal as a function of the Lanczos steps.
 \\
The convergence is achieved in $N_{\text{step}} = 500$ (Fig.\ \ref{fig:hydrogen IR convergence}) steps which are half of those employed in Ref.\ \cite{TDSCHA_monacelli} ($N_{\text{step}} = 1000$). This is due to a more stable Lanczos algorithm thanks to the symmetry of $\boldsymbol{\mathcal{L}}(\omega)$ in the Wigner formalism.

Panel (a) and (b) of Figure \ref{fig:hydrogen IR} show the effect of adding the second-order IR effects using the non-interacting SCHA phonons. The position modulation of effective charges generates a signal between $2000$-$4000$ cm$^{-1}$ and around $5000$ cm$^{-1}$. In panels (c) of Figure \ref{fig:hydrogen IR}  we add all the anharmonic interactions contained in TDSCHA. Notably, the two phonon processes at high frequency are stable after adding the anharmonic scattering of two phonons. This feature is in agreement with the overtone observed in the experiments by Goncharov et al.\ \cite{Goncharov_hydrogen}, confirming that it is a high-order IR process.

\section{Conclusions}
The Wigner picture simplifies the TDSCHA equations improving the physical intuition of the method. This allows us to discuss the equivalence of quantum and classical dynamics and rewrite the equations of motion in terms of position and momentum correlators.

\replaced{We have established a direct relationship between the response function and the diagrammatic expression of the interacting Green's function, which has allowed us to build a bridge to many-body perturbation theory.}{The response function is directly related to the diagrammatic expression of the interacting Green's function, through which we have built a bridge to many-body perturbation theory.} In the context of linear response theory, we clarified which diagrams and scattering processes are included in the method. 

The TDSCHA infrared spectra of high-pressure hydrogen phase III showed that only two phonon effects explain the overtone experimentally observed in Ref.\ \cite{Goncharov_hydrogen}.

\section*{Acknowledgements}
The authors acknowledge support by \replaced{European Union}{EU} under project ERC-SYN MORE-TEM (grant agreement No 951215) and the CINECA award under the ISCRA initiative, for the availability of high-performance computing resources and support. We also acknowledge PRACE for awarding us access to Joliot-Curie Rome at TGCC, France.

\clearpage

\appendix
\begin{widetext}

\section{Equations of motion}
\label{Equations of motion}
In this Appendix, we prove the Wigner-TDSCHA equations of motion Eqs \eqref{dynamical eqs motion delta P delta R}. For compactness, we define the mass-rescaled free parameters
\begin{equation}
\label{mass rescaled quantities TDSCHA}
\begin{aligned}
     & \widetilde{\alpha}(t)_{ab} = \frac{\alpha(t)_{ab}}{\sqrt{m_a m_b}} \quad \widetilde{\beta}(t)_{ab} = \sqrt{m_a m_b}\beta(t)_{ab}  \quad \widetilde{\gamma}(t)_{ab} = \sqrt{\frac{m_b}{m_a}} \gamma(t)_{ab} \\
     & \widetilde{R}_a = \sqrt{m_a} R_a \quad \widetilde{\mathcal{R}}(t)_a = \sqrt{m_a} \mathcal{R}(t)_a \\
     &  \widetilde{P}_a = \frac{  P_a}{\sqrt{m_a}} \quad \widetilde{\mathcal{P}}(t)_a = \frac{\mathcal{P}(t)_a}{\sqrt{m_a}}
\end{aligned}
\end{equation}
The equation of motion for the free parameters $\widetilde{\bm \alpha}(t), \widetilde{\bm \beta}(t), \widetilde{\bm \gamma}(t),\widetilde{\boldsymbol{\mathcal{R}}}(t),\widetilde{\boldsymbol{\mathcal{P}}}(t)$ are found with Eq.\ \eqref{Liouville equation for Wigner TDSCHA}:
\begin{equation}
\small
\label{appendix: Lioville equation TDSCHA}
\begin{aligned}
     & \pdv{\widetilde{\rho}(\bm R, \bm P,t)}{t} = \pdv{\mathcal{H}(\widetilde{\rho})}{\widetilde{\bm R}} \cdot\pdv{\widetilde{\rho}(\bm R, \bm P,t)}{\widetilde{\bm P}} - 
     \pdv{\mathcal{H}(\widetilde{\rho})}{\widetilde{\bm P}} \cdot\pdv{\widetilde{\rho}(\bm R, \bm P,t)}{\widetilde{\bm R}} = \\
     & = \left(\left\langle \pdv{V^{(\text{tot})}}{\widetilde{\bm R}} \right\rangle_{\widetilde{\rho}(t)}  +
     \delta \widetilde{\bm R}(t)\cdot\left\langle \pdv{V^{(\text{tot})}}{\widetilde{\bm R}}{\widetilde{\bm R}} \right\rangle_{\widetilde{\rho}(t)} 
      \right)\cdot
     \pdv{\widetilde{\rho}(\bm R, \bm P,t)}{\widetilde{\bm P}} - \left(\delta \widetilde{\bm P}(t) + \widetilde{\boldsymbol{\mathcal{P}}}(t)     \right)\cdot
     \pdv{\widetilde{\rho}(\bm R, \bm P, t)}{\widetilde{\bm R}}
\end{aligned}
\end{equation}
with $\delta\widetilde{\bm R}(t) = \widetilde{\bm R} - \widetilde{\boldsymbol{\mathcal{R}}}(t)$ and $\delta\widetilde{\bm P}(t) = \widetilde{\bm P} - \widetilde{\boldsymbol{\mathcal{P}}}(t)$.
The gradient of $\widetilde{\rho}(\bm R, \bm P, t)$, defined in Eq.\ \eqref{rho(R,P,t) gaussian TDSCHA}, is
\begin{subequations}
\small
\label{appendix: gradient rho TDSCHA}
\begin{align}
    &\pdv{\log\left(\widetilde{\rho}(\bm R, \bm P, t)\right)}{\widetilde{\bm P}}
      = - \widetilde{\bm\beta}(t)\cdot\delta \widetilde{\bm P}(t)
      + \widetilde{\bm\gamma}^T(t)\cdot\delta \widetilde{\bm R}(t), \\
      &\pdv{\log\left(\widetilde{\rho}(\bm R, \bm P, t)\right)}{ \widetilde{\bm R}}
      = -  \widetilde{\bm\alpha}(t)\cdot\delta \widetilde{\bm R}(t) + \widetilde{\bm\gamma}(t)\cdot\delta \widetilde{\bm P}(t) .
\end{align}
\end{subequations}
 The time derivative of $\widetilde{\rho}(\bm R, \bm P, t)$ gives
\begin{equation}
\label{appendix: time derivative rho TDSCHA}
\begin{aligned}
      &\frac{\partial \log\left(\widetilde{\rho}(\bm R, \bm P, t)\right)}{\partial t}
      = \frac{\dot{\mathcal{N}}(t)}{\mathcal{N}(t)}
      -\frac{1}{2}\delta \widetilde{\bm R}(t)\cdot 
      \dot{ \widetilde{\bm \alpha}} (t) \cdot \delta \widetilde{\bm R}(t)
      - \frac{1}{2}\delta \widetilde{\bm P}(t)\cdot
      \dot{ \widetilde{\bm \beta}}(t)\cdot \delta \bm P(t)
      +\delta \widetilde{\bm R}(t) \cdot\dot{ \widetilde{\bm \gamma}} (t)\cdot\delta \widetilde{\bm P}(t)
      +\\
      & + \dot{\widetilde{\boldsymbol{\mathcal{R}}}}(t)\cdot\bm \alpha(t) \cdot
      \delta \widetilde{\bm R}(t)
      + \dot{\widetilde{\boldsymbol{\mathcal{P}}}}(t)\cdot \widetilde{\bm \beta}(t)\cdot \delta \widetilde{\bm P}(t)
      - \dot{\widetilde{\boldsymbol{\mathcal{R}}}}(t) \cdot \widetilde{\bm \gamma}(t) \cdot\delta \widetilde{\bm P}(t)
      - \dot{\widetilde{\boldsymbol{\mathcal{P}}}}(t) \cdot \widetilde{\bm \gamma}^T(t) \cdot \delta \widetilde{\bm R}(t),
\end{aligned}
\end{equation}
where $\dot{\circ}$ denotes the time derivative. With Eqs \eqref{appendix: gradient rho TDSCHA} and Eq.\ \eqref{appendix: time derivative rho TDSCHA} the Wigner-Liouville equation Eq.\ \eqref{appendix: Lioville equation TDSCHA} becomes a polynomial in $\delta \widetilde{\bm R}(t)$ $\delta \widetilde{\bm P}(t)$ then, setting to zero the coefficients, we get the equations of motion for the free parameters
\begin{subequations}
\small
\label{dynamical eqs motion}
\begin{align}
    \frac{d}{d t} \widetilde{\boldsymbol{\mathcal{R}}}(t) = &\widetilde{\boldsymbol{\mathcal{P}}}(t) \label{Newton eq R}\\
     \frac{d}{d t} \widetilde{\boldsymbol{\mathcal{P}}}(t) =& - \left\langle \frac{\partial V^{(\text{tot})}}{\partial \widetilde{  \bm R}}\right\rangle_{\widetilde{\rho}(t)}\label{Newton eq P}\\ 
     \frac{d}{d t}\widetilde{ \boldsymbol{\alpha}}(t) =&-\left\langle\frac{\partial^2 V^{(\text{tot})}}{\partial \widetilde{\bm R} \partial \widetilde{ \bm R}}  \right\rangle_{\widetilde{\rho}(t)}\hspace{-0.3cm} \cdot \widetilde{\bm\gamma}^\dagger (t)  - \widetilde{\bm \gamma}(t) \cdot \left\langle\frac{\partial^2 V^{(\text{tot})}}{\partial \widetilde{ \bm R} \partial \widetilde{\bm  R}}\right\rangle_{\widetilde{\rho}(t)} \label{alpha equation motion}\\ 
    \frac{d}{d t}\widetilde{ \bm \beta}(t)=&  \widetilde{\bm \gamma}^\dagger(t)+ \widetilde{\bm  \gamma}(t)  \label{beta equation motion}\\ 
     \frac{d}{d t}\widetilde{ \bm \gamma}(t) = &\widetilde{\bm \alpha}(t)  - \left\langle\frac{\partial^2 V^{(\text{tot})}}{\partial \widetilde{ \bm R} \partial \widetilde{ \bm R}}\right\rangle_{\widetilde{\rho}(t)}\hspace{-0.3cm} \cdot \widetilde{ \bm \beta}(t)\label{gamma equation motion}
\end{align}
\end{subequations}
where $\circ^\dagger$ denotes the hermitian conjugate of a matrix. 
The equations of motion for the tensors keep the distribution normalized. 

The equal-time position and momentum correlators can be written in terms of $\widetilde{\bm \alpha}(t), \widetilde{\bm \beta}(t), \widetilde{\bm \gamma}(t)$
\begin{equation}
\label{correlators definition TDSCHA}
    \left\langle \delta \widetilde{X}(t)_a \delta \widetilde{Y}(t)_b\right\rangle_{\widetilde{\rho}(t)} = \int d\bm R \int d\bm P \widetilde{\rho}(\bm R, \bm P, t) \delta \widetilde{X}(t)_a \delta \widetilde{Y}(t)_b ,
    \qquad 
    \delta \widetilde{\bm X}(t), \delta \widetilde{\bm Y}(t) = \delta \widetilde{\bm R}(t), \delta \widetilde{\bm P}(t),
\end{equation}
and using Gaussian integration we get the following expressions
\begin{subequations}
\label{appendix: correlators R-P}
\begin{align}
      &\left\langle \delta \widetilde{\bm R}(t) \delta \widetilde{\bm R}(t) \right\rangle_{\widetilde{\rho}(t)} = \left(\widetilde{\bm \alpha}(t) - \widetilde{\bm \gamma}(t) \cdot \widetilde{\bm \beta}^{-1}(t) \cdot \widetilde{\bm \gamma}^T(t) \right)^{-1} ,\\
      &\left\langle \delta \widetilde{\bm R}(t) \delta \widetilde{\bm P}(t) \right\rangle_{\widetilde{\rho}(t)}  = 
      \widetilde{\bm \alpha}^{-1}(t) \cdot \widetilde{\bm \gamma}(t) \cdot\left(\widetilde{\bm \beta}(t) - \widetilde{\bm \gamma}^T(t)\cdot\widetilde{\bm \alpha}^{-1}(t)\cdot \widetilde{\bm \gamma}(t)\right)^{-1}, \\
      &\left\langle \delta \widetilde{\bm P}(t) \delta \widetilde{\bm P}(t) \right\rangle_{\widetilde{\rho}(t)}  = 
      \left(\widetilde{\bm \beta}(t) - \widetilde{\bm \gamma}^T(t)\cdot\widetilde{\bm \alpha}^{-1}(t)\cdot \widetilde{\bm \gamma}(t)\right)^{-1}.
\end{align}
\end{subequations}
Then deriving with respect to time Eqs. \eqref{appendix: correlators R-P} and using the equations of motion Eqs. \eqref{dynamical eqs motion} we prove  Eqs. \eqref{dynamical eqs motion delta P delta R}.

\section{Equivalence with Time-Dependent Self-consistent Harmonic Approximation}
\label{Equivalence with Time-Dependent Self-consistent Harmonic Approximation}
In this Appendix, we show that our method is a Wigner reformulation of TDSCHA presented in \cite{TDSCHA_monacelli}. We compute the matrix elements of the von Neumann density operator corresponding to the Wigner distribution, Eq.\ \eqref{rho(R,P,t) gaussian TDSCHA}. To do this we need the inverse of the Wigner transformation which is defined as (see Ref.\ \cite{imre1967wigner})
\begin{equation}
\label{def inverse rho Wigner}
\begin{aligned}
    & \hat{\rho} = 
    \int  \frac{d \bm R  d\bm R' d \bm P d\bm P' }{(2\pi\hbar)^{3N}}
    \rho(\bm R', \bm P') \exp{-\frac{i}{\hbar}\left[\bm P \cdot \left(\hat{\bm R} - \bm R'\right)  + \bm R \cdot\left( \hat{\bm P} - \bm P' \right) \right]},
\end{aligned}
\end{equation}
where $\rho(\bm R', \bm P')$ is the Wigner quasi-distribution. The $\hat{\circ}$ indicates quantum operators.

Inserting the Wigner distribution of Eq.\ \eqref{rho(R,P,t) gaussian TDSCHA} in Eq.\ \eqref{def inverse rho Wigner} we get a Gaussian integral for the density operator matrix elements 
\begin{equation}
\label{rho(R,R',t) TDSCHA}
\begin{aligned}
    \bra{\bm R }\hat{\widetilde{\rho}}(t)\ket{\bm R'} =& \int d\bm P \exp\left(\frac{i}{\hbar}(\bm R - \bm R') \cdot \bm P\right)  \widetilde{\rho}\left(\frac{\bm R + \bm R'}{2}, \bm P,t \right)  \\
     = &\mathcal{N}(t) \exp\left[-\frac{1}{8}\left(\delta \bm R(t) + \delta \bm R'(t)\right)\cdot\bm \alpha(t)\cdot\left(\delta \bm R(t) + \delta \bm R'(t)\right) + \frac{i}{\hbar}(\bm R - \bm R') \cdot \boldsymbol{\mathcal{P}}(t)\right] \\
    & \int d\bm P  \exp\left\{-\frac{1}{2}\delta \bm P(t)\cdot \bm \beta(t)\cdot \delta\bm P(t) +\frac{1}{2}\left[\frac{2i}{\hbar}(\bm R - \bm R') + (\delta \bm R(t) + \delta \bm R'(t))\cdot \bm \gamma(t) \right]\cdot \delta\bm P(t)\right\}  \\
    = & \sqrt{\det\left(\frac{\bm \Upsilon(t)}{2\pi}\right)}
    \exp\biggl\{i\bm Q(t) \cdot (\bm R- \bm R')  \\
    &- \left(\bm R - \boldsymbol{\mathcal{R}}(t)\right)\cdot\left(\frac{1}{4}\replaced{\boldsymbol{\theta}(t)}{\boldsymbol{\Theta}(t)} + i \bm C(t)\right)\cdot \left(\bm R - \boldsymbol{\mathcal{R}}(t)\right)
    - \left(\bm R' - \boldsymbol{\mathcal{R}}(t)\right)\cdot  \left(\frac{1}{4}\replaced{\boldsymbol{\theta}(t)}{\boldsymbol{\Theta}(t)} - i \bm C(t)\right)\cdot  \left(\bm R' - \boldsymbol{\mathcal{R}}(t)\right) \\
    &
     +\left(\bm R - \boldsymbol{\mathcal{R}}(t)\right)\cdot \left(\Re \bm A(t) + i \Im \bm A(t)\right)\cdot \left(\bm R' - \boldsymbol{\mathcal{R}}(t)\right)\biggl\}.
\end{aligned}
\end{equation}
The last line of Eq.\ \eqref{rho(R,R',t) TDSCHA} is the trial density operator used in \cite{TDSCHA_monacelli}.
The free parameters used in \cite{TDSCHA_monacelli}  $\bm Q (t)$, $\replaced{\boldsymbol{\theta}(t)}{\boldsymbol{\Theta}(t)}$, $\bm C (t)$, $\Re \bm A(t)$, $\Im \bm A(t)$ are related to the ones used in the Wigner formalism
\begin{subequations}
\label{Theta C A Upsilon -> alpha beta gamma}
\begin{align}
       & \bm Q (t)= \frac{1}{\hbar} \boldsymbol{\mathcal{P}}(t),\\
    & \replaced{\boldsymbol{\theta}(t)}{\boldsymbol{\Theta}(t)}= \frac{1}{2} \left(\bm \alpha(t) - \bm \gamma(t)\cdot \bm \beta^{-1}(t) \cdot\bm \gamma^T(t)\right)  + \frac{2}{\hbar^2}\bm \beta^{-1}(t) ,\\
    & \bm{C}(t) = -\frac{1}{4\hbar^2}\left(\bm{\beta}^{-1}(t)\cdot\bm{\gamma}^T(t) + \bm{\gamma}(t)\cdot\bm{\beta}^{-1}(t) \right),\\
    & \Re \bm A(t) = -\frac{1}{4}\left(\bm \alpha(t) - \bm \gamma(t)\cdot \bm \beta^{-1}(t) \cdot\bm \gamma^T(t)\right)  +\frac{1}{\hbar^2}\bm \beta^{-1}(t),\\
    & \Im \bm A(t) = \frac{1}{2\hbar}
    \left(\bm \gamma(t)\cdot \bm \beta^{-1}(t)   - \bm \beta^{-1}(t) \cdot\bm \gamma^T(t) \right) \\
    & \bm \Upsilon(t) = \replaced{\boldsymbol{\theta}(t)}{\boldsymbol{\Theta}(t)} - 2\Re \bm A(t) = \bm \alpha(t) - \bm \gamma(t) \cdot\bm \beta^{-1}(t)\cdot \bm \gamma^T(t),
\end{align}
\end{subequations}
where $\circ^{-1}$ denotes the inverse of a matrix. The tensor $\bm \Upsilon(t)$ is a linear combination of $\replaced{\boldsymbol{\theta}(t)}{\boldsymbol{\Theta}(t)}$ and $\Re \bm A(t)$.
The same notation for the average position $\boldsymbol{\mathcal{R}}(t)$ is adopted. Using the relations between free parameters, Eqs \eqref{Theta C A Upsilon -> alpha beta gamma}, it is easy to prove that the equations of motion Eqs \eqref{dynamical eqs motion} are equivalent to the TDSCHA ones reported in \cite{TDSCHA_monacelli}.

Here, we also prove that Eq.\ \eqref{rho(R,P) gaussian SCHA} is the Wigner transform of the SCHA equilibrium density matrix $\hat{\widetilde{\rho}}^{(0)}$ \cite{SCHA_main,TDSCHA_monacelli}:
\begin{equation}
\begin{aligned}
    \bra{\bm R}\hat{\widetilde{\rho}} ^{(0)}\ket{\bm R'} = &\sqrt{\det\left(\frac{\boldsymbol{\Upsilon}^{(0)}}{2\pi}\right)}\exp\left[-\frac{1}{4}\sum_{ab=1}^{3N}\replaced{\theta^{(0)}_{ab}}{\Theta^{(0)}_{ab}}(R_a - \mathcal{R}^{(0)}_a)(R_b - \mathcal{R}^{(0)}_b) -\frac{1}{4}\sum_{ab=1}^{3N}\replaced{\theta^{(0)}_{ab}}{\Theta^{(0)}_{ab}}(R'_a - \mathcal{R}^{(0)}_a)(R'_b - \mathcal{R}^{(0)}_b) \right.\\
    &\left. +\sum_{ab=1}^{3N}A^{(0)}_{ab}
    (R_a - \mathcal{R}^{(0)}_a)(R'_b - \mathcal{R}^{(0)}_b)\right]
\end{aligned}
\end{equation}
with $\boldsymbol{\Upsilon}^{(0)} = \replaced{\boldsymbol{\theta}^{(0)}}{\boldsymbol{\Theta}^{(0)}} - 2\boldsymbol{A}^{(0)}$, where $\boldsymbol{\Upsilon}^{(0)}$ and $\boldsymbol{A}^{(0)}$ are defined as
\begin{equation}
    \Upsilon^{(0)}_{ab} =\sqrt{m_a m_b} \sum_{\mu=1}^{3N}\frac{2\omega_\mu}{\hbar(1 + 2n_\mu)} e^a_\mu e^b_\mu,
    \qquad 
    A^{(0)}_{ab} =\sqrt{m_a m_b} \sum_{\mu=1}^{3N}\frac{2\omega_\mu n_\mu(1 + n_\mu)}{\hbar(1 + 2 n_\mu)} e^a_\mu e^b_\mu,
\end{equation}
where ${\omega^2_\mu}$ and $\{\bm e_\mu\}$ are the auxiliary SCHA modes Eq.\ \eqref{SCHA phonons}. The Wigner quasi-distribution, according to Eq.\ \eqref{def rho Wigner}, is obtained in the following way
\begin{equation}
\begin{aligned}
   \widetilde{\rho}^{(0)}(\bm R, \bm P) =& \sqrt{\det\left(\frac{\boldsymbol{\Upsilon}^{(0)}}{2\pi}\right)} \exp\left[-\frac{1}{2}\sum_{ab=1}^{3N}(R_a - \mathcal{R}_a)\Upsilon^{(0)}_{ab}(R_b - \mathcal{R}_b)\right] \\
     &\int \frac{d^{3N}\bm R' }{(4\pi\hbar^2)^{3N/2}}\exp\left[i \sum_{a=1}^{3N} \frac{ P_a R'_a}{\hbar}-\frac{1}{8}\sum_{ab=1}^{3N}(\replaced{\theta^{(0)}_{ab}}{\Theta^{(0)}_{ab}} + 2 A^{(0)}_{ab})R'_a R'_b\right]\\
    =&\sqrt{\det\left(\frac{\boldsymbol{\Upsilon}^{(0)}}{2\pi}\right)}
     \sqrt{\det\left[\frac{1}{2\pi\hbar^2}\left(\boldsymbol{A}^{(0)} + \frac{1}{4}\boldsymbol{\Upsilon}^{(0)}\right)^{-1}\right]}\\ &\exp\left[-\frac{1}{2}\sum_{ab=1}^{3N}(R_a - \mathcal{R}^{(0)}_a)\Upsilon^{(0)}_{ab}(R_b - \mathcal{R}^{(0)}_b)-\frac{1}{2}\sum_{ab=1}^{3N}P_a \left(\hbar^2\boldsymbol{A}^{(0)} + \frac{\hbar^2}{4}\boldsymbol{\Upsilon}^{(0)}\right)^{-1}_{ab} P_b\right].
\end{aligned}
\end{equation}
The final result is a positive-definite Gaussian Wigner distribution which coincides with Eq.\ \eqref{rho(R,P) gaussian SCHA}
\begin{equation}
\label{rho(R,P) eq with alpha beta}
    \widetilde{\rho}^{(0)}(\bm R, \bm P) = \sqrt{\det\left(\frac{\boldsymbol{\alpha}^{(0)}}{2\pi}\right)\det\left(\frac{\boldsymbol{\beta}^{(0)}}{2\pi}\right)} \exp\left[-\frac{1}{2}\sum_{ab=1}^{3N}(R_a - \mathcal{R}^{(0)}_a)\alpha^{(0)}_{ab}(R_b - \mathcal{R}^{(0)}_b)-\frac{1}{2}\sum_{ab=1}^{3N}P_a \beta^{(0)}_{ab} P_b\right],
\end{equation}
once we recognize that
\begin{equation}
\label{alpha static R-R}
        \left\langle \delta \bm R \delta \bm R \right\rangle_{(0)} = {\bm \alpha^{(0)}}^{-1} = {\bm \Upsilon^{(0)}}^{-1},
\end{equation}
and
\begin{equation}
\label{beta static P-P}
        \left\langle  \bm P \bm P \right\rangle_{(0)} = {\bm \beta^{(0)}}^{-1} = \hbar^2\left(\boldsymbol{A}^{(0)} + \frac{1}{4}\boldsymbol{\Upsilon}^{(0)}\right),
\end{equation}
where the equilibrium correlators are defined in Eqs \eqref{RR PP static}.

\section{Energy conservation}
\label{Energy conservation}
In this Appendix, we show that the TDSCHA equations of motion Eqs \eqref{dynamical eqs motion delta P delta R} satisfy the energy conservation principle. The Wigner quantum time-dependent Hamiltonian has the same form as the classical one
\begin{equation}
    H(t) = \sum_{a=1}^{3N} \frac{ P^2_a}{2m_a} + V^{(\text{BO})}(\bm R) + V^{(\text{ext})}(\bm R, t).
\end{equation}
We compute the total time-derivative of $\left\langle H(t)\right\rangle_{\widetilde{\rho}(t)}$ where $\widetilde{\rho}(t)$ is defined in Eq.\ \eqref{rho(R,P,t) gaussian TDSCHA}:
\begin{equation}
    \dv{\left\langle H(t)\right\rangle_{\widetilde{\rho}(t)}}{t} = \frac{d}{d t}\left\{
     \sum_{a=1}^{3N}\frac{1}{2}\left[\left\langle\delta \widetilde{P}(t)_a \delta \widetilde{P}(t)_a\right\rangle_{\widetilde{\rho}(t)} + \widetilde{\mathcal{P}}(t)_a^2\right] + 
    \left\langle V^{(\text{tot})} \right\rangle_{\widetilde{\rho}(t)}\right\}.
\end{equation}
The time derivative of the kinetic energy gives:
\begin{equation}
\begin{aligned}
    & \frac{d}{d t}
    \sum_{a=1}^{3N} \frac{1}{2}\left[\left\langle\delta \widetilde{P}(t)_a \delta \widetilde{P}(t)_a\right\rangle_{\widetilde{\rho}(t)} + \widetilde{\mathcal{P}}(t)_a^2\right]   = \frac{1}{2}  \sum_{a=1}^{3N} \dv{\left\langle\delta \widetilde{P}(t)_a \delta \widetilde{P}(t)_a\right\rangle_{\widetilde{\rho}(t)}}{t}
     + \sum_{a=1}^{3N} \widetilde{\mathcal{P}}(t)_a  \dv{\widetilde{\mathcal{P}}(t)_a}{t} \\
    & = \frac{1}{2} \Tr\left[\dv{\left\langle\delta \widetilde{\bm P}(t) \delta \widetilde{\bm P}(t)\right\rangle_{\widetilde{\rho}(t)}}{t}\right] - \widetilde{\boldsymbol{\mathcal{P}}}(t)\cdot  \left\langle \pdv{ V^{(\text{tot})}}{ \widetilde{\bm R}} \right \rangle_{\widetilde{\rho}(t)}.
\end{aligned}
\end{equation}
The derivative of the total potential average is more involved since the position probability distribution depends on time through $\boldsymbol{\mathcal{R}}(t)$ and $\left\langle \delta \widetilde{\bm R}(t) \delta \widetilde{\bm R}(t) \right\rangle_{\widetilde{\rho}(t)}$. The derivative is worked out using the formulas proved in Ref.\ \cite{Bianco}:
\begin{equation}
\begin{aligned}
    & \dv{ \left\langle V^{(\text{tot})} \right\rangle_{\widetilde{\rho}(t)}}{t}= \left\langle \pdv{ V^{(\text{ext})}}{ t} \right \rangle + \sum_{a=1}^{3N} \dv{\widetilde{\mathcal{R}}(t)_a}{t} \left\langle \pdv{ V^{(\text{tot})}}{ \widetilde{\mathcal{R}}(t)_a} \right \rangle_{\widetilde{\rho}(t)}  + \frac{1}{2}\sum_{ab=1}^{3N} \dv{\left\langle\delta \widetilde{ R}(t)_a \delta \widetilde{ R}(t)_b\right\rangle_{\widetilde{\rho}(t)}}{t}\left\langle\pdv{ V^{(\text{tot})}}{ \widetilde{R}_b}{ \widetilde{R}_a} \right\rangle_{\widetilde{\rho}(t)} \\
    &= \left\langle \pdv{ V^{(\text{ext})}}{ t} \right \rangle_{\widetilde{\rho}(t)} +  \widetilde{\boldsymbol{\mathcal{P}}}(t) \cdot \left\langle \pdv{ V^{(\text{tot})}}{ \widetilde{\bm R}} \right \rangle_{\widetilde{\rho}(t)}  + \frac{1}{2}\Tr\left[ \dv{\left\langle\delta \widetilde{\bm R}(t) \delta \widetilde{\bm R}(t)\right\rangle_{\widetilde{\rho}(t)}}{t}\cdot\left\langle\pdv{ V^{(\text{tot})}}{ \widetilde{\bm R}}{ \widetilde{\bm R}} \right\rangle_{\widetilde{\rho}(t)}\right] .
\end{aligned}
\end{equation}
Then using Eqs \eqref{delta R delta R equation}-\eqref{delta P delta P equation} and the permutation properties of the trace it is shown that:
\begin{equation}
    \frac{1}{2} \Tr\left[ \frac{d}{d t}  \left\langle\delta \widetilde{\bm P}(t) \delta \widetilde{\bm P}(t)\right\rangle\right] = - \frac{1}{2}\Tr\left[ \dv{\left\langle\delta \widetilde{\bm R}(t) \delta \widetilde{\bm R}(t)\right\rangle_{\widetilde{\rho}(t)}}{t}\cdot\left\langle\pdv{ V^{(\text{tot})}}{ \widetilde{\bm R}}{ \widetilde{\bm R}} \right\rangle_{\widetilde{\rho}(t)}\right].
\end{equation}
So in the end, we found:
\begin{equation}
    \dv{\left\langle H(t)\right\rangle_{\widetilde{\rho}(t)}}{t} =  \left\langle \pdv{ V^{(\text{ext})}}{ t} \right \rangle_{\widetilde{\rho}(t)}.
\end{equation}
This derivation is more compact than the one presented in Ref.\ \cite{TDSCHA_monacelli}.

\section{Expansion of the probability distribution}
\label{Expansion of the probability distribution}
In this Appendix, we show how to expand at first order the TDSHCA position probability distribution and how the anharmonic vertices, Eqs \eqref{def D3 main} \eqref{def D4 main}, emerge.
All the free parameters are perturbed with respect to their static value (denoted by $(0)$)
\begin{subequations}
\label{pert parameters app}
\begin{align}
    \boldsymbol{\mathcal{R}}(t) = & \boldsymbol{\mathcal{R}}^{(0)} + \boldsymbol{\mathcal{R}}^{(1)}(t) \\
    \boldsymbol{\mathcal{P}}(t) = & \boldsymbol{\mathcal{P}}^{(1)}(t) \\
    \bm \alpha(t) = & \bm \alpha^{(0)} + \bm \alpha^{(1)}(t) \\
    \bm \beta(t) = & \bm \beta^{(0)} + \bm \beta^{(1)}(t) \\
    \bm \gamma(t) = &  \bm \gamma^{(1)}(t)
\end{align}
\end{subequations}
The first thing to do is to expand at first order the position probability distribution in the perturbative free parameters, i.e. those denoted by the superscript $(1)$. Before performing the expansion, we report the full position probability distribution obtained from Eq.\ \eqref{rho(R,P,t) gaussian TDSCHA}
\begin{equation}
\begin{aligned}
    & \widetilde{\rho}(\bm R, t) =  \sqrt{\det\left(\frac{\bm \alpha(t) - \bm\gamma(t)\cdot \overset{-1}{\bm \beta}(t)\cdot\bm \gamma(t)^T}{2\pi}\right)} \\
    & \exp\left[ - \frac{1}{2}(\bm R - \boldsymbol{\mathcal{R}}(t))\cdot\left(\bm \alpha(t) - \bm\gamma(t) \cdot\overset{-1}{\bm \beta}(t)\cdot\bm \gamma(t)^T\right)\cdot(\bm R - \boldsymbol{\mathcal{R}}(t))\right].
\end{aligned}
\end{equation}
The leading order is controlled by $\bm \alpha^{(0)} + \bm \alpha^{(1)}(t)$. We define the displacements with respect the equilibrium position as $\delta \widetilde{\bm R}^{(0)} = \widetilde{\bm R} - \widetilde{\boldsymbol{\mathcal{R}}}^{(0)}$. The expansion gives
\begin{equation}
\label{weights perurbed}
\begin{aligned}
    & \widetilde{\rho}(\bm R, t)  = \widetilde{\rho}^{(0)}(\bm R) +   \widetilde{\rho}^{(1)}(\bm R,t),
\end{aligned}
\end{equation}
where $\widetilde{\rho}^{(0)}(\bm R)$ is the equilibrium probability distribution (see Eq.\ \eqref{rho(R,P) gaussian SCHA})
\begin{equation}
\label{rho 0}
\begin{aligned}
    &\widetilde{\rho}^{(0)}(\bm R) = \sqrt{\det\left(\frac{\bm \alpha^{(0)}}{2\pi}\right)} \exp\left(-\frac{1}{2}\delta \widetilde{\bm R}^{(0)} \cdot \widetilde{\bm \alpha}^{(0)}\cdot \delta \widetilde{\bm R}^{(0)} \right).
\end{aligned}
\end{equation}
The explicit expression for $\widetilde{\rho}^{(1)}(\bm R,t)$ in Eq.\ \eqref{weights perurbed}, following \cite{TDSCHA_monacelli}, is
\begin{equation}
\label{rho 1}
    \widetilde{\rho}^{(1)}(\bm R,t) = \widetilde{\rho}^{(0)}(\bm R) \left\{\frac{1}{2}Tr\left[{\bm \alpha^{(0)}}^{-1} \cdot \bm \alpha^{(1)}(t)\right]   -\frac{1}{2}\delta \bm R^{(0)}\cdot\bm \alpha^{(1)}(t)\cdot \delta \bm R^{(0)} 
    +\delta \bm R^{(0)}\cdot \bm \alpha^{(0)}\cdot\boldsymbol{\mathcal{R}}^{(1)}(t) \right\}.
\end{equation}
Next, we derive an expression for the perturbed averages of a position-dependent observable $O(\bm R)$. 
Using the expression for $\widetilde{\rho}^{(1)}(\bm R,t)$, Eq.\ \eqref{rho 1}, and integration by parts we get
\begin{equation}
\label{O on rho 1}
\begin{aligned}
    & \left\langle O \right\rangle_{(1)} (t) = \int d\bm R \widetilde{\rho}^{(1)}(\bm R, t) O(\bm R)  \\
    & = -\frac{1}{2}\sum_{ab=1}^{3N} \widetilde{\alpha}^{(1)}(t)_{ab} 
    \left( \left\langle \delta \widetilde{R}^{(0)}_a \delta \widetilde{R}^{(0)}_b O \right\rangle_{(0)} - ({\widetilde{\alpha}^{(0)}})^{-1}_{ba}
     \left\langle O \right\rangle_{(0)}\right) 
     + \sum_{a=1}^{3N} \widetilde{\mathcal{R}}^{(1)}_a(t) 
    \left\langle \frac{\partial O}{\partial \widetilde{R}_a} \right\rangle_{(0)}.
\end{aligned}
\end{equation}
Note that now all the averages have to be performed on the equilibrium ensemble.

We introduce the equilibrium three and four phonon scattering vertices as in \cite{Bianco}
\begin{subequations}
\label{D3 D4 app}
\small
\begin{align}
    & \overset{(3)}{D}_{abc} = \left\langle \frac{\partial V^{(\text{BO})}}{\partial \widetilde{R}_{a} \partial \widetilde{R}_{b} \partial \widetilde{R}_{c}} \right\rangle_{(0)} =
    \sum_{mn=1}^{3N} \widetilde{\alpha}^{(0)}_{an}  \widetilde{\alpha}^{(0)}_{bm} 
    \left\langle \delta \widetilde{R}_n \delta \widetilde{R}_m
    \frac{\partial V^{(\text{BO})}}{ \partial \widetilde{R}_c}\right\rangle_{(0)}- \widetilde{\alpha}^{(0)}_{ab} \left\langle  \frac{\partial V^{(\text{BO})}}{\partial \widetilde{R}_c}\right\rangle_{(0)}
    \label{D3}\\
    & \overset{(4)}{D}_{abcd} = \left\langle \frac{\partial V^{(\text{BO})}}{\partial \widetilde{R}_{a} \partial \widetilde{R}_{b} \partial \widetilde{R}_{c} \partial \widetilde{R}_{d}} \right\rangle_{(0)}= 
    \sum_{nm=1}^{3N} \widetilde{\alpha}^{(0)}_{an} \widetilde{\alpha}^{(0)}_{bm}  \left\langle \delta \widetilde{R}_n \delta \widetilde{R}_m \frac{\partial ^2V^{(\text{BO})}}{\partial \widetilde{R}_c \partial \widetilde{R}_d}\right\rangle_{(0)}
     - \widetilde{\alpha}^{(0)}_{ab} \left\langle \frac{\partial ^2V^{(\text{BO})}}{\partial \widetilde{R}_c \partial \widetilde{R}_d}\right\rangle_{(0)}  . \label{D4}
\end{align}
\end{subequations}
It is convenient also to introduce the potential $\mathbb{V}(\bm R)$ as the difference between the \text{BO} potential and the harmonic auxiliary potential obtained at equilibrium
\begin{equation}
\label{mathbb(V) = V_BO - V_SCHA}
    \mathbb{V}(\bm R) = V^{(\text{BO})}(\bm R) - \frac{1}{2} \delta \widetilde{\bm R}^{(0)} \cdot  \overset{(2)}{ \bm D}\cdot \delta \widetilde{\bm R}^{(0)},
\end{equation}
where $\overset{(2)}{\bm D}$ defines the SCHA phonons, Eq.\ \eqref{SCHA phonons}. 
Using Eq.\ \eqref{O on rho 1}, we relate the perturbed averages of $\mathbb{V}(\bm R)$ to the scattering vertices of Eqs \eqref{D3 D4 app}
\begin{subequations}
\label{derivative of V_bo - V_scha explicit}
\begin{align}
   & \left\langle \frac{\partial \mathbb{V}}{\partial \widetilde{R}_a}\right\rangle_{(1)} =  \frac{1}{2}\sum_{bcde=1}^{3N}\widetilde{\alpha}^{(1)}(t)_{de}(\widetilde{\alpha}^{(0)}\,\!^{-1})_{db}
   (\widetilde{\alpha}^{(0)}\,\!^{-1})_{ec}\overset{(3)}{D}_{abc} ; \label{dV1}\\
    & \left\langle \frac{\partial^2 \mathbb{V}}{\partial \widetilde{R}_a \partial \widetilde{R}_b}\right\rangle_{(1)} = 
    \sum_{c=1}^{3N} \overset{(3)}{D}_{abc} \widetilde{\mathcal{R}}^{(1)}(t)_c  -\frac{1}{2}\sum_{cdef=1}^{3N}\widetilde{\alpha}^{(1)}(t)_{ef}(\widetilde{\alpha}^{(0)}\,\!^{-1})_{ec}(\widetilde{\alpha}^{(0)}\,\!^{-1})_{fd}\overset{(4)}{D}_{abcd}. \label{dV2}
\end{align}
\end{subequations}
At this point it is straightforward to get the following perturbed averages for the \text{BO} potential:
\begin{subequations}
\label{derivative of V_bo - Vscha pert average}
\begin{align}
    \left\langle \frac{\partial \mathbb{V}}{\partial \widetilde{R}_a}\right\rangle_{(1)} =  &\left\langle \frac{\partial V^{(\text{BO})}}{\partial \widetilde{R}_a}\right\rangle_{(1)} - \sum_{b=1}^{3N} \overset{(2)}{D}_{ab} \widetilde{\mathcal{R}}^{(1)}_b(t) , \label{pertrubed average anharmonic 1} \\
     \left\langle \frac{\partial^2 \mathbb{V}}{\partial \widetilde{R}_a \partial \widetilde{R}_b}\right\rangle_{(1)}
     =  & \left\langle \frac{\partial^2 V^{(\text{BO})}}{\partial \widetilde{R}_a \partial \widetilde{R}_b}\right\rangle_{(1)}
    \label{pertrubed average anharmonic 2}.
\end{align}
\end{subequations}

\section{Derivation of the linear response system}
\label{Derivation of the linear response system}
In this Appendix, we prove the linearized equations of motion discussed in Section \ref{Linearized equations of motion and general response function}. To do this we write all the supercell tensors in the static equilibrium polarization basis $\{\bm e_\mu\}$ defined in Eq.\ \eqref{SCHA phonons}. So a multi-indices tensor $A(t)_{a_1,..,a_N}$ defined in the supercell can be written in the polarization basis as
\begin{equation}
\label{pol basis}
     A(t)_{\mu_1,..,\mu_N} = \sum_{a_1,..,a_N=1}^{3N} e^{a_1}_{\mu_1} .. e^{a_N}_{\mu_N} A(t)_{a_1,..,a_N}.
\end{equation}
From now on all the quantities are written in this basis. 

All the supercell tensors are written in the equilibrium polarization basis, see Eq.\ \eqref{pol basis}.
The equations of motion (Eqs \eqref{dynamical eqs motion}) expanded at first order are
\begin{subequations}
\label{linearized equation in alpha beta gamma}
\begin{align}
   & \frac{d^2}{d t^2} \widetilde{\mathcal{R}}^{(1)}(t)_\alpha = - \omega^2_\alpha \widetilde{\mathcal{R}}^{(1)}(t)_\alpha 
    - \left\langle \frac{\partial \mathbb{V}}{\partial \widetilde{R}_\alpha}\right\rangle_{(1)}  - \left\langle \frac{\partial V^{(\text{ext})}(t)}{\partial \widetilde{R}_\alpha}\right\rangle_{(0)}; \label{perturbed R}\\
    & \frac{d}{d t}\widetilde{ \alpha}^{(1)}(t)_{\alpha\beta}=-2\sum_{\mu\nu=1}^{3N}
    S_{\alpha\beta\mu\nu} \left( \widetilde{\gamma}^{(1)}(t)_{\mu\nu} \omega^2_\nu\right); \label{perturbed alpha}\\
    & \frac{d}{d t}\widetilde{ \beta}^{(1)}(t)_{\alpha\beta}=2 \sum_{\mu\nu=1}^{3N}
    S_{\alpha\beta\mu\nu} \left(\widetilde{\gamma}^{(1)}(t)_{\mu\nu}\right); \label{perturbed beta}\\
    & \frac{d}{d t} \widetilde{ \gamma}^{(1)}(t)_{\alpha\beta}
    = \widetilde{\alpha}^{(1)}(t)_{\alpha\beta} 
    -\omega^2_\alpha \widetilde{\beta} ^{(1)}(t)_{\alpha\beta}   - \left[
    \left\langle \frac{\partial^2 V^{(\text{ext})}(t)}{\partial \widetilde{R}_\alpha \partial \widetilde{R}_\beta}\right\rangle_{(0)} 
    +\left\langle \frac{\partial^2 \mathbb{V}}{\partial \widetilde{R}_\alpha \partial \widetilde{R}_\beta}\right\rangle_{(1)} 
     \right] \widetilde{\beta}^{(0)}_{\beta\beta} \label{perturbed gamma},
\end{align}
\end{subequations}
where 
\begin{equation}
\label{def S}
    S_{\alpha\beta\mu\nu} = \frac{1}{2}(\delta_{\alpha\mu} \delta_{\beta\nu} + \delta_{\alpha\nu}\delta_{\beta\mu})
\end{equation}
and $\mathbb{V}(\bm R)$ is the difference between the exact BO energy surface and the SCHA auxiliary potential, Eq.\ \eqref{mathbb(V) = V_BO - V_SCHA}. The averages of $\mathbb{V}(\bm R)$ are defined in Eqs. \eqref{derivative of V_bo - V_scha explicit} and contain anharmonic corrections.
Now we make three more steps. 

First, we derive with respect to time Eqs \eqref{linearized equation in alpha beta gamma} to delete the equation for $\widetilde{\bm \gamma}^{(1)}(t)$ since the perturbed averages do not depend on this parameter, see Eq.\ \eqref{O on rho 1}. 

Secondly, we take the Fourier transform of the second order set of differential equations for $\widetilde{\boldsymbol{\mathcal{R}}}^{(1)}(t)$
$\widetilde{ \bm \alpha}^{(1)}(t)$ and $\widetilde{ \bm \beta}^{(1)}(t)$. 

The third and last step is to perform a change of variables.
Instead of using the basis $\{\widetilde{\bm \alpha}^{(1)}(\omega), \widetilde{\bm \beta}^{(1)}(\omega)\}$, we work with $\{\widetilde{\bm a}'^{(1)}(\omega), \widetilde{ \bm b}'^{(1)}(\omega)\}$ which is defined as a linear combination of the original free parameters
\begin{equation}
\label{change of variables}
    \begin{bmatrix}
    \widetilde{a}'^{(1)}(\omega)_{\mu\nu}\\ 
    \widetilde{b}'^{(1)}(\omega)_{\mu\nu}
    \end{bmatrix} = M_{\mu\nu}
    \begin{bmatrix}
    \widetilde{\alpha}^{(1)}(\omega)_{\mu\nu}\\ 
    \widetilde{\beta}^{(1)}(\omega)_{\mu\nu}
    \end{bmatrix},
\end{equation}
where we define $\bm M$ as
\begin{equation}
\label{change of variables}
    M_{\mu\nu} = 
    \begin{bmatrix}
     \frac{K^-_{\mu\nu}}{\omega_{\mu} \omega_{\nu}} &  K^-_{\mu\nu}\\ \\
    -\frac{K^+_{\mu\nu}}{\omega_{\mu} \omega_{\nu}} &  K^+_{\mu\nu}
    \end{bmatrix}.
\end{equation}
The coefficients in Eqs \eqref{change of variables} are functions of the equilibrium auxiliary frequencies $\{\omega^2_\mu\}$ defined in Eq.\ \eqref{SCHA phonons} and
\begin{subequations}
\label{change of variables parameters}
\begin{align}
    K^{\pm}_{\mu\nu} = & \frac{\hbar^2n_{\mu\nu}}{2X^{\pm}_{\mu\nu}} 
    \label{K_+-}  ,\\
    X^{\pm}_{\mu\nu} = & \sqrt{\pm \frac{1}{2} \frac{\hbar\left[\omega_\mu \pm \omega_\nu\right]\left[(1\pm 1) + 2(n_\mu \pm n_\nu)\right]}{4\omega_\mu\omega_\nu}} \label{def X + -} ,\\
    n_{\mu\nu} = &\frac{1}{8}(1 + 2n_\nu)(1 + 2n_\mu)  
    \label{def nmunu}.
\end{align}
\end{subequations}
Using the basis defined in Eq.\ \eqref{change of variables} we get the following equations of motion for $\{\widetilde{\boldsymbol{\mathcal{R}}}^{(1)}(\omega), \widetilde{\bm a}'^{(1)}(\omega), \widetilde{ \bm b}'^{(1)}(\omega)\}$
\begin{subequations}
\label{linearized equation in a b}
\begin{align}
    & \left(\omega^2 - \omega^2_\alpha\right)\widetilde{\mathcal{R}}^{(1)}(\omega)_\alpha 
    -\left\langle \frac{\partial \mathbb{V}}{\partial \widetilde{R}_\alpha}\right\rangle_{(1)} = \left\langle \frac{\partial V^{(\text{ext})}(\omega)}{\partial \widetilde{R}_\alpha}\right\rangle_{(0)}
    , \\
    & \left(\omega^2 -\omega^{-}\,\!_{\alpha\beta}^2\right)
    \widetilde{a}'^{(1)}(\omega)_{\alpha\beta} 
    +X^{-}_{\alpha\beta}
    \left\langle \frac{\partial^2 \mathbb{V}}{\partial \widetilde{R}_\alpha \partial \widetilde{R}_\beta}\right\rangle_{(1)}  = -X^{-}_{\alpha\beta}
    \left\langle \frac{\partial^2 V^{(\text{ext})}(\omega)}{\partial \widetilde{R}_\alpha \partial \widetilde{R}_\beta}\right\rangle_{(0)}, 
    \\
    & \left(\omega^2 -\omega^{+}\,\!_{\alpha\beta}^2\right)
    \widetilde{b}'^{(1)}(\omega)_{\alpha\beta}
    -X^+_{\alpha\beta}
    \left\langle \frac{\partial^2 \mathbb{V}}{\partial \widetilde{R}_\alpha \partial \widetilde{R}_\beta}\right\rangle_{(1)} = 
     X^+_{\alpha\beta}
     \left\langle \frac{\partial^2 V^{(\text{ext})}(\omega)}{\partial \widetilde{R}_\alpha \partial \widetilde{R}_\beta}\right\rangle_{(0)},
\end{align}
\end{subequations}
where $\omega^2$ comes from the Fourier transform of a second order derivative with respect to time.

Eqs. \eqref{linearized equation in a b} are written in terms of a matrix vector product in the space of the perturbative free parameters. Recalling the definition of $V^{(\text{ext})}(\bm R, \omega)$ given in Eq.\ \eqref{V ext definition}, we write the linearized equations of motion as a matrix-vector product  
\begin{equation}
    (\omega^2 \bm 1 + \boldsymbol{\mathcal{L}}')   \cdot
    \begin{bmatrix}
    \widetilde{\boldsymbol{\mathcal{R}}}^{(1)}(\omega)\\
    \widetilde{\bm a}'^{(1)}(\omega)\\
    \widetilde{\bm b}'^{(1)}(\omega)
    \end{bmatrix} = 
    \begin{bmatrix}
    \left\langle \frac{\partial \mathcal{B}}{\partial \widetilde{\bm R}}\right\rangle_{(0)}\\ 
    -\overset{(4)}{\bm X^-} : 
    \left\langle \frac{\partial^2 \mathcal{B}}{\partial \widetilde{\bm R} \partial \widetilde{\bm R}}\right\rangle_{(0)} \\ 
    \overset{(4)}{\bm X^+} : 
    \left\langle \frac{\partial^2 \mathcal{B}}{\partial \widetilde{\bm R}\partial \widetilde{\bm R}}\right\rangle_{(0)}
    \end{bmatrix}
    \mathcal{V}(\omega),
\end{equation}
where
\begin{equation}
    \overset{(4)}{X^{\pm}}_{\alpha\beta\mu\nu} = X^{\pm}_{\alpha\beta}  S_{\alpha\beta\mu\nu}
\end{equation}
with $X^{\pm}_{\alpha\beta}$ defined in Eq.\ \eqref{def X + -} and $S_{\alpha\beta\mu\nu}$ in Eq.\ \eqref{def S}. We define the RHS vector as the perturbation vector $\perturbationvector'$
\begin{equation}
    \label{app: perturbation vector Lanczos}
    \perturbationvector'^\dagger =
    \begin{bmatrix}
    \left\langle \frac{\partial \mathcal{B}}{\partial \widetilde{\bm R}}\right\rangle_{(0)}, &
    -\overset{(4)}{\bm X^-} : 
    \left\langle \frac{\partial^2 \mathcal{B}}{\partial \widetilde{\bm R} \partial \widetilde{\bm R}}\right\rangle_{(0)},  &
    \overset{(4)}{\bm X^+} : 
    \left\langle \frac{\partial^2 \mathcal{B}}{\partial \widetilde{\bm R}\partial \widetilde{\bm R}}\right\rangle_{(0)}
    \end{bmatrix}.
\end{equation}
The matrix $\boldsymbol{\mathcal{L}}'$ acts in the space of the perturbative parameters. It is symmetric and contains two terms, the harmonic and anharmonic contribution
\begin{equation}
    \boldsymbol{\mathcal{L}}' = \boldsymbol{\mathcal{L}}'_{\text{harm}} + \boldsymbol{\mathcal{L}}'_{\text{anh}}.
\end{equation}
The harmonic part $\boldsymbol{\mathcal{L}}'_{\text{harm}}$ is diagonal in our basis
\begin{equation}
\label{L harm r-a-b}
\begin{aligned}
    \boldsymbol{\mathcal{L}}'_{\text{harm}} \cdot
    \begin{bmatrix}
    \widetilde{\boldsymbol{\mathcal{R}}}^{(1)}(\omega)\\
    \widetilde{\bm a}'^{(1)}(\omega)\\
    \widetilde{\bm b}'^{(1)}(\omega)
    \end{bmatrix}
    = -\begin{bmatrix}
      \overset{(2)}{\bm D} \cdot & 0 & 0 \\ 
    0 &  \overset{(4)}{\bm \omega_- } {}^2  : & 0 \\ 
    0 & 0 &  \overset{(4)}{\bm \omega_+ }{}^2   : 
    \end{bmatrix}
    \begin{bmatrix}
    \widetilde{\boldsymbol{\mathcal{R}}}^{(1)}(\omega)\\ 
    \widetilde{\bm a}'^{(1)}(\omega)\\ 
    \widetilde{\bm b}'^{(1)}(\omega)
    \end{bmatrix}.
\end{aligned}
\end{equation}
The matrix $\boldsymbol{\mathcal{L}}'_{\text{harm}} $ depends only on the equilibrium auxiliary frequencies of Eq.\ \eqref{SCHA phonons}.
We introduced a four indices tensor
\begin{equation}
    \overset{(4)}{\omega_{\pm}}{} ^2_{\alpha\beta\mu\nu}= (\omega^{\pm}_{\alpha\beta}) ^2 S_{\alpha\beta\mu\nu},
\end{equation}
with $\bm S$ defined in Eq.\ \eqref{def S} and
\begin{equation}
    \omega^{\pm}_{\mu\nu} = \omega_\mu \pm \omega_\nu.
\end{equation}
The RHS of Eq.\ \eqref{L harm r-a-b} should be read as a standard matrix-vector product. The matrix element contains also information on how to contract the indices, the operation $\cdot$ is defined in general as the contraction of the last and first index of two tensors
\begin{equation}
\label{. def}
    \bm A \cdot \bm B = \sum_{\mu=1}^{3N} A_{...\mu} B_{\mu...}
\end{equation}
and $:$ is defined as
\begin{equation}
\label{: def}
    \bm C : \bm D = \sum_{\mu\nu=1}^{3N} C_{...\mu\nu} D_{\mu\nu...}
\end{equation}
For example the first line of Eq.\ \eqref{L harm r-a-b} is
\begin{equation}
    -\overset{(2)}{\bm D} \cdot \widetilde{\boldsymbol{\mathcal{R}}}^{(1)}(\omega) = 
    -\sum_{\nu = 1}^{3N} \overset{(2)}{ D}_{\mu\nu} \widetilde{\mathcal{R}}^{(1)}(\omega)_{\nu},
\end{equation}
and returns a tensor of rank 1. The same holds for the other lines. As an example, consider
\begin{equation}
    -\overset{(4)}{\bm \omega_-} {}^2 :\widetilde{\bm a}'^{(1)}(\omega) = 
    -\sum_{\mu\nu = 1}^{3N} (\omega^{-}_{\mu\nu})^2\widetilde{a}'^{(1)}(\omega)_{\mu\nu},
\end{equation}

The application of $\boldsymbol{\mathcal{L}}'_{\text{anh}}$ gives
\begin{equation}
\label{L anharm r-a-b}
\begin{aligned}
    & \boldsymbol{\mathcal{L}}'_{\text{anh}} \cdot
    \begin{bmatrix}
    \widetilde{\boldsymbol{\mathcal{R}}}^{(1)}(\omega)\\ 
    \widetilde{\bm a}'^{(1)}(\omega)\\ 
    \widetilde{\bm b}'^{(1)}(\omega)
    \end{bmatrix} = \begin{bmatrix}
    -\left\langle \frac{\partial \mathbb{V}}{\partial \widetilde{\bm R}}\right\rangle_{(1)}\\
    \overset{(4)}{\bm X^-} : 
    \left\langle \frac{\partial^2 \mathbb{V}}{\partial \widetilde{\bm R} \partial \widetilde{\bm R}}\right\rangle_{(1)}\\
    -\overset{(4)}{\bm X^+} : 
    \left\langle \frac{\partial^2 \mathbb{V}}{\partial \widetilde{\bm R}\partial \widetilde{\bm R}}\right\rangle_{(1)}
    \end{bmatrix}.
\end{aligned}
\end{equation}
Writing the perturbed averages of $\mathbb{V}(\bm R)$ in terms of the scattering tensors, as in Eq.\ \eqref{derivative of V_bo - V_scha explicit}, and using the change of variables of Eq.\ \eqref{change of variables}, it is trivial to prove that in the new basis $\boldsymbol{\mathcal{L}}'_{\text{anh}}$ is symmetric and has the following form
\begin{equation}
\label{L anharm matrix}
\begin{aligned}
    & \boldsymbol{\mathcal{L}}'_{\text{anh}}  = 
    \begin{bmatrix}
    0 
    &  \overset{(3)}{\bm D}  : \overset{(4)}{\bm X^-}
    &  -\overset{(3)}{\bm D} : \overset{(4)}{\bm X^+}  \\ 
       \overset{(4)}{\bm X^-} : \overset{(3)}{\bm D} 
    & -\overset{(4)}{\bm X^-} : \overset{(4)}{\bm D}:\overset{(4)}{\bm X^-} 
    & \overset{(4)}{\bm X^-} : \overset{(4)}{\bm D}:\overset{(4)}{\bm X^+}   \\ 
      -\overset{(4)}{\bm X^+} : \overset{(3)}{\bm D} 
    &  \overset{(4)}{\bm X^+} : \overset{(4)}{\bm D} : \overset{(4)}{\bm X^-} 
    & -\overset{(4)}{\bm X^+} : \overset{(4)}{\bm D} : \overset{(4)}{\bm X^+}  
    \end{bmatrix} .
\end{aligned}
\end{equation}
Again, the matrix contains \deleted{the} information on how to contract the indices.
This term contains information on the anharmonicity of the system through the third and fourth phonon scattering tensors, defined in Eq.\ \eqref{D3 D4 app}.

Now that we have the linearized equations of motion, we present the general response function Eq.\ \eqref{general response formula}. To do this we need the correction of a position-dependent observable $\mathcal{A}(\bm R)$ in the new basis Eq.\ \eqref{change of variables}
\begin{equation}
\label{pertub formula basis diagonalizing the harmonic part}
\begin{aligned}
    & \left\langle \mathcal{A} \right\rangle_{(1)}(\omega) = 
    \sum_{\alpha=1}^{3N} \frac{\partial \left\langle \mathcal{A} \right\rangle_{(0)}}{\partial \widetilde{\mathcal{R}}_{\alpha}}\widetilde{\mathcal{R}}^{(1)}(\omega)_\alpha+
    \sum_{\alpha\beta=1}^{3N} \frac{\partial \left\langle \mathcal{A} \right\rangle_{(0)}}{\partial \widetilde{a}'^{(0)}_{\alpha\beta}} \widetilde{a}'^{(1)}(\omega)_{\alpha\beta}
    + 
    \sum_{\alpha\beta=1}^{3N} \frac{\partial \left\langle \mathcal{A} \right\rangle_{(0)}}{\partial \widetilde{b}'^{(0)}_{\alpha\beta}} \widetilde{b}'^{(1)}(\omega)_{\alpha\beta} = \\
    & = \sum_{\alpha=1}^{3N} \left\langle \frac{\partial  \mathcal{A}}{\partial \widetilde{R}_{\alpha}}  \right\rangle_{(0)}\widetilde{\mathcal{R}}^{(1)}(\omega)_\alpha
    -\sum_{\alpha\beta=1}^{3N}X^-_{\alpha\beta} \left\langle \frac{\partial^2  \mathcal{A}}{\partial \widetilde{R}_\alpha \partial \widetilde{R}_{\beta}} \right\rangle_{(0)} \widetilde{a}'^{(1)}(\omega)_{\alpha\beta} +  \sum_{\alpha\beta=1}^{3N}X^+_{\alpha\beta}  \left\langle \frac{\partial^2  \mathcal{A}}{\partial \widetilde{R}_\alpha \partial \widetilde{R}_{\beta}} \right\rangle_{(0)} \widetilde{b}'^{(1)}(\omega)_{\alpha\beta}.
\end{aligned}
\end{equation}
The previous expression can be demonstrated using the change of variable definition, Eq.\ \eqref{change of variables}, the chain rule and the following relations in the original basis (i.e. the one used in Appendix \ref{Expansion of the probability distribution})
\begin{subequations}
\label{derivatives wrt eq parameters}
\begin{align}
    &  \frac{\partial \left\langle \mathcal{A} \right\rangle_{(0)}}{\partial \widetilde{\mathcal{R}}^{(0)}_{\alpha}} = \left\langle \frac{\partial  \mathcal{A}}{\partial \widetilde{R}_{\alpha}}  \right\rangle_{(0)}, \\
    & \frac{\partial \left\langle \mathcal{A}\right\rangle_{(0)}}{\partial \widetilde{\alpha}^{(0)}_{\alpha\beta}}  = - 
     \frac{1}{2\widetilde{\alpha}^{(0)}_{\alpha\alpha} \widetilde{\alpha}^{(0)}_{\beta\beta}}\left\langle \frac{\partial^2  \mathcal{A}}{\partial \widetilde{R}_\alpha
    \partial \widetilde{R}_\beta } \right\rangle_{(0)},\\
    & \frac{\partial \left\langle \mathcal{A}\right\rangle_{(0)}}{\partial \widetilde{\beta}^{(0)}_{\alpha\beta}}  = 0.
\end{align}
\end{subequations}
The derivative with respect to $\widetilde{\alpha}^{(0)}_{\alpha\beta}$ is obtained using the formalism of \cite{Bianco}.

We define the response vector $\responsevector'$ similarly to $\perturbationvector'$ (Eq.\ \eqref{app: perturbation vector Lanczos})
\begin{equation}
    \label{app: response vector Lanczos}
    \responsevector'^\dagger =
    \begin{bmatrix}
    \left\langle \frac{\partial \mathcal{A}}{\partial \widetilde{\bm R}}\right\rangle_{(0)}  ,&
    -\overset{(4)}{\bm X^-} : 
    \left\langle \frac{\partial^2 \mathcal{A}}{\partial \widetilde{\bm R} \partial \widetilde{\bm R}}\right\rangle_{(0)} ,&
    \overset{(4)}{\bm X^+} : 
    \left\langle \frac{\partial^2 \mathcal{A}}{\partial \widetilde{\bm R}\partial \widetilde{\bm R}}\right\rangle_{(0)}
    \end{bmatrix}
\end{equation}
so from Eq.\ \eqref{pertub formula basis diagonalizing the harmonic part} we can extract the response formula (Eq.\ \eqref{general response formula})
\begin{equation}
\label{chi code}
    \frac{\left\langle \mathcal{A} \right\rangle_{(1)}(\omega)}{ \mathcal{V}(\omega) } = \frac{1}{\mathcal{V}(\omega)} \responsevector' \cdot 
    \begin{bmatrix}
    \widetilde{\boldsymbol{\mathcal{R}}}^{(1)}(\omega)\\
    \widetilde{\bm a}'^{(1)}(\omega)\\
    \widetilde{\bm b}'^{(1)}(\omega)
    \end{bmatrix}
    = \responsevector '\cdot \left(\boldsymbol{\mathcal{L}}' + \omega^2\right)^{-1} \cdot \perturbationvector ' = \chi(\omega)_{\mathcal{A}, \mathcal{B}}
\end{equation}
where $\left\langle \mathcal{A} \right\rangle_{(1)}(\omega)$ is expressed as a scalar product in the space of the perturbative parameters. 
This is the expression of $\chi(\omega)_{\mathcal{A}, \mathcal{B}}$ implemented in the code.

\section{Lanczos algorithm}
\label{appendix: Lanczos algorithm}
In this Appendix, we discuss the Lanczos implementation \cite{TURBOLancz,TDSCHA_monacelli} of the general response function Eq.\ \eqref{chi code}. Both for infrared and Raman calculations, we can always work with $\perturbationvector' = \responsevector'$ setting $\mathcal{A} = \mathcal{B}$ (see Eqs \eqref{app: perturbation vector Lanczos} \eqref{app: response vector Lanczos}) so Eq.\ \eqref{chi code} becomes
\begin{equation}
\label{chi implemented}
    \chi(\omega)_{\mathcal{A}, \mathcal{A}} = (\perturbationvector'\cdot\perturbationvector')\overline{\perturbationvector}' \cdot \left(\boldsymbol{\mathcal{L}}' + \omega^2\right)^{-1} \cdot \overline{\perturbationvector}' 
\end{equation}
where we normalize the vector $\perturbationvector'$ (Eq.\ \eqref{app: perturbation vector Lanczos})
\begin{equation}
\label{pert vector normalized}
\overline{\perturbationvector} ' = \frac{\perturbationvector'}{ \sqrt{\perturbationvector' \cdot \perturbationvector'}}.
\end{equation}

To get in one shot for all values of $\omega$ the response formula, Eq.\ \eqref{chi implemented}, we modified the Lanczos algorithm presented in \cite{TDSCHA_monacelli} exploiting that $\boldsymbol{\mathcal{L}}' = \boldsymbol{\mathcal{L}}'^\dagger$. This algorithm allows to find a basis in which $\boldsymbol{\mathcal{L}}'$ is tridiagonal
\begin{equation}
    \bm P'^{-1} \cdot \boldsymbol{\mathcal{L}}' \cdot \bm P' = \bm T' ,
\end{equation}
where $\bm T'$ has the following form
\begin{equation}
    \bm T' = \begin{bmatrix} 
    t_{1} & r_1 & \dots  & \dots & 0\\
    r_1 & t_2 & \ddots &  & \vdots \\
     & \ddots & \ddots & \ddots & &\\
     \vdots& &  \ddots & \ddots & r_{n-1} \\
     0&  &  & r_{n-1} & t_{n}
    \end{bmatrix}
\end{equation}
where $n$ is the size of $\bm{\mathcal{L}}'$.
The change of basis matrix $\bm P'$ is
\begin{equation}
    \bm P' = \begin{bmatrix}
        \bm p'_1 \quad \bm p'_2\quad  ... \quad \bm p'_n
    \end{bmatrix}
\end{equation}
and it is unitary
\begin{equation}
    \bm P'^{-1} = \bm P'^\dagger.
\end{equation}
The coefficients of $\bm T'$ can be found following this iterative procedure \cite{TDSCHA_monacelli,TURBOLancz}
\begin{subequations}
\begin{align}
    t_k = & \bm p'_k \cdot \boldsymbol{\mathcal{L}}' \cdot \bm p'_k \\
    r_k \bm p'_{k+1} =  &  \bm v_{k}= (\boldsymbol{\mathcal{L}}' - t_k) \cdot \bm p'_k - r_{k-1} \bm p'_{k-1} \\
    r_k = & \sqrt{\bm v_k \cdot \bm v_k} \\
    \bm p'_{k+1} = & \bm v_k /r_k 
\end{align}
\end{subequations}
with the initial vector equal to the normalized perturbation vector, $\bm p'_1 = \overline{\perturbationvector}'$. This procedure ends when either $\bm p'_k$ is a linear combination of the previous vectors or $\bm p'_k \cdot \bm p'_k = 0$. Unless the system is perfectly harmonic, this
condition is usually never reached in practical runs, and
the algorithm is truncated after a maximum number of
steps $N_{\text{steps}}$.

After we build the change of variables matrix $\bm P'$ we can use it in Eq.\ \eqref{chi implemented}
\begin{equation}
    \chi(\omega)_{\mathcal{A}, \mathcal{A}} = (\perturbationvector'\cdot\perturbationvector')\overline{\perturbationvector}' \cdot \bm P' \cdot \left[\bm P'^{-1} \cdot \left(\boldsymbol{\mathcal{L}}' + \omega^2\right)^{-1} \cdot \bm P' \right] \cdot \bm P'^{-1} \cdot \overline{\perturbationvector}' =(\perturbationvector'\cdot\perturbationvector')\overline{\perturbationvector}' \cdot \bm P' \cdot \left(\bm T' + \omega^2\right)^{-1} \cdot \bm P'^{-1} \cdot \overline{\perturbationvector} '
\end{equation}
then noting that
\begin{equation}
    \bm P'^{-1} \cdot \overline{\perturbationvector}' = \bm P'^\dagger \cdot \overline{\perturbationvector}' = \begin{bmatrix}
        1 \\ 0 \\ 0 \\ \vdots
    \end{bmatrix}
\end{equation}
we get that the response function is given by
\begin{equation}
    \chi(\omega)_{\mathcal{A}, \mathcal{A}} = 
    (\perturbationvector'\cdot\perturbationvector') \left(\bm T' + \omega^2\right)^{-1}_{11}
\end{equation}
where $ \left(\bm T'+ \omega^2\right)^{-1}_{11}$ can be written as a continuous  fraction using the coefficients obtained up to $N_{\text{steps}}$
\begin{equation}
    \left(\bm T' + \omega^2\right)^{-1}_{11} = \frac{1}{\omega^2 + t_1 - \frac{r_1^2}{\omega^2 + t_2 - \frac{r_2^2}{\omega^2 + ...}}}.
\end{equation}

At each Lanczos step, we have to apply $\boldsymbol{\mathcal{L}}'$ to a given vector $\bm w$ in the space of the perturbed free parameters.
As showed in Appendix \ref{Derivation of the linear response system} $\boldsymbol{\mathcal{L}}'$ contains two terms
\begin{equation}
    \boldsymbol{\mathcal{L}}' \cdot \bm w = \boldsymbol{\mathcal{L}}'_{\text{harm}} \cdot \bm w+
    \boldsymbol{\mathcal{L}}'_{\text{anh}}\cdot \bm w.
\end{equation}
The application of the harmonic part is done using Eq.\ \eqref{L harm r-a-b}, while the anharmonic part is done using Eq.\ \eqref{L anharm r-a-b} applying a reweighting procedure to compute the perturbed average as explained in \cite{TDSCHA_monacelli}.

\section{Symbolic inversion}
\label{Symbolic inversion}
In this Appendix we describe the symbolic inversion of a symmetric square super-tensor with this form
\begin{equation}
\label{L super-matrix}
    \bm L = \begin{bmatrix}
     \bm A & \bm C\\ 
     \bm C^\dagger & \bm B
    \end{bmatrix}
\end{equation}
where $\bm A$, $\bm B$, $\bm C$, $\bm C^\dagger$ are tensors.
Using Gaussian reduction we get the inverse
\begin{equation}
\label{inverse of L symbolic 3x3}
    \bm L^{-1} = \begin{bmatrix}
    \bm D^{-1} & \quad  -  \bm D^{-1}\cdot \bm C\cdot  \bm B^{-1} \\
     -\bm B^{-1} \cdot \bm C^\dagger \cdot \bm D^{-1} &
     \quad \bm B^{-1}  +   \bm B^{-1}\cdot  \bm C^\dagger \cdot \bm D^{-1} \added{\cdot}
     \bm C \cdot \bm B^{-1}.
    \end{bmatrix}
\end{equation}
where $\bm D = \bm A -  \bm C \cdot\bm B^{-1} \cdot\bm C^\dagger$. It is trivial to check that $\bm L \cdot\bm L^{-1} = \bm L^{-1}\cdot\bm L = \bm 1$. For what follows we need $\bm C = \bm 1$ so Eq.\ \eqref{inverse of L symbolic 3x3} becomes
\begin{equation}
\label{inverse of L symbolic 2x2 C = 1}
     \bm L^{-1} = \begin{bmatrix}
     \bm A^{-1} - \bm A^{-1} \cdot \left(1 - \bm A\cdot\bm B\right)^{-1} &
     \left(1 - \bm B\cdot\bm A\right)^{-1}\\
     \left(1 - \bm A\cdot\bm B\right)^{-1}& -\bm A\cdot\left(1 - \bm B\cdot\bm A\right)^{-1}.
    \end{bmatrix}
\end{equation}
Again, for our purposes (see next Appendix \ref{Derivation of the interacting Green function}), we need to find a formula for the sum of entries of Eq.\ \eqref{inverse of L symbolic 2x2 C = 1}. Summing the coefficients of Eq.\ \eqref{inverse of L symbolic 2x2 C = 1} we get
\begin{equation}
\label{sum of inverse entries}
\begin{aligned}
    \bm L^{-1}_{11} + \bm L^{-1}_{21} + \bm L^{-1}_{12} + \bm L^{-1}_{22} = & \bm A^{-1} - \bm A^{-1} \cdot \left(1 - \bm A\cdot\bm B\right)^{-1} +
    \left(\bm 1 - \bm B\cdot\bm A\right)^{-1}+
    \left(\bm 1 - \bm A\cdot\bm B\right)^{-1}  -\bm A\cdot\left(\bm 1 - \bm B\cdot\bm A\right)^{-1}\\
    = &  \bm A^{-1}\cdot \left[(\bm 1 - \bm A\cdot \bm B) - (\bm 1-\bm A)\right]\cdot(1 - \bm A \cdot\bm B)^{-1}
    +(1 - \bm A) \cdot\left(\bm 1 - \bm B\cdot\bm A\right)^{-1} \\
     =&  \bm A^{-1}\cdot \left[\bm A\cdot(\bm 1 - \bm B)\right]\cdot (1 - \bm A \cdot\bm B)^{-1}
    +(1 - \bm A) \cdot \left(\bm 1 - \bm B\cdot\bm A\right)^{-1} \\
     = & (1 - \bm B)\cdot\left(1 - \bm A\cdot\bm B\right)^{-1}
    +(1 - \bm A)\cdot \left(\bm 1 - \bm B\cdot\bm A\right)^{-1}.
\end{aligned}
\end{equation}
Now we set $\bm A = \bm 1 + \widetilde{\bm A}$ and $\bm B = \bm 1 + \widetilde{\bm B}$ so we have that Eq.\ \eqref{sum of inverse entries} is
\begin{equation}
\label{formula for symbolic inversion sum of entries}
\begin{aligned}
    &  \widetilde{\bm B}\cdot \left(\widetilde{\bm A} + \widetilde{\bm B} + \widetilde{\bm A}\cdot  \widetilde{\bm B}\right)^{-1}
    +\widetilde{\bm A} \cdot \left(\widetilde{\bm A} + \widetilde{\bm B} + \widetilde{\bm B} \cdot \widetilde{\bm A}\right)^{-1}  \\
    & = \left( 1 +\widetilde{\bm A} \cdot\widetilde{\bm B}^{-1}+ \widetilde{\bm A} \right)^{-1}
    + \left(1 + \widetilde{\bm B}\cdot\widetilde{\bm A}^{-1} + \widetilde{\bm B} \right)^{-1} \\
    & =  \left(\widetilde{\bm A}^{-1} + \widetilde{\bm B}^{-1}+ \bm 1\right)^{-1} \cdot  \widetilde{\bm A}
    + \left(\widetilde{\bm B}^{-1} + \widetilde{\bm A}^{-1} + \bm 1\right)^{-1} \cdot \widetilde{\bm B} \\
    & = \left(1 + \widetilde{\bm B}^{-1} + \widetilde{\bm A}^{-1}\right)^{-1}\cdot(\widetilde{\bm A}^{-1}+\widetilde{\bm B}^{-1}).
\end{aligned}
\end{equation}
We will use this formula in Appendix \ref{Derivation of the interacting Green function}. 

\section{Derivation of the interacting Green's function}
\label{Derivation of the interacting Green function}
The easiest way to get the interacting Green's function is to use another change of variables in Eqs \eqref{linearized equation in alpha beta gamma}
\begin{equation}
\label{change of variables for Green functions}
\begin{bmatrix}
\widetilde{a}^{(1)}_{\mu\nu}(\omega) \\ 
\widetilde{b}^{(1)}_{\mu\nu}(\omega)
\end{bmatrix} = -\frac{\hbar^2 n_{\mu\nu}}{2}
\begin{bmatrix}
\frac{1}{\omega_\mu \omega_\nu} & 1 \\
\frac{1}{\omega_\mu \omega_\nu} & -1 
\end{bmatrix}
\begin{bmatrix}
\widetilde{\alpha}^{(1)}_{\mu\nu}(\omega) \\
\widetilde{\beta}^{(1)}_{\mu\nu}(\omega)
\end{bmatrix}
\end{equation}
where $n_{\mu\nu}$ is defined in Eq.\ \eqref{def nmunu} and $\{\omega^2_\mu\}$ in Eq.\ \eqref{SCHA phonons}. As done in Appendix \ref{Derivation of the linear response system}, we write Eqs \eqref{linearized equation in alpha beta gamma} in this new basis switching to second-order time-derivatives
\begin{equation}
\small
\label{linearize system 2ph}
   \begin{bmatrix}
    \left(\boldsymbol{\mathcal{G}}^{(0)}(\omega)\right)^{-1}\cdot & 0 & 0 \\ 
    0 & \left(\bm \chi^{(0)}_-(\omega)\right)^{-1}:  & 0 \\ 
    0 & 0 & -\left(\bm \chi^{(0)}_+(\omega)\right)^{-1}: 
    \end{bmatrix} \begin{bmatrix}
    \widetilde{\boldsymbol{\mathcal{R}}}^{(1)}(\omega)\\ 
    \widetilde{\bm a}^{(1)}(\omega)\\ 
    \widetilde{\bm b}^{(1)}(\omega)
    \end{bmatrix} = \begin{bmatrix}
    \left\langle \frac{\partial \mathbb{V}}{\partial \widetilde{\bm R}}\right\rangle_{(1)}\\ 
    \left\langle \frac{\partial^2 \mathbb{V}}{\partial \widetilde{\bm R} \partial \widetilde{\bm R}_\beta}\right\rangle_{(1)}\\ 
    \left\langle \frac{\partial^2 \mathbb{V}}{\partial \widetilde{\bm R} \partial \widetilde{\bm R}_\beta}\right\rangle_{(1)}
    \end{bmatrix}
    +
    \begin{bmatrix}
    \left\langle \frac{\partial V^{(\text{ext})}}{\partial \widetilde{\bm R}}\right\rangle_{(0)}\\ 
    \left\langle \frac{\partial^2 V^{(\text{ext})}}{\partial \widetilde{\bm R} \partial \widetilde{\bm R}}\right\rangle_{(0)}\\ 
     \left\langle \frac{\partial^2 V^{(\text{ext})}}{\partial \widetilde{\bm R} \partial \widetilde{\bm R}}\right\rangle_{(0)}
    \end{bmatrix}
\end{equation}
where we recognize the resonant and anti-resonant terms of the two-phonon propagator
\begin{equation}
\label{chi plus chi minus}
\begin{aligned}
    & \chi^{(0)}_{-}(\omega)_{\mu\nu\sigma\pi} =\delta_{\mu\sigma}\delta_{\nu\pi} \frac{\hbar\left[\omega_\mu - \omega_\nu\right]\left[n_\mu - n_\nu\right]}{4\omega_\mu\omega_\nu[(\omega_\mu -\omega_\nu)^2 - \omega^2]} ;\\
    & \chi^{(0)}_{+}(\omega)_{\mu\nu\sigma\pi}  = \delta_{\mu\sigma}\delta_{\nu\pi}
    \frac{\hbar\left[\omega_\mu + \omega_\nu\right]\left[1+n_\mu + n_\nu\right]}{4\omega_\mu\omega_\nu [(\omega_\mu + \omega_\nu)^2 - \omega^2]} .
\end{aligned}
\end{equation}
The anharmonic vector in this basis is simply
\begin{equation}
    \begin{bmatrix}
    \left\langle \frac{\partial \mathbb{V}}{\partial \widetilde{\bm R}}\right\rangle_{(1)}\\ 
    \left\langle \frac{\partial^2 \mathbb{V}}{\partial \widetilde{\bm R} \partial \widetilde{\bm R}_\beta}\right\rangle_{(1)}\\ 
    \left\langle \frac{\partial^2 \mathbb{V}}{\partial \widetilde{\bm R} \partial \widetilde{\bm R}_\beta}\right\rangle_{(1)}
    \end{bmatrix} = 
    \begin{bmatrix}
    0 & \overset{(3)}{\bm D}: &
    \overset{(3)}{\bm D}: \\ 
    \overset{(3)}{\bm D} \cdot &
    \overset{(4)}{\bm D}  : & 
    \overset{(4)}{\bm D} :\\ 
    \overset{(3)}{\bm D} \cdot
    & \overset{(4)}{\bm D} :
    & \overset{(4)}{\bm D} :
    \end{bmatrix} 
    \begin{bmatrix}
    \widetilde{\boldsymbol{\mathcal{R}}}^{(1)}\\ 
    \widetilde{\bm a}^{(1)}\\ 
    \widetilde{\bm b}^{(1)}
    \end{bmatrix}.
\end{equation}
In a compact form, the linearized equations of motion are
\begin{equation}
\label{linearized equation for Green function}
    \boldsymbol{\mathcal{L}}(\omega) \cdot \begin{bmatrix}
    \widetilde{\boldsymbol{\mathcal{R}}}^{(1)}(\omega)\\ 
    \widetilde{\bm a}^{(1)}(\omega)\\ 
    \widetilde{\bm b}^{(1)}(\omega)
    \end{bmatrix} =
    \begin{bmatrix}
    \left\langle \frac{\partial V^{(\text{ext})}}{\partial \widetilde{\bm R}}\right\rangle_{(0)}\\ 
    \left\langle \frac{\partial^2 V^{(\text{ext})}}{\partial \widetilde{\bm R} \partial \widetilde{\bm R}}\right\rangle_{(0)}\\ 
     \left\langle \frac{\partial^2 V^{(\text{ext})}}{\partial \widetilde{\bm R} \partial \widetilde{\bm R}}\right\rangle_{(0)}
    \end{bmatrix}
\end{equation}
The tensor $\boldsymbol{\mathcal{L}}(\omega)$ describes the evolution in the linear regime and it is
\begin{equation}
\label{L matrix def}
     \boldsymbol{\mathcal{L}}(\omega) = \begin{bmatrix}
    \left(\boldsymbol{\mathcal{G}}^{(0)}(\omega)\right)^{-1} & -\overset{(3)}{\bm D} &
    -\overset{(3)}{\bm D} \\ 
    -\overset{(3)}{\bm D}  &
     \left(\bm \chi^{(0)}_-(\omega)\right)^{-1} -\overset{(4)}{\bm D}   & 
    -\overset{(4)}{\bm D} \\ 
    -\overset{(3)}{\bm D} 
    & -\overset{(4)}{\bm D} 
    &   -\left(\bm \chi^{(0)}_+(\omega)\right)^{-1}-\overset{(4)}{\bm D}
    \end{bmatrix} 
\end{equation}

The correction to the average of an observable $\mathcal{A}$ is
\begin{equation}
\label{A 1 in R a' b'}
 \left\langle \mathcal{A} \right\rangle_{(1)}  =  \left\langle\frac{\partial \mathcal{A} }{\partial \widetilde{\bm R}} \right\rangle_{(0)} \cdot \widetilde{\boldsymbol{\mathcal{R}}}^{(1)}(\omega)
    +  \left\langle \frac{\partial^2  \mathcal{A}}{\partial \widetilde{\bm R} \partial \widetilde{\bm R}} \right\rangle_{(0)} : \bm S : \widetilde{\bm a}^{(1)}(\omega)
    +  \left\langle \frac{\partial^2  \mathcal{A}}{\partial \widetilde{\bm R} \widetilde{\bm R}} \right\rangle_{(0)} : \bm S :  \widetilde{\bm b}^{(1)}(\omega)
\end{equation}
where $\bm S$ is defined in Eq.\ \eqref{def S}. So, following the procedure described in Appendix \ref{Derivation of the linear response system}, the response function is
\begin{equation}
\label{response formula new basis appendix}
    \chi_{\mathcal{A}, \mathcal{B}}(\omega) =  
    \bm r^\dagger \cdot
     \boldsymbol{\mathcal{L}}(\omega)^{-1}\cdot
   \bm p
\end{equation}
defining the response vector $\bm r$ as
\begin{equation}
\label{appendix response vector new basis}
    \bm r^\dagger = \begin{bmatrix}
    \left\langle \frac{\partial \mathcal{A}}{\partial \widetilde{\bm R}}\right\rangle_{(0)}, & 
    \left\langle \frac{\partial^2 \mathcal{A}}{\partial \widetilde{\bm R} \partial \widetilde{\bm R}}\right\rangle_{(0)}, &
     \left\langle \frac{\partial^2 \mathcal{A}}{\partial \widetilde{\bm R} \partial \widetilde{\bm R}}\right\rangle_{(0)}
    \end{bmatrix},
\end{equation}
and the perturbation vector $\bm p$ as
\begin{equation}
\label{appendix perturbation vector new basis}
    \bm p^\dagger =  \begin{bmatrix}
    \left\langle \frac{\partial \mathcal{B}}{\partial \widetilde{\bm R}}\right\rangle_{(0)}, &
    \left\langle \frac{\partial^2 \mathcal{B}}{\partial \widetilde{\bm R} \partial \widetilde{\bm R}}\right\rangle_{(0)}, &
     \left\langle \frac{\partial^2 \mathcal{B}}{\partial \widetilde{\bm R} \partial \widetilde{\bm R}}\right\rangle_{(0)}
    \end{bmatrix}.
\end{equation}

First we discuss the non-interacting case setting $\overset{(3)}{\bm D} = \bm 0$ $\overset{(4)}{\bm D} = \bm 0$. Using $\mathcal{A}$ and $\mathcal{B}$ as in Eq.\ \eqref{1 ph pertubation} we get
\begin{equation}
    \bm r^\dagger  = \begin{bmatrix}
    \bm \delta &
    \bm 0 & \bm 0 
    \end{bmatrix} \qquad  \bm p^\dagger = \begin{bmatrix}
    \bm \delta &
    \bm 0 & \bm 0 
    \end{bmatrix}.
\end{equation}
with $\bm \delta = \bm \delta_\mu$ as in Eq.\ \eqref{r p 1 ph}.
The free phonon propagator is
\begin{equation}
    \mathcal{G}^{(0)}(\omega)_{\mu\nu} = \frac{\delta_{\mu\nu}}{\omega^2 - \omega^2_\mu}.
\end{equation}
Then we chose $\mathcal{A}$ and $\mathcal{B}$ according to Eq.\ \eqref{2 ph pertubation} so
\begin{equation}
   \bm r^\dagger = \begin{bmatrix}
    \bm 0 &
    \bm S & \bm S
    \end{bmatrix} \qquad  \bm p^\dagger = \begin{bmatrix}
    \bm 0 &
    \bm S & \bm S
    \end{bmatrix}, 
\end{equation}
where $\bm S$ is defined in Eq.\ \eqref{def S}.
The two-phonon free propagator is
\begin{equation}
    \chi^{(0)}(\omega)_{\mu\nu\sigma\pi} = -\left(\chi^{(0)}_{+}(\omega)_{\mu\nu\sigma\pi} - \chi^{(0)}_{-}(\omega)_{\mu\nu\sigma\pi}.
\right)
\end{equation}
We chose $\mathcal{A}$ and $\mathcal{B}$ according to Eq.\ \eqref{1 2 ph pertubation} \begin{equation}
    \bm r^\dagger =  \begin{bmatrix}
    \bm \delta &
    \bm 0 & \bm 0 
    \end{bmatrix}  \quad  \bm p^\dagger = \begin{bmatrix}
    \bm 0 &
    \bm S & \bm S 
    \end{bmatrix} 
\end{equation}
$\boldsymbol{\mathcal{L}}(\omega)$ is diagonal in the case $\overset{(3)}{\bm D} = \bm 0$ $\overset{(4)}{\bm D} = \bm 0$ so we get that the one-two phonon free propagator is zero
\begin{equation}
    \Gamma^{(0)}(\omega)_{\mu\sigma\pi} = 0.
\end{equation}

Now we derive the one-phonon interacting Green's function ($\overset{(3)}{\bm D} \neq \bm 0$ $\overset{(4)}{\bm D} \neq \bm 0$). Following \cite{TDSCHA_monacelli} we chose the observables $\mathcal{A}$ and $\mathcal{B}$ as in Eq.\ \eqref{1 ph pertubation}
\begin{equation}
\label{appendix one ph int green function}
    \boldsymbol{\mathcal{G}}(\omega) = \begin{bmatrix}
    \bm \delta &
    \bm 0 & \bm 0 
    \end{bmatrix} \cdot
     \boldsymbol{\mathcal{L}}(\omega)^{-1}\cdot
    \begin{bmatrix}
    \bm \delta\\
    \bm 0 \\
    \bm 0 \\
    \end{bmatrix} = \left( \boldsymbol{\mathcal{L}}(\omega)^{-1}\right)_{11}.
\end{equation}
We use Eq \eqref{inverse of L symbolic 3x3} to get
\begin{equation}
    \boldsymbol{\mathcal{G}}(\omega)^{-1} = \boldsymbol{\mathcal{G}}^{(0)}(\omega)^{-1}
    - \begin{bmatrix}
     \overset{(3)}{\bm D}: & \overset{(3)}{\bm D}:
    \end{bmatrix}
     \begin{bmatrix}
    \left(\bm \chi^{(0)}_-(\omega):\overset{(4)}{\bm D}\right)^{-1}
    -\bm 1 & -\bm 1\\ 
    -\bm 1 
    & -\left(\bm \chi^{(0)}_+(\omega):\overset{(4)}{\bm D}\right)^{-1} -\bm 1
    \end{bmatrix}^{-1} 
    \begin{bmatrix}
     :\overset{(4)}{\bm D}^{-1}:\overset{(3)}{\bm D} \\
     :\overset{(4)}{\bm D}^{-1}:\overset{(3)}{\bm D}
    \end{bmatrix} 
\end{equation}
now Eq.\  \eqref{formula for symbolic inversion sum of entries} comes in help since we just need the sum of the inverse tensor's entries
\begin{equation}
\begin{aligned}
     \boldsymbol{\mathcal{G}}(\omega)^{-1} & =\boldsymbol{\mathcal{G}}^{(0)}(\omega)^{-1}
    -
     \overset{(3)}{\bm D} :
     \left[\bm 1 - \bm \chi^{(0)}(\omega):\overset{(4)}{\bm D}\right]^{-1} :
    \left(\bm \chi^{(0)}(\omega):\overset{(4)}{\bm D} \right):\overset{(4)}{\bm D}^{-1}: \overset{(3)}{\bm D} \\
    & =\boldsymbol{\mathcal{G}}^{(0)}(\omega)^{-1} - \bm \Pi(\omega)
\end{aligned}
\end{equation}
where $\bm \Pi(\omega)$ coincides with the one presented in \cite{TDSCHA_monacelli,Bianco,LihmTDSCHA}. 

Now we derive the two phonons interacting Green's function $\bm \chi(\omega)$ which is obtained by choosing $\mathcal{A}$ and $\mathcal{B}$ according to Eq.\ \eqref{2 ph pertubation}. In this basis this means computing
\begin{equation}
\label{eq for two phonon response complicated}
    \bm \chi(\omega) = \begin{bmatrix}
    \bm 0 &
    \bm  S & \bm S
    \end{bmatrix} \cdot
     \boldsymbol{\mathcal{L}}(\omega)^{-1}\cdot
    \begin{bmatrix}
    \bm 0\\
    \bm  S \\
    \bm  S \\
    \end{bmatrix}.
\end{equation}

The perturbation $\mathcal{B}$ chosen (Eq.\ \eqref{2 ph pertubation}) leads to the following linearized equations of motion (see Eq.\ \eqref{linearized equation for Green function})
\begin{equation}
\label{linearized equations r s 1-2 ph GF}
     \boldsymbol{\mathcal{L}}(\omega)\cdot\begin{bmatrix}
    \widetilde{\boldsymbol{\mathcal{R}}}^{(1)}(\omega)\\ 
    \widetilde{\bm a}^{(1)}(\omega)\\ 
    \widetilde{\bm b}^{(1)}(\omega)
    \end{bmatrix} = 
    \begin{bmatrix}
    \bm 0\\
    \bm S \\
    \bm S
    \end{bmatrix}.
\end{equation}
Using the expression of $\boldsymbol{\mathcal{L}}(\omega)$, Eq.\ \eqref{L matrix def}, we find that the first free parameter is related to the other two
\begin{equation}
\label{R = g0 d3 r + s}
    \widetilde{\boldsymbol{\mathcal{R}}}^{(1)}(\omega) = \boldsymbol{\mathcal{G}}^{(0)}(\omega) \cdot \overset{(3)}{\bm D} : \left(\widetilde{\bm a}^{(1)}(\omega) +  \widetilde{\bm b}^{(1)}(\omega)\right).
\end{equation}
So instead of having to invert the full $\boldsymbol{\mathcal{L}}(\omega)$ we reduce Eq.\ \eqref{eq for two phonon response complicated} to
\begin{equation}
\small
\begin{aligned}
       \bm \chi(\omega) &=   \begin{bmatrix}
    \bm  S: & \bm S:
    \end{bmatrix} 
    \begin{bmatrix}
     +\left(\bm \chi^{(0)}_-(\omega)\right)^{-1}
    -\overset{(4)}{\bm D} -\overset{(3)}{\bm D}\cdot \boldsymbol{\mathcal{G}}^{(0)}(\omega) \cdot \overset{(3)}{\bm D}
    & -\overset{(4)}{\bm D} -\overset{(3)}{\bm D}\cdot \boldsymbol{\mathcal{G}}^{(0)}(\omega) \cdot \overset{(3)}{\bm D}\\ \\
    -\overset{(4)}{\bm D}-\overset{(3)}{\bm D}\cdot \boldsymbol{\mathcal{G}}^{(0)}(\omega) \cdot \overset{(3)}{\bm D} & -\left(\bm \chi^{(0)}_+(\omega)\right)^{-1} -
    \overset{(4)}{\bm D}-\overset{(3)}{\bm D}\cdot \boldsymbol{\mathcal{G}}^{(0)}(\omega) \cdot \overset{(3)}{\bm D}
    \end{bmatrix}^{-1} 
    \begin{bmatrix}
    :\bm  S \\
    :\bm  S \\
    \end{bmatrix} \\
    &=   \begin{bmatrix}
    \bm  S: & \bm S:
    \end{bmatrix} 
    \begin{bmatrix}
     +\left(\bm \chi^{(0)}_-(\omega)\right)^{-1}
    -\bm \Sigma(\omega)
    & -\bm \Sigma(\omega)\\ \\
    -\bm \Sigma(\omega) & -\left(\bm \chi^{(0)}_+(\omega)\right)^{-1} -\bm \Sigma(\omega)
    \end{bmatrix}^{-1} 
    \begin{bmatrix}
    :\bm  S \\
    :\bm  S \\
    \end{bmatrix} \\
    & = -\begin{bmatrix}
    \bm  S: & \bm S:
    \end{bmatrix} 
    \begin{bmatrix}
     +\bm 1-\left(\bm \chi^{(0)}_-(\omega):\bm \Sigma(\omega)\right)^{-1}
    & \bm 1\\ \\
    \bm 1 & \left(\bm \chi^{(0)}_+(\omega):\bm \Sigma(\omega)\right)^{-1} +
    \bm 1
    \end{bmatrix}^{-1}  
    \begin{bmatrix}
    :\bm \Sigma(\omega)^{-1} : \bm  S \\
    :\bm \Sigma(\omega)^{-1} : \bm  S \\
    \end{bmatrix}
\end{aligned}
\end{equation}
where we define the two phonon self-energy $\bm \Sigma(\omega)$ as in  Eq.\ \eqref{2 ph self energy}
\begin{equation}
    \bm \Sigma(\omega) = \overset{(4)}{\bm D} +\overset{(3)}{\bm D}\cdot \boldsymbol{\mathcal{G}}^{(0)}(\omega) \cdot \overset{(3)}{\bm D}.
\end{equation}
Again we can use Eq.\ \eqref{formula for symbolic inversion sum of entries} to get the two-phonon interacting Green's function
\begin{equation}
\begin{aligned}
    \bm \chi(\omega) & =  \bm S : \left[\bm 1 - \bm \chi^{(0)}(\omega) : \bm \Sigma(\omega)\right]^{-1} : \bm \chi^{(0)}(\omega): \bm \Sigma(\omega)  : \bm \Sigma^{-1}(\omega) : \bm S \\
     & =  \left[\bm 1 - \bm \chi^{(0)}(\omega) : \bm \Sigma(\omega)\right]^{-1} : \bm \chi^{(0)}(\omega) \\
     & = \bm \chi^{(0)}(\omega) : \left[\bm 1 - \bm \Sigma(\omega):\bm \chi^{(0)}(\omega) \right]^{-1}  
\end{aligned}
\end{equation}
proving Eq.\ \eqref{2 ph Green function}.

The last Green's function to discuss is the one-two phonon $\bm \Gamma(\omega)$ obtained setting $\mathcal{A}$ and $\mathcal{B}$ as in Eq.\ \eqref{1 2 ph pertubation}
\begin{equation}
\label{Gamma definition}
    \bm \Gamma(\omega) =
    \begin{bmatrix}
    \bm \delta\\
    \bm 0 \\
    \bm 0
    \end{bmatrix}\cdot
     \boldsymbol{\mathcal{L}}(\omega)^{-1}\cdot\begin{bmatrix}
    \bm 0\\
    \bm S \\
    \bm S
    \end{bmatrix}.
\end{equation}
We simplify this inversion using again Eq.\ \eqref{R = g0 d3 r + s}.
Now $\widetilde{\bm a}^{(1)}(\omega) +  \widetilde{\bm b}^{(1)}(\omega)$ are found considering the reduced linear system extracted from Eq.\ \eqref{linearized equations r s 1-2 ph GF}
\begin{equation}
\label{a' b' linear system}
\begin{aligned}
    & \bm \Sigma(\omega) : \begin{bmatrix}
     +\left(\bm \chi^{(0)}_-(\omega):\bm \Sigma(\omega)\right)^{-1}
    -\bm 1
    & -\bm 1\\ 
    -\bm 1 & -\left(\bm \chi^{(0)}_+(\omega):\bm \Sigma(\omega)\right)^{-1} -\bm 1
    \end{bmatrix}
    \begin{bmatrix}
     :\widetilde{\bm a}^{(1)}(\omega) \\ 
     :\widetilde{\bm b}^{(1)}(\omega)
    \end{bmatrix}
    = \begin{bmatrix}
     \bm S \\
     \bm S
    \end{bmatrix} \\
    &
    \begin{bmatrix}
     \widetilde{\bm a}^{(1)}(\omega) \\ 
     \widetilde{\bm b}^{(1)}(\omega)
    \end{bmatrix}
    = -\begin{bmatrix}
     \bm 1-\left(\bm \chi^{(0)}_-(\omega):\bm \Sigma(\omega)\right)^{-1}
    & \bm 1\\ 
    \bm 1 &  \bm 1 +\left(\bm \chi^{(0)}_+(\omega):\bm \Sigma(\omega)\right)^{-1} 
    \end{bmatrix}^{-1}\begin{bmatrix}
     :\bm \Sigma^{-1}(\omega) \\ 
     :\bm \Sigma^{-1}(\omega)
    \end{bmatrix} \\
\end{aligned}
\end{equation}
From Eq.\ \eqref{a' b' linear system} we get the sum $\widetilde{\bm a}^{(1)}(\omega) + \widetilde{\bm b}^{(1)}(\omega)$
\begin{equation}
\begin{aligned}
    \widetilde{\bm a}^{(1)}(\omega) + \widetilde{\bm b}^{(1)}(\omega) = -\begin{bmatrix}
    \bm S: & \bm S:
    \end{bmatrix}\begin{bmatrix}
     \bm 1-\left(\bm \chi^{(0)}_-(\omega):\bm \Sigma(\omega)\right)^{-1}
    & \bm 1\\ \\
    \bm 1 &  \bm 1 +\left(\bm \chi^{(0)}_+(\omega):\bm \Sigma(\omega)\right)^{-1} 
    \end{bmatrix}^{-1}\begin{bmatrix}
     :\bm S : \bm \Sigma^{-1}(\omega) \\ \\
     :\bm S : \bm \Sigma^{-1}(\omega)
    \end{bmatrix} 
    .
\end{aligned}
\end{equation}
Again Eq.\ \eqref{formula for symbolic inversion sum of entries} comes in help so
\begin{equation}
\label{a' + b' = chi}
     \widetilde{\bm a}^{(1)}(\omega) +
     \widetilde{\bm b}^{(1)}(\omega) = \left[\bm 1 -\bm \chi^{(0)}(\omega):\bm \Sigma(\omega)\right]^{-1}: \bm \chi^{(0)}(\omega):\bm \Sigma(\omega) :\bm \Sigma^{-1}(\omega)  = \bm \chi(\omega)
\end{equation}
Since the response vector $\responsevector $ is just $ [\bm \delta\quad \bm 0 \quad \bm 0]$ the one-two phonon Green's function $\bm \Gamma(\omega)$ is given by $\boldsymbol{\mathcal{R}}^{(1)}(\omega)$ (see Eq.\ \eqref{A 1 in R a' b'}), so, with Eq.\ \eqref{R = g0 d3 r + s}, we end up with
\begin{equation}
    \bm \Gamma(\omega) = \boldsymbol{\mathcal{R}}^{(1)}(\omega) = \boldsymbol{\mathcal{G}}^{(0)}(\omega) \cdot \overset{(3)}{\bm D} : \bm \chi(\omega).
\end{equation}
This proves Eq.\ \eqref{1 2 ph Green function}.

We discuss also the three phonon Green's function obtained with $\mathcal{A}$
and $\mathcal{B}$ as in Eq.\ \eqref{three phonon A B}
\begin{equation}
    \mathcal{A} = \delta\widetilde{R}^{(0)}_{\alpha} \delta\widetilde{R}^{(0)}_{\beta} \delta\widetilde{R}^{(0)}_{\gamma} \qquad
    \mathcal{B} = \delta\widetilde{R}^{(0)}_{\alpha'}\delta\widetilde{R}^{(0)}_{\beta'}\delta\widetilde{R}^{(0)}_{\gamma'}
\end{equation}
In this case we have
\begin{equation}
    \left\langle \frac{\partial^2 \mathcal{A}}{\partial \widetilde{\bm R} \partial \widetilde{\bm R}}\right\rangle_{(0)} = 
    \left\langle \frac{\partial^2 \mathcal{B}}{\partial \widetilde{\bm R} \partial \widetilde{\bm R}}\right\rangle_{(0)} = \bm 0
\end{equation}
and
\begin{equation}
    \left\langle \frac{\partial \mathcal{A}}{\partial\widetilde{R}_\mu }\right\rangle_{(0)} = 
     \delta_{\mu\alpha}\left(\widetilde{\alpha}^{(0)}\,\!^{-1}\right)_{\beta\gamma}
    +\delta_{\mu\beta}\left(\widetilde{\alpha}^{(0)}\,\!^{-1}\right)_{\alpha\gamma}
    +\delta_{\mu\gamma}\left(\widetilde{\alpha}^{(0)}\,\!^{-1}\right)_{\alpha\beta}
\end{equation}
so only the first entries of $\bm r'$ and $\bm p'$ are non zero. The response calculation is formally identical to the one-phonon interacting Green's function one Eq.\ \eqref{appendix one ph int green function}.
Using, Eq.\ \eqref{response formula new basis appendix}, we get the three phonon propagator
\begin{equation}
\begin{aligned}
&\chi_{\text{3ph}}(\omega) = \sum_{\mu\nu=1}^{3N} 
\left(\delta_{\mu\alpha}\left(\widetilde{\alpha}^{(0)}\,\!^{-1}\right)_{\beta\gamma}
+\delta_{\mu\beta}\left(\widetilde{\alpha}^{(0)}\,\!^{-1}\right)_{\alpha\gamma}
+\delta_{\mu\gamma}\left(\widetilde{\alpha}^{(0)}\,\!^{-1}\right)_{\alpha\beta}\right)\mathcal{G}(\omega)_{\mu\nu} \\
&
\left(\delta_{\nu\alpha'}\left(\widetilde{\alpha}^{(0)}\,\!^{-1}\right)_{\beta'\gamma'}
+\delta_{\nu\beta'}\left(\widetilde{\alpha}^{(0)}\,\!^{-1}\right)_{\alpha'\gamma'}
+\delta_{\nu\gamma'}\left(\widetilde{\alpha}^{(0)}\,\!^{-1}\right)_{\alpha'\beta'}\right) 
\end{aligned}
\end{equation}
The diagrammatic interpretation is straightforward once we use Eq.\ \eqref{flower definition SCHA}. The three-phonon response is
\begin{equation}
\begin{aligned}
   \chi_{\text{3ph}}(\omega) =  &  \mathcal{G}^{(0)}(t = 0^-)_{\beta\gamma}
    \mathcal{G}^{(0)}(t = 0^-)_{\beta'\gamma'}\mathcal{G}(\omega)_{\alpha\alpha'} \\
    &
    +\text{permutations of $(\alpha\beta\gamma)$ and $(\alpha'\beta'\gamma')$ separately}
\end{aligned}
\end{equation}
This proves the diagrammatic expression of Fig.\ \ref{fig:bunny}.

\textcolor{black}{In closing, we present the proof of Eqs \eqref{G chi Gamma Theta compact}. The Dyson equations for the one and two-phonon Green's functions in TDSCHA are respectively
\begin{equation} 
    \label{one ph green appendix}
    \boldsymbol{\mathcal{G}}(\w)^{-1} = \boldsymbol{\mathcal{G}}^{(0)}(\w)^{-1} - \overset{(3)}{\bm D} :\left(\bm 1 -\bm \chi^{(0)}(\omega) :\overset{(4)}{\bm D}\right)^{-1} \hspace{-0.2cm}: \bm \chi^{(0)}(\omega):\overset{(3)}{\bm D}
\end{equation}
and
\begin{equation}
    \bm \chi (\w)^{-1} =  \bm \chi ^{(0)}(\w) ^{-1} - {\bDfour} - \bDthree \cdot \boldsymbol{\mathcal{G}}^{(0)}(\w) \cdot \bDthree \:.
\end{equation}
We rewrite the above definitions using the partially screened two-phonon propagator, defined as
\begin{equation}
    \bm \Theta(\w)^{-1} = \bm \chi ^{(0)}(\w) ^{-1} - \bDfour .
\end{equation}
One sees immediately that the full two-phonon propagator is
\begin{equation} \label{two ph green appendix}
     \bm \chi (\w)^{-1} =  \bm \Theta(\w)^{-1}-\bDthree \cdot \boldsymbol{\mathcal{G}}^{(0)}(\w) \cdot \bDthree  
\end{equation}
and that the one phonon propagator is
\begin{equation}
\begin{split}
    \boldsymbol{\mathcal{G}}(\w)^{-1} & = \boldsymbol{\mathcal{G}}^{(0)}(\w)^{-1} - \overset{(3)}{\boldsymbol{D}} :
    \left(\boldsymbol{\chi}^{(0)}(\omega)^{-1} 
    -\overset{(4)}{\boldsymbol{D}}\right)^{-1} \hspace{-0.2cm}
    : \overset{(3)}{\boldsymbol{D}}= 
    \boldsymbol{\mathcal{G}}^{(0)}(\w)^{-1}
    - \overset{(3)}{\boldsymbol{D}} :\bm \Theta(\w): \overset{(3)}{\boldsymbol{D}} .
\end{split}
\end{equation}
Below we show that the two phonon propagator Eq.\ \eqref{two ph green appendix} is given by
\begin{equation} 
\label{chi = w + w g w appendix}
    \bm \chi (\w) = \bm \Theta(\w) + \bm \Theta(\w) : \bDthree \cdot  \boldsymbol{\mathcal{G}}(\w) \cdot \bDthree : \bm \Theta(\w) \:.
\end{equation}
We express the one phonon propagator as
\begin{equation}
    \boldsymbol{\mathcal{G}}(\omega) = 
    \left(\boldsymbol{\mathcal{G}}^{(0)}(\omega)^{-1}
    - \overset{(3)}{\boldsymbol{D}} 
    :\boldsymbol{\Theta}(\omega):  \overset{(3)}{\boldsymbol{D}} 
    \right)^{-1}
     = \left\{\overset{(3)}{\bm{D}}:\left[\left(\overset{(3)}{\bm{D}}\cdot\boldsymbol{\mathcal{G}}^{(0)}(\omega){}\cdot\overset{(3)}{\bm{D}}\right)^{-1} - \bm{\Theta}(\omega) \right]  :\overset{(3)}{\bm{D}}\right\}^{-1}
\end{equation}
and plugging the latter expression in Eq.\ \eqref{chi = w + w g w appendix} we get
\begin{subequations}
\begin{align}
    \bm \chi(\omega)  &= \bm \Theta(\omega) + \bm \Theta(\omega) :\left[\left(\bDthree\cdot\boldsymbol{\mathcal{G}}^{(0)}(\omega){}\cdot\bDthree\right)^{-1} - \bm \Theta(\omega) \right]^{-1} : \bm \Theta(\omega)\\
    &= \bm \Theta(\omega) + \bm \Theta(\omega) :\left(\bm 1 - \bDthree\cdot\boldsymbol{\mathcal{G}}^{(0)}(\omega)\cdot\bDthree:\bm \Theta(\omega) \right)^{-1} :\bDthree\cdot\boldsymbol{\mathcal{G}}^{(0)}(\omega)\cdot\bDthree: \bm \Theta(\omega)\\
    & = \bm \Theta(\omega) +  \left(\bm \Theta(\omega)^{-1} - \bDthree\cdot\boldsymbol{\mathcal{G}}^{(0)}(\omega)\cdot\bDthree\right)^{-1} :\bDthree\cdot\boldsymbol{\mathcal{G}}^{(0)}(\omega)\cdot\bDthree: \bm \Theta(\omega) \\
    &= \bm \Theta(\omega) +  \bm \chi(\omega) :\bDthree\cdot\boldsymbol{\mathcal{G}}^{(0)}(\omega)\cdot\bDthree: \bm \Theta(\omega)  .
\end{align}
\end{subequations}
Now moving the last term on the left-hand side and inverting we  get
\begin{subequations}
\begin{align}
    & \bm \chi(\omega) :(\bm 1 - \bDthree\cdot\boldsymbol{\mathcal{G}}^{(0)}(\omega)\cdot\bDthree: \bm \Theta(\omega) ) = \bm \Theta(\omega),\\
    & \bm \chi(\omega)  = \bm \Theta(\omega) :\left(\bm 1 - \bDthree\cdot\boldsymbol{\mathcal{G}}^{(0)}(\omega)\cdot\bDthree: \bm \Theta(\omega) \right)^{-1} = \left(\bm \Theta(\omega)^{-1} - \bDthree\cdot\boldsymbol{\mathcal{G}}^{(0)}(\omega)\cdot\bDthree\right)^{-1},
\end{align}
\end{subequations}
and finally, we recover the standard expression for the two phonon propagator
\begin{equation}
    \boldsymbol{\chi}(\w) 
    = \left(\boldsymbol{\chi}^{(0)}(\w)^{-1} - \bDfour - \bDthree \cdot 
    \boldsymbol{\mathcal{G}}^{(0)}(\w) \cdot \bDthree  \right)^{-1} 
\end{equation}
which proves that Eq.\ \eqref{two ph green appendix} and Eq.\ \eqref{chi = w + w g w appendix} are equivalent expressions of the anharmonic two-phonon propagator.}

\section{Scattering vertices}
\label{Scattering vertices}
In this Appendix, we present the diagrammatic expression of the scattering vertices in TDSCHA, Eqs \eqref{def D3 main} \eqref{def D4 main}. We consider first the three-phonon term Eq.\ \eqref{def D3 main} since the same holds for Eq.\ \eqref{def D4 main}.

We average the third-derivative of the BO potential on the equilibrium SCHA distribution $\widetilde{\rho}^{(0)}(\bm R)$ (see Eq.\ \eqref{rho(R,P) eq with alpha beta})
\begin{equation}
\label{D3 def app}
    \overset{(3)}{D}_{ijk} = \int d\bm R  \widetilde{\rho}^{(0)}(\bm R)\frac{\partial^3 V^{(\text{BO})}(\bm R)}{\partial \widetilde{R}_{i} \partial \widetilde{R}_{j} \partial \widetilde{R}_{k}} .
\end{equation}
Starting from Eq.\ \eqref{D3 def app} we perform the change of variables $\widetilde{u}_a = \widetilde{R}_a - \widetilde{\mathcal{R}}^{(0)}_a$
and we expand in $\bm u$
\begin{equation}
\label{D3 def app bis}
  \overset{(3)}{D}_{ijk} 
    =\int d\bm u \widetilde{\rho}^{(0)}(\bm u)\left[\sum_{n=0}^{+\infty} \frac{1}{n!}\sum_{a_1..a_n}\overset{(3 + n)}{D^{(0)}}_{ijka_1..a_{n}}  \widetilde{u}_{a_1}...\widetilde{u}_{a_n}\right],
\end{equation}
where $\overset{(n)}{\bm D^{(0)}}$ is defined in Eq.\ \eqref{bare anharmonic tensor}.
Note that $\overset{(n)}{\bm D^{(0)}}$ differs in general from $ \overset{(n)}{\bm d}$, see Eq.\ \eqref{anh vertex on harm position}, since the minimum of the Born-Oppenheimer potential $\bm{\mathcal{R}}_{\text{BO}}$ does not coincide with the SCHA centroid $\bm{\mathcal{R}}^{(0)}$.

Only even terms in Eq.\ \eqref{D3 def app bis} are non-zero 
\begin{equation}
\begin{aligned}
    \overset{(3)}{D}_{ijk} 
    =& \sum_{n=0}^{+\infty} \frac{1}{(2n)!}\sum_{a_1..a_{2n}}\overset{(3 + 2n)}{D^{(0)}}_{ijka_1..a_{2n}} \left\langle \widetilde{u}_{a_1}...\widetilde{u}_{a_{2n}}\right\rangle_{(0)}
    \\
     = & \sum_{n=0}^{+\infty} \frac{1}{(2n)!}\sum_{a_1..a_{2n}}\overset{(3 + 2n)}{D^{(0)}}_{ijka_1..a_{2n}} \sum_{P} P\left[\left(\widetilde{\alpha}^{(0)}\,\!^{-1}\right)_{a_1a_2}...\left(\widetilde{\alpha}^{(0)}\,\!^{-1}\right)_{a_{2n-1}a_{2n}}\right] \\
     =& \sum_{n=0}^{+\infty} \frac{1}{2^{n}n!}\sum_{a_1..a_{2n}}\overset{(3 + 2n)}{D^{(0)}}_{ijka_1..a_{2n}} \left(\widetilde{\alpha}^{(0)}\,\!^{-1}\right)_{a_1a_2}...\left(\widetilde{\alpha}^{(0)}\,\!^{-1}\right)_{a_{2n-1}a_{2n}} 
\end{aligned}
\end{equation}
where $P$ denotes the permutations of the indices according to the Wick theorem. In the last line, we use the symmetry properties of the anharmonic vertices and the fact that the number of contractions for a $2n$ multivariate Gaussian expectation value is $(2n -1)!!$, where $!!$ is the double factorial. In polarization the final result is
\begin{equation}
\label{diagramm expr for d3}
    \overset{(3)}{D}_{\mu\nu\phi} = \sum_{n=0}^{+\infty} \frac{(-1)^n}{2^{n}n!} \sum_{\alpha_1..\alpha_{2n}}\overset{(3 + 2n)}{D^{(0)}}_{\mu\nu\phi\alpha_1..\alpha_{2n}}\underbrace{\mathcal{G}^{(0)}(t=0^-)_{\alpha_1\alpha_2}...\mathcal{G}^{(0)}(t=0^-)_{\alpha_{2n-1}\alpha_{2n}} }_{\text{n}},
\end{equation}
where we use $\widetilde{\bm \alpha}^{(0)}\,\!^{-1} = \left\langle \delta \widetilde{\bm R} \delta \widetilde{\bm R} \right\rangle_{(0)}$, see Eq.\ \eqref{alpha static R-R} and Eq.\ \eqref{flower definition SCHA}. The same holds for the fourth-order scattering vertex
\begin{equation}
\label{diagramm expr for d4}
    \overset{(4)}{D}_{\mu\nu\phi\psi} = \sum_{n=0}^{+\infty} \frac{(-1)^n}{2^{n}n!}\sum_{\alpha_1..\alpha_{2n}}\overset{(4 + 2n)}{D^{(0)}}_{\mu\nu\phi\psi\alpha_1..\alpha_{2n}}
     \underbrace{\mathcal{G}^{(0)}(t=0^-)_{\alpha_1\alpha_2}...\mathcal{G}^{(0)}(t=0^-)_{\alpha_{2n-1}\alpha_{2n}} }_{\text{n}}.
\end{equation}
Eq.\ \eqref{diagramm expr for d3} Eq.\ \eqref{diagramm expr for d4} give a diagrammatic expression for the TDSCHA scattering vertices, see Fig. \ref{fig:d3 d4 flower diagrams}.

\section{Momentum Green's function}
\label{Momentum Green function}
In this Appendix, we discuss the momentum Green's function using the many-body formalism for bosons.
The interacting Green's function with imaginary time $\tau\in [-\beta, +\beta]$ ($\beta^{-1}=k_b T$ with $k_b$ the Boltzmann constant) is defined as
\begin{equation}
    G^{AB}(\tau) = -\left\langle T_\tau \left(\hat{S}(\beta,0)\hat{A}(\tau) \hat{B}(0)\right)\right\rangle_0
\end{equation}
where only the connected diagrams are included. The average $\left\langle ... \right\rangle_{0}$ is performed on the harmonic system defined by
\begin{equation}
    \hat{H}_{\text{harm}} = \sum_{\mu=1}^{3N}\hbar\Omega_\mu \left(\hat{a}^\dagger_\mu \hat{a}_\mu + \frac{1}{2}\right),
\end{equation}
where $\{\Omega^2_\mu\}$ are the harmonic frequencies, i.e. the poles of the harmonic propagator Eq.\ \eqref{static harm prop}.
The scattering matrix is
\begin{equation}
    \hat{S}(\tau) = \hat{S}(\tau,0) =T_\tau e^{-\int_0^\tau d\tau' \hat{H}_{\text{anh}}(\tau')},
\end{equation}
here $\hat{H}_{\text{anh}}(\tau)$ is the anharmonic part of the BO energy surface in the interacting picture. 
The Matsubara transform is
\begin{equation}
    G^{AB}(i\Omega_n) = \frac{1}{2}\int_{-\beta}^{+\beta} d\tau e^{i\Omega_n \tau} G^{AB}(\tau)
\end{equation}
with $\Omega_n = \frac{2\pi n}{\beta}$ with $n$ integer.

First, we define the harmonic (non-interacting) Green's function for position and momentum. In the harmonic polarization basis we have
\begin{equation}
  \delta   \hat{\widetilde{R}}(\tau)_\mu  =  \hat{\widetilde{R}}(\tau)_\mu -  \widetilde{\mathcal{R}}_\mu  =\sqrt{\frac{\hbar}{2\Omega_\mu}}\left[\hat{a}(\tau)_\mu + \hat{a}^\dagger(\tau)_\mu\right] \qquad
    \hat{\widetilde{P}}(\tau)_\mu  = -i\sqrt{\frac{\hbar\Omega_\mu}{2}}\left[\hat{a}(\tau)_\mu - \hat{a}^\dagger(\tau)_\mu\right]
\end{equation}
The Green's functions in Matsubara frequencies are
\begin{subequations}
\label{GF omega}
\begin{align}
    & {G^{(0)}}^{RR}_{\mu\nu}(i\Omega_n)  = \delta_{\mu\nu}\frac{\hbar^2}{(i\Omega_n)^2 - (\hbar\Omega_\mu)^2} \label{many body G rr appendix}\\
    & {G^{(0)}}^{PP}_{\mu\nu}(i\Omega_n) = \Omega^2_\mu {G^{(0)}}^{RR}_{\mu\nu}(i\Omega_n) \\
    & {G^{(0)}}^{PR}_{\mu\nu}(i\Omega_n) = i\delta_{\mu\nu}\hbar\frac{i\Omega_n}{(i\Omega_n)^2 - (\hbar\Omega_\mu)^2} = \frac{i}{\hbar}(i\Omega_n){G^{(0)}}^{RR}_{\mu\nu}(i\Omega_n)  \label{GF PR}\\
    & {G^{(0)}}^{RP}_{\mu\nu}(i\Omega_n) =   {G^{(0)}}^{PR}_{\mu\nu}(-i\Omega_n) = -\frac{i}{\hbar}(i\Omega_n){G^{(0)}}^{RR}_{\mu\nu}(i\Omega_n) \label{GF RP}.
\end{align}
\end{subequations}
Note that the analytical continuation of ${G^{(0)}}^{RR}_{\mu\nu}(i\Omega_n)$ gives
\begin{equation}
        G^{(0)}\,\!^{RR}_{\mu\nu}(i\Omega_n \rightarrow \hbar\omega + i0^+) = \frac{\delta_{\mu\nu}}{(\omega + i0^+)^2 - \Omega_\mu^2}
\end{equation}
which coincides with our definition of harmonic free propagators, see Eq.\ \eqref{static harm prop}.
The interacting momentum Green's function is
\begin{equation}
\begin{aligned}
     & G^{PP}_{\mu\nu}(\tau) = -\left\langle T_{\tau}\left(\hat{S}(\beta,0) \hat{\widetilde{P}}(\tau)_{\mu} \hat{\widetilde{P}}_{\nu}(0)\right) \right\rangle_0 = {G^{(0)}}^{PP}_{\mu\nu}(\tau) -\left\langle T_{\tau}\left(\hat{S}_{\text{anh}}(\beta,0) \hat{\widetilde{P}}(\tau)_{\mu} \hat{\widetilde{P}}_{\nu}(0)\right) \right\rangle_0
\end{aligned}
\end{equation}
where $\hat{S}_{\text{anh}}(\beta,0) = \hat{S}(\beta,0) - \hat{1}$. The anharmonic correction is proportional to terms like
\begin{equation}
\label{S anh}
     \int_{0}^{\beta} d\tau_1 .. \int_{0}^{\beta} d\tau_m \left\langle T_{\tau}\left( \hat{\widetilde{P}}(\tau)_{\mu} \hat{\widetilde{R}}(\tau_1)_{\alpha_1}
     \hat{\widetilde{R}}(\tau_1)_{\alpha_2}.. . \hat{\widetilde{R}}(\tau_m)_{\alpha_m} \hat{\widetilde{P}}_{\nu}(0)\right) \right\rangle_0
\end{equation}
where all the indices, except for $\mu\nu$, will be contracted with anharmonic vertices contained in the full BO energy surface.
Eq.\ \eqref{S anh} is computed using the Wick theorem and contains terms that have the following form
\begin{equation}
\label{intermediate GF interacting tau}
\begin{aligned}
     &\int_{0}^{\beta} d\tau_1 .. \int_{0}^{\beta} d\tau_m \left\langle T_{\tau}\left( \hat{\widetilde{P}}(\tau)_{\mu} \hat{\widetilde{R}}(\tau_1)_{\alpha_1}\right) \right\rangle_0
     \left\langle T_{\tau}\left( \hat{\widetilde{R}}(\tau_1)_{\alpha_2} \hat{\widetilde{R}}(\tau_2)_{\alpha_3}\right) \right\rangle_0 ...
     \left\langle T_{\tau}\left( \hat{\widetilde{R}}(\tau_m)_{\alpha_m}\hat{\widetilde{P}}(0)_{\nu} \right) \right\rangle_0
      \\
    & = \int_{0}^{\beta} d\tau_1 .. \int_{0}^{\beta} d\tau_m
    {G^{(0)}}^{PR}_{\mu\alpha_1}(\tau-\tau_1)
    {G^{(0)}}^{RR}_{\alpha_2\alpha_3}(\tau_1 -\tau_2)
      ...G^{(0)}\,\!^{RP}_{\alpha_m\nu}(\tau_m)
\end{aligned}
\end{equation}
When doing the contraction of the momentum variables we use ${G^{(0)}}^{PR}_{\mu\nu}(\tau) = {G^{(0)}}^{RP}_{\mu\nu}(-\tau)$ and we take into account the multiplicity of the diagrams which cancels the $n!$ coming from the scattering matrix.

Knowing that the Matsubara frequencies are conserved in all the diagrams and the relation between ${G^{(0)}}^{RP/PR}_{\mu\nu}(i\Omega_n)$ and ${G^{(0)}}^{RR}_{\mu\nu}(i\Omega_n)$ (Eq.\ \eqref{GF omega}), Eq.\ \eqref{intermediate GF interacting tau} becomes simply proportional to the anharmonic correction of the one-phonon Green's function
\begin{equation}
\begin{aligned}
    {G^{(0)}}^{PR}_{\mu\alpha_1}(i\Omega_n) \pi(i\Omega_n)_{\alpha_1 .. \alpha_{m}} {G^{(0)}}^{RP}_{\alpha_m\nu}(i\Omega_n)  = &(i\Omega_n)^2 {G^{(0)}}^{RR}_{\mu\alpha_1}(i\Omega_n) \pi(i\Omega_n)_{\alpha_1 .. \alpha_{m}} 
    {G^{(0)}}^{RR}_{\alpha_m\nu}(i\Omega_n) 
    \deleted{\\
    = & (i\Omega_n)^2\left[G^{RR}_{\mu\nu}(i\Omega_n)  - {G^{(0)}}^{RR}_{\mu\nu}(i\Omega_n)\right]}
\end{aligned}
\end{equation}
where $\pi(i\Omega_n)_{\alpha_2 .. \alpha_{m-1}}$ is the Matsubara transform of the terms that contain only products of ${G^{(0)}}^{RR}_{\alpha_i\alpha_j}(\tau_i - \tau_j)$ in Eq.\ \eqref{intermediate GF interacting tau}.

In the end, using Eqs \eqref{GF PR} \eqref{GF RP}, we get the following result for the interacting momentum Green's function
\begin{equation}
\begin{aligned}
     G^{PP}_{\mu\nu}(i\Omega_n)   &=  {G^{(0)}}^{PP}_{\mu\nu}(i\Omega_n) + \frac{(i\Omega_n)^2}{\hbar^2
     }\left[G^{RR}_{\mu\nu}(i\Omega_n)  - {G^{(0)}}^{RR}_{\mu\nu}(i\Omega_n)\right] \\
      & = \Omega^2_\mu {G^{(0)}}^{RR}_{\mu\nu}(i\Omega_n) + \frac{(i\Omega_n)^2}{\hbar^2
      }\left[G^{RR}_{\mu\nu}(i\Omega_n)  - {G^{(0)}}^{RR}_{\mu\nu}(i\Omega_n)\right] \\
       & = -\delta_{\mu\nu} + \frac{(i\Omega_n)^2}{\hbar^2
      }G^{RR}_{\mu\nu}(i\Omega_n)
\end{aligned}
\end{equation}
Performing the analytical continuation $i\Omega_n \rightarrow \hbar\omega + i0^+$ and using the TDSCHA one-phonon Green's function, Eq.\ \eqref{1 ph Green function}, we prove Eq.\ \eqref{momentum Green function TDSCHA}.

\section{Prepare IR/Raman spectra calculation}
\label{Prepare IR/Raman spectra calculation}
To compute IR spectra we need Eqs \eqref{Ir 1 component} \eqref{Ir 2 component} in polarization basis Eq.\ \eqref{SCHA phonons}.
The first component of the response/perturbation vector is the one phonon vertex and contains equilibrium averages of the effective charges
\begin{equation}
    \overline{Z}_{\mu,a\alpha} = \left\langle \frac{\partial  \mathrm{p}_{\mu}(\bm R)}{\partial R_{a\alpha} } \right \rangle_{(0)} =  \left\langle Z^*(\bm R)_{\mu,a\alpha} \right\rangle_{(0)},
\end{equation}
where $\mu$ indicates the direction of the electric field, $R_{a,\alpha}$ is the position of atom $a$ along the $\alpha$ coordinate and $\bm Z^*(\bm R)$ is the effective charges tensor for a given configuration $\bm R$. 
The second and third components of the response/perturbation vector contain the two-phonon vertex, i.e. first derivatives of the effective charges. Integration by parts leads to
\begin{equation}
    \overline{\overline{Z}}_{\mu,a\alpha,b\beta} =\left\langle \frac{\partial^2  \mathrm{p}_{\mu}(\bm R)}{\partial R_{a\alpha}  \partial R_{b\beta}} \right \rangle_{(0)} 
    =  \sum_{c=1}^N\sum_{\gamma=1}^3 \alpha^{(0)}_{b\beta,c\gamma} \left\langle \delta R^{(0)}_{c\gamma}  \left(Z^*(\bm R)_{\mu,a\alpha}  - Z^*(\boldsymbol{\mathcal{R}}^{(0)})_{\mu,a\alpha} \right) \right\rangle_{(0)}. 
\end{equation}
We subtract the equilibrium effective charges to reduce the noise in the average.

To compute Raman spectra we need Eqs \eqref{RAMAN 1 component} \eqref{RAMAN 2 component}.
The first component of the response/perturbation vector contains equilibrium averages of the Raman tensor which give one-phonon processes
\begin{equation}
\begin{aligned}
    & \overline{\Xi}_{\mu,\nu,a\alpha}=\left\langle \frac{\partial \alpha(\bm R)_{\mu\nu}}{\partial R_{a\alpha} } \right \rangle_{(0)} = \left\langle \Xi(\bm R)_{\mu\nu,a\alpha} \right\rangle_{(0)},
\end{aligned}
\end{equation}
where $\mu$ $\nu$ indicates the photon polarization and $\bm \Xi(\bm R)$ is the Raman tensor for a configurations $\bm R$.
The two phonon channel depends on the Raman tensor first derivatives and using integration by parts we have
\begin{equation}
    \overline{\overline{\Xi}}_{\mu,\nu,a\alpha,b\beta} = \left\langle \frac{\partial^2 \alpha(\bm R)_{\mu\nu}}{\partial R_{a\alpha} \partial R_{b\beta} } \right \rangle_{(0)} = \sum_{c=1}^N\sum_{\gamma=1}^3 \alpha^{(0)}_{b\beta,c\gamma } \left\langle \delta R^{(0)}_{c\gamma} \left(\Xi(\bm R)_{\mu,\nu,a\alpha} -
    \Xi(\boldsymbol{\mathcal{R}}^{(0)})_{\mu,\nu,a\alpha} 
    \right) \right\rangle_{(0)}.
\end{equation}
So to prepare the response and perturbation vector $\responsevector$ and $\perturbationvector$ we can use a stochastic approach as in \cite{SCHA_main} since all the averages have to be done on the equilibrium ensemble.

We can enforce symmetries both for effective charges/Raman tensors and for their second-order counterparts.  To symmetrize $\overline{\bm Z}$ we note that the dipole $\bm p$ is related to the effective charge
\begin{equation}
    p_\mu = \sum_{a=1}^{N} \sum_{\alpha=1}^3\overline{Z}_{\mu,a\alpha} u_{a\alpha}
\end{equation}
where $u_{a\alpha}$ is a displacement of atom $a$ in the direction $\alpha$. If we apply a symmetry on $\bm u$ (defined in the supercell), the dipole will change according to the symmetry $\bm \sigma$ ($3\times 3$ unitary matrix)
\begin{equation}
\label{symm transf of P}
    \sum_{\nu=1}^3 \sigma_{\mu\nu}p_{\nu} = \sum_{ab=1}^{N}\sum_{\alpha\beta=1}^3\overline{Z}_{\mu,a\alpha}  S^\sigma_{a\alpha,b\beta} u_{b\beta}
\end{equation}
where $\bm S^\sigma$ ($3N\times 3N$ matrix) is the symmetry operation associated with $\bm \sigma$ in the supercell
\begin{equation}
    S^\sigma_{a\alpha,b\beta} = \sigma_{\alpha\beta}\delta_{a\sigma(b)}.
\end{equation}
$j=\sigma(i)$ indicates that the symmetry $\sigma$ maps $i$ into $j$.
So using Eq.\ \eqref{symm transf of P} we get
\begin{equation}
    \overline{Z}_{\mu',a\alpha'} =  \frac{1}{N_{\text{s}}} \sum_{\sigma=1}^{N_{\text{s}}} \sum_{\mu\alpha=1}^{3} \left(\sigma^\dagger_{\mu'\mu}\overline{Z}_{\mu,\sigma(a)\alpha} \sigma_{\alpha\alpha'}\right) 
\end{equation}
where $N_\text{s}$ is the number of symmetries. The symmetries for the second-order dipole moment are extracted noting that
\begin{equation}
    P_\mu = \sum_{ab=1}^{N} \sum_{\alpha\beta=1}^3 \overline{\overline{Z}}_{\mu,a\alpha,b\beta} u_{a\alpha} u_{b\beta} .
\end{equation}
Since we know how the effective charges transform under a symmetry operation we can symmetrize $\overline{\overline{\bm Z}}$
\begin{equation}
    \overline{\overline{Z}}_{\nu',a\alpha',b\beta'} = \frac{1}{N_\text{s}}\sum_{\sigma=1}^{N_\text{s}}\sum_{\nu=1}^{3}\sum_{\alpha\beta=1}^3
    \left(\sigma^\dagger_{\nu'\nu}
    \overline{\overline{Z}}_{\nu,\sigma(a)\alpha,\sigma(b)\beta} \sigma_{\alpha\alpha'} \sigma_{\beta\beta'} \right).
\end{equation}

We do the same for the Raman tensors. Similarly to what we do before,  the polarizability $\bm \alpha$ is related to the Raman tensors
\begin{equation}
    \alpha_{\mu,\nu} = \sum_{a=1}^N \sum_{\alpha=1}^3\overline{\Xi}_{\mu,\nu,a\alpha}  u_{a,\alpha},
    \qquad
    \alpha_{\mu,\nu} = \sum_{ab=1}^N \sum_{\alpha\beta=1}^3\overline{\overline{\Xi}}_{\mu,\nu,a\alpha,b\beta}  u_{a,\alpha}  u_{b,\beta}
\end{equation}
and we end up with the rules to symmetrize the averages of Raman-tensors
\begin{subequations}
\begin{align}
    \overline{\Xi}_{\chi,\phi,m\mu'} =& \frac{1}{N_\text{s}}\sum_{\sigma=1}^{N_\text{s}}
    \left(\sum_{\alpha\beta=1}^3 \sum_{\nu'=1}^3\sigma_{\chi\alpha} \sigma_{\phi\beta}  \overline{\Xi}_{\alpha,\beta,\sigma(m),\nu'} \sigma_{\nu'\mu'}   \right), \\
    \overline{\overline{\Xi}}_{\chi,\phi,p\pi,r\rho} =& \frac{1}{N_\text{s}}\sum_{\sigma=1}^{N_{\text{s}}}\left(\sum_{\alpha\beta=1}^3\sum_{\mu\nu=1}^3\sigma_{\chi\alpha}\sigma_{\phi\beta}\overline{\overline{\Xi}}_{\alpha,\beta,\sigma(p),\mu,\sigma(r),\nu}
    \sigma_{\nu\rho} \sigma_{\mu\pi}
    \right).
\end{align}
\end{subequations}

\end{widetext}

\bibliography{apssamp}

\end{document}